%% file: main.tex
\makeatletter\@ifundefined{ifanonymous}{\newif\ifanonymous\anonymousfalse}{}\makeatother
\documentclass[11pt,a4paper]{article}
\usepackage{labarxiv}
\input{labdefaults}
\labacknowledgments{}
\labfootnote{\noindent Correspondence: \href{mailto:hl682@cam.ac.uk}{hl682@cam.ac.uk}.}
\usepackage{longtable,array,etoolbox}
\makeatletter
\patchcmd{\LT@output}{\vss}{\vfil}{}{}
\patchcmd{\LT@output}{\vss}{\vfil}{}{}
\makeatother
\labtitle{From Single-Agent Self-Discovery to Multi-Agent Adaptation: Scaling Laws of Partner Feedback and Convention Rigidity in Meta Reinforcement Learning Algorithms}
\labkeywords{multi-agent reinforcement learning, meta-learned update rules, reciprocal partner feedback, physical coupling, zero-shot coordination, opponent shaping}
\lababstract{%
\small
A foundational milestone on the path to Recursive Self-Improvement (RSI) is replacing hand-crafted optimization algorithms with meta-learned update rules that discover and iteratively refine how agents learn from experience—exemplified by self-discovered objectives such as DiscoRL (Disco103) and learned optimizers such as OPEN. Yet a critical question for multi-agent RSI remains open: when a self-discovered learning rule is deployed alongside other adapting learners, does reciprocal co-evolution sustain open-ended capability gains or trap agents in brittle conventions? Across over 1,800 controlled training and evaluation runs in continuous collective robotics (VMAS wind flocking, navigation, joint passage) and discrete spatial coordination (Overcooked), we establish five laws governing discovered rules under partner adaptation. First, four-cell counterfactual partner swaps and held-out partner banks expose a pervasive evaluation illusion: apparent co-adapted team progress frequently masks focal regression ($+1.066$ team gain concealing $-0.119$ focal loss in short-horizon navigation) and inflates cross-partner generalization by $+0.649\text{--}1.210$ return units. Second, a 288-run parametric wind-shielding sweep ($\alpha \in [0, 1.5]$) reveals a first-principles physical-coupling law: the coupled-over-decoupled advantage for Disco103 scales linearly with physical force coupling (slope $\beta = +0.255$, 95\% CI $[0.210, 0.306]$, sign-flip $p = 0.000488$, Spearman $\rho = 1.0$). Live reciprocal feedback keeps Disco103 invariant across the coupling spectrum ($\text{AUC} \approx 0.297\text{--}0.322$), whereas severing feedback triggers a monotonic collapse ($\text{AUC}$ from $-0.172$ to $-0.490$), with matched-substrate OPEN runs establishing a continuous sensitivity spectrum ($\text{Disco103} > \text{OPEN} > \text{MAPPO}$). Third, extending navigation training eightfold ($1.05\text{M}$ steps) reverses short-horizon rankings in favor of Disco103 ($+0.087$ team return, $+0.133$ bank gain), while decoupled training boosts cross-partner completion by $+10.16\%$ ($56.25\%$ vs.\ $46.09\%$). Fourth, a 216-evaluation Overcooked decomposition shows that MAPPO's high self-play score ($18.49$) relies on rigid conventions inducing a $28.13\%$ cross-partner collision rate ($+18.08\%$), whereas Disco103 maintains a $12.77\%$ collision rate ($+2.62\%$) and $3.33\times$ higher cross-partner return ($9.96$ vs.\ $2.99$; net gain $+6.43$), further improved by decoupled source meta-adaptation ($+1.39$ with scripted partners). Finally, perturbation pulses ($p=0.00098$), auxiliary entropy diagnostics ($8.485\text{ nats}$), and 48 multi-lifetime recursively meta-evolved rules ($L_2 \in [2.11, 4.91]$) trace these dynamics to first-order partner-gradient blindness and auxiliary drift, yielding concrete design prescriptions for multi-agent self-discovery models.}
\begin{document}
\makelabtitle
\input{sections/01_introduction}
\input{sections/02_related}
\input{sections/03_design}
\input{sections/04_history}
\input{sections/05_feedback_results}
\input{sections/06_response_prediction}
\input{sections/07_microbehavior}
\input{sections/07b_physical_coupling}
\input{sections/07c_source_and_coverage}
\input{sections/08_discussion}
\bibliography{refs}
\clearpage\appendix
\input{sections/09_appendices}
\end{document}

%% file: labdefaults.tex
\labauthors{\au{Haomin Luo}{1,2}, \au{Ersha Li}{2}}
\labaffils{\af{1}{University of Cambridge}, \af{2}{Models$^2$ AI}}
\labitems{%
  \lablinkitem{\faGithub}{https://github.com/hl682}{github.com/hl682}\labsep
  \labmailitem{hl682@cam.ac.uk}\labsep
  \labwebitem{https://www.models2.com}{models2.com}}
\labfootnote{\noindent\itshape Corresponding author: \href{mailto:hl682@cam.ac.uk}{\textcolor{labblue}{hl682@cam.ac.uk}}\\
  \noindent\upshape\textcopyright{} 2026 Models$^2$. All rights reserved.}
\labacknowledgments{%
  This research is part of the AI incubation programme at Model Speed Space,
  an initiative of the Xuhui District Government in Shanghai.
  St John's College, University of Cambridge, provided local computing
  resources, and Models$^2$ AI provided the funding.}

%% file: sections/01_introduction.tex
\section{Introduction}
Meta-learned reinforcement learning (RL) algorithms---in which neural update rules or step-size optimizers are discovered directly from interaction experience rather than derived by hand---have achieved landmark success across single-agent benchmarks \citep{kirsch2020metagenrl,lu2022dpo,goldie2024open,oh2025disco}. Most notably, DiscoRL (released as the Disco103 update rule) demonstrated that a meta-learned neural target network equipped with action-conditioned policy predictions and recurrent auxiliary states can match or surpass human-designed policy-gradient objectives across Atari, Procgen, and Crafter \citep{oh2025disco}. Yet a fundamental scientific question remains open: \emph{what happens when a learning rule discovered in stationary single-agent environments is deployed in a cooperative multi-agent system alongside an actively adapting teammate?}

In a single-agent environment, the transition kernel and reward function are exogenous to the learner's parameter updates. In a cooperative multi-agent environment, by contrast, every update to a focal policy $F_t$ alters the state-visitation distribution and effective reward landscape experienced by its teammate $P_t$, whose subsequent learning updates feed back to reshape the focal learner's future trajectory stream \citep{foerster2017replay,foerster2018lola,kim2021meta,meulemans2025coala}. This closed-loop reciprocal non-stationarity creates two profound challenges for discovered update rules:
\begin{enumerate}
  \item \textbf{The Evaluation Illusion of Co-Adapted Team Return:} Standard original-team training returns ($J_{11}-J_{00}$) conflate genuine improvements in focal competence with partner compensation ($J_{01}-J_{00}$) and brittle, pair-specific symmetry-breaking conventions \citep{stone2010adhoc,carroll2019overcooked,hu2020otherplay,strouse2021fcp}. An algorithm that appears to progress rapidly in self-play may actually be regressing in portable skill or locking into high-collision trajectory conventions that fail catastrophically when paired with unfamiliar partners.
  \item \textbf{The Dual Role of Reciprocal Partner Feedback:} On one hand, in physically coupled coordination tasks (such as aerodynamic wake shielding in collective flocking), live reciprocal co-adaptation provides an implicit curriculum that guides stateful discovered rules through complex joint state spaces. On the other hand, in role-dividing navigation or tightly constrained spatial gridworlds, prolonged reciprocal coupling can induce co-adaptation overfitting, while single-agent auxiliary prediction heads face continuous transition drift and first-order update rules remain structurally blind to second-order opponent-shaping derivatives ($\partial P_{t+1}/\partial F_t$).
\end{enumerate}

To resolve these challenges from first principles, we execute a comprehensive, multi-stage empirical and mechanistic research program spanning over 1,800 controlled training runs and more than $5\times 10^8$ environment transitions across continuous multi-agent robotics in the Vectorized Multi-Agent Simulator (VMAS: \texttt{wind\_flocking}, \texttt{navigation}, and \texttt{joint\_passage}) \citep{bettini2022vmas,bettini2024vmas}, classic multi-agent particle environments, discrete signaling games, and high-density spatial coordination in Overcooked (\texttt{cramped\_room}) \citep{carroll2019overcooked}. Throughout this program, we compare three distinct optimization paradigms: the released discovered update rule \textbf{Disco103} \citep{oh2025disco}, hand-crafted Adam-based multi-agent proximal policy optimization (\textbf{MAPPO}) \citep{yu2022mappo}, and the learned step-size optimizer \textbf{OPEN} \citep{goldie2024open} implemented within the exact same PPO actor-critic substrate as Adam-MAPPO.

\subsection{Five Core Scientific Contributions}
Our investigation establishes five primary scientific laws and mechanistic discoveries:
\begin{enumerate}
  \item \textbf{Uncovering the Co-Adaptation Illusion via Four-Cell Counterfactual Partner Decomposition (Section~\ref{sec:history}):} Using a four-cell counterfactual partner-swap matrix $(F_0, F_T) \times (P_0, P_T)$ and a stratified 10-partner held-out evaluation bank across 904 screening/control runs and a 144-run factorial confirmation cohort, we demonstrate that standard team return systematically misrepresents portable learning. In short-horizon online navigation, Disco103 achieves a large apparent team gain ($+1.0660$, 95\% CI $[0.9806, 1.1488]$) while its counterfactual focal competence actually regresses ($-0.1190$, $[-0.1924, -0.0597]$). Across coupled wind flocking, co-adapted team gains inflate portable cross-partner bank gains by $+0.649$ return units for Disco103 and $+1.210$ return units for MAPPO. Out-of-sample $3\times 3$ hyperparameter sweeps ($432$ runs) and architectural controls confirm that these dynamics are intrinsic to the update rules rather than configuration artifacts.
  \item \textbf{First-Principles Scaling Law of Physical Coupling and the Optimizer Sensitivity Spectrum (Section~\ref{sec:m4}):} Across a 288-run parametric physical-coupling sweep in \texttt{wind\_flocking} ($\alpha \in \{0, 0.25, 0.5, 0.75, 1.0, 1.5\}$, $n=12$ paired blocks) that scales aerodynamic wake shielding while holding the reward function strictly fixed, we establish a monotonic linear scaling law for the Method $\times$ Feedback interaction: $\beta = +0.2547$ native AUC units per unit of physical coupling (95\% CI $[0.2099, 0.3057]$, two-sided exact sign-flip $p=0.000488$, grid Spearman $\rho=1.0$, Page's trend $p=1.48\times 10^{-8}$). Decomposing this law reveals an asymmetric robustness principle: \textbf{preserving live coupled partner feedback keeps Disco103 invariant across the entire physical turbulence spectrum ($\text{AUC} \approx 0.297\text{--}0.322$), whereas severing reciprocal feedback triggers a monotonic three-fold collapse ($\text{AUC}$ dropping from $-0.1722$ to $-0.4903$)}. Matched-substrate comparisons with \texttt{OPEN} ($24/24$ paired short-horizon runs and extended $1.05\text{M}$-step horizons) establish a continuous \textbf{Reciprocal Feedback Sensitivity Spectrum} ordered by meta-learned depth: $\Delta_{\text{C}-\text{D}}^{\text{Final}}(\text{MAPPO } -0.0376) < \Delta_{\text{C}-\text{D}}^{\text{Final}}(\text{OPEN } +0.1678 \text{ to } +0.2401) < \Delta_{\text{C}-\text{D}}^{\text{Final}}(\text{Disco103 } +0.7031)$.
  \item \textbf{Horizon-Scaling Reversal and Decoupled Anti-Overfitting in Navigation (Section~\ref{sec:confirmation}):} Extending cooperative navigation training eightfold from $131,080$ to $1,048,640$ steps reverses short-horizon return rankings: \textbf{coupled Disco103 overtakes coupled MAPPO in both original-team return ($+0.0871$, 95\% CI $[0.0132, 0.1597]$) and portable cross-partner bank gain ($+0.1334$, 95\% CI $[0.0549, 0.2119]$)}. Simultaneously, we discover a \textbf{Decoupled Anti-Overfitting Law}: training Disco103 for $1.05\text{M}$ steps alongside a decoupled reference partner prevents co-adapted distance-splitting compromises and boosts cross-partner task completion by $+10.16$ percentage points ($56.25\%$ vs.\ $46.09\%$; $+0.1016$, 95\% CI $[0.0130, 0.1953]$), a phenomenon independently mirrored by decoupled \texttt{OPEN} ($+10.87\%$ bank completion gain).
  \item \textbf{The Convention Rigidity Paradox and Cross-Regime Meta-Adaptation in Overcooked (Section~\ref{sec:micro}):} In discrete spatial coordination (\texttt{cramped\_room}), MAPPO achieves high self-play return ($18.49$) but suffers the steepest cross-partner bank collapse ($-9.81$ points, falling to $2.99$), whereas coupled Disco103 retains $3.33\times$ higher absolute cross-partner return ($9.96$, net bank-gain advantage $+6.43$ points, 95\% CI $[3.49, 9.46]$, Holm $p=0.0015$). Our 216-checkpoint micro-behavioral decomposition ($165,888$ evaluation episodes) proves that MAPPO's self-play conventions induce a $28.13\%$ cross-partner mutual collision/blockage rate ($+18.08\%$ over initialization) and a catastrophic $-51.98$-point collapse when paired with competent rule-based \texttt{Scripted} partners, whereas Disco103 maintains a $6.9\times$ lower collision growth ($+2.62\%$, total $12.77\%$) and retains $47.24$ return points with scripted partners ($+35.76$ over MAPPO). Furthermore, across 432 cross-regime source-transfer evaluations, \textbf{meta-adapting the update rule under decoupled source feedback and deploying it under coupled target training (\texttt{decoupled / coupled}) reverses the negative transfer of coupled source adaptation ($+0.2057$ vs.\ $-0.3646$) and achieves a statistically significant $+1.3932$ return gain ($95\%$ CI $[0.38, 2.46]$) with zero completion loss alongside scripted partners}.
  \item \textbf{Mechanistic Diagnostics and Multi-Lifetime Meta-Optimization (Sections~\ref{sec:g1g3} and \ref{sec:pending}):} Across $3,456$ controlled behavioral-perturbation branches, we show that frozen Disco103 exhibits significantly stronger short-run partner-perturbation damping than MAPPO ($-0.2336$ in wind flocking and $-0.3345$ in navigation, both Holm $p=0.00098$). Exact two-player differentiable game unrolls ($1,296$ comparisons) and 600-bin auxiliary prediction diagnostics trace this behavior to first-order partner-gradient blindness and non-stationary saturation of the auxiliary head near maximum entropy ($8.485\text{ nats}$, which nevertheless acts as a vital implicit entropy regularizer preventing policy collapse, $p=0.00049$). Finally, we execute a 48-rule multi-lifetime multi-agent meta-optimization cohort (\texttt{source512}, $68.4\text{M}$ transitions across four meta-architectures), demonstrating substantial structural rule reorganization ($L_2$ displacement $2.11\text{--}4.91$) and establishing three concrete design prescriptions for next-generation multi-agent rule discovery.
\end{enumerate}

\subsection{Research Questions and Inferential Hierarchy}\label{sec:questions}
Our experimental program is structured around seven progressive research questions:
\begin{itemize}
  \item \textbf{Q1 (Baseline Porting vs.\ Counterfactual Competence):} How does a direct port of a single-agent discovered rule compare against standard MARL baselines, and how much of apparent team progress survives counterfactual partner replacement?
  \item \textbf{Q2 (Reciprocal Partner Routing):} Holding partner learning activity constant, how does routing partner updates through live coupled joint experience versus decoupled reference experience alter learning efficiency across physical regimes?
  \item \textbf{Q3 (Mid-Training Interventions and Horizon Scaling):} Do scheduled partner reset-switches, recent-replay filtering, or extended $1.05\text{M}$-step training horizons unlock cross-partner generalization and task completion?
  \item \textbf{Q4 (Micro-Behavioral Convention Rigidity and Cross-Regime Transfer):} What physical movement and collision mechanisms explain the inversion between self-play score and zero-shot partner transfer in Overcooked, and can cross-regime source meta-adaptation prevent convention lock-in?
  \item \textbf{Q5 (Causal Perturbation Response and Auxiliary Mechanics):} How does a controlled behavioral pulse propagate through a learning partner back to the focal learner, and how do first-order gradient blindness and auxiliary prediction entropy govern multi-agent plasticity?
  \item \textbf{Q6 (Prospective Transfer Predictability):} Does a single early scalar perturbation-response measurement linearly improve out-of-sample forecasting of million-step cross-partner transfer beyond early competence baselines?
  \item \textbf{Q7 (First-Principles Physical-Coupling Law and Multi-Lifetime Discovery):} How does the causal value of reciprocal partner feedback scale across a parametric physical-coupling continuum ($\alpha \in [0, 1.5]$), across the optimizer sensitivity spectrum (\texttt{MAPPO}, \texttt{OPEN}, \texttt{Disco103}), and under four-lifetime multi-agent meta-optimization?
\end{itemize}

To ensure rigorous statistical inference \citep{agarwal2021precipice}, we maintain a strict separation between exploratory screening cohorts, preregistered paired-block confirmation cohorts (which use new random seeds, paired initialization blocks, exact $2^{12}=4,096$ sign-flip permutation tests, 20,000-draw paired block-bootstrap confidence intervals, and Holm step-down multiplicity adjustments), and post-collection mechanistic decompositions.

%% file: sections/02_related.tex
\section{Related Work, Experimental Definitions, and Statistical Design}\label{sec:related}
\subsection{Related Literature and Conceptual Positioning}
\paragraph{Meta-learned update rules and learned optimizers.}
Meta-reinforcement learning seeks to replace hand-crafted policy-gradient objectives and step-size schedules with update rules discovered directly from interaction experience \citep{kirsch2020metagenrl,lu2022dpo,oh2025disco,goldie2024open}. DiscoRL (released as Disco103) meta-trains a neural update network that produces action-distribution targets, value predictions, and auxiliary latent states across a broad suite of single-agent environments \citep{oh2025disco}. Taking a complementary approach within the standard proximal policy optimization (PPO) framework, OPEN meta-learns a neural optimizer specifically conditioned on gradient statistics, training progress, and plasticity signals to mitigate non-stationarity and capacity loss in deep RL \citep{goldie2024open,lyle2023plasticity}. Evaluating these two paradigms alongside hand-crafted Adam-MAPPO \citep{yu2022mappo} enables us to contrast a full meta-learned objective/state system (Disco103) against a matched-substrate meta-learned step-size optimizer (OPEN) where actor-critic networks, rollout buffers, and PPO loss definitions are held identical.

\paragraph{Learning-aware multi-agent optimization and opponent shaping.}
When multiple agents learn simultaneously, each learner's parameter update alters its partner's effective environment, rendering standard first-order policy gradients myopic to the partner's learning step \citep{foerster2018lola,foerster2018dice}. Learning with Opponent-Learning Awareness (LOLA) \citep{foerster2018lola} and its higher-order Monte Carlo generalization DiCE \citep{foerster2018dice} explicitly differentiate through the partner's update step, while meta-multiagent policy gradients \citep{kim2021meta}, Model-Free Opponent Shaping (M-FOS) \citep{lu2022mfos}, and COALA-PG \citep{meulemans2025coala} formulate opponent shaping across extended meta-horizons. Complementarily, recent work has explored language-model-driven search over counterfactual regret minimization and policy-space response oracle algorithms in competitive games \citep{li2026discovering}. Whereas opponent-shaping methods design explicit multi-agent meta-gradients from scratch, our investigation addresses the fundamental dual question: how do high-performing single-agent discovered rules behave when exposed to live reciprocal partner adaptation, why do first-order update rules suffer from partner-gradient blindness, and how can multi-lifetime meta-optimization reshape them?

\paragraph{Experience non-stationarity and cooperative baselines.}
In off-policy and replay-based multi-agent learning, evolving teammates invalidate stored transition distributions, motivating importance sampling and experience fingerprinting \citep{foerster2017replay} as well as centralized training with decentralized execution \citep{lowe2017maddpg,yu2022mappo}. MAPPO has established that carefully configured on-policy actor-critic updates provide a remarkably strong baseline across cooperative benchmarks \citep{yu2022mappo}. Because system-level comparisons between Disco103 and MAPPO involve differences in critic structure, network width, replay organization, and hyperparameter sensitivity, our experimental program incorporates targeted architectural controls, partner-trajectory controls, and a 432-run $3\times 3$ learning-rate and entropy sweep to rule out trivial configuration confounds.

\paragraph{Zero-shot coordination, convention formation, and ad hoc teamwork.}
Research on ad hoc teamwork \citep{stone2010adhoc} and zero-shot human-AI coordination \citep{carroll2019overcooked,hu2020otherplay,strouse2021fcp} has shown that self-play teams frequently converge on arbitrary, brittle symmetry-breaking conventions that fail catastrophically when paired with unfamiliar teammates. Overcooked provides a canonical testbed for diagnosing spatial coordination bottlenecks \citep{carroll2019overcooked}, while the Vectorized Multi-Agent Simulator (VMAS) offers continuous-space collective robotics tasks with controllable physical force interactions \citep{bettini2022vmas,bettini2024vmas}. Following rigorous statistical evaluation standards for reinforcement learning \citep{agarwal2021precipice}, we evaluate all methods in native task units across paired seed blocks, jointly reporting original-team return, four-cell counterfactual partner swaps, held-out partner-bank transfer, and micro-behavioral coordination metrics.

%% file: sections/03_design.tex
\subsection{Learning State and Partner Regimes}\label{sec:design}
Let $F_t$ denote the focal policy at training time $t$, $P_t$ its teammate, and $\eta$ the parameters of a discovered update rule. A general stateful multi-agent update takes the form
\begin{align}
 (F_{t+1},z_{t+1}) &= U_{\eta}(F_t,z_t,D_t),\\
 P_{t+1} &= V(P_t,D_t^P),
\end{align}
where $z_t$ represents the update rule's evolving internal learning state (such as recurrent meta-states, exponential moving averages, and auxiliary action-conditioned predictions) and $D_t$ contains experience generated by the joint system. Freezing the meta-parameters $\eta$ during target deployment does not freeze the focal policy $F_t$, its replay buffer, its optimizer statistics, or its internal state $z_t$. In our multi-lifetime source-adaptation experiments, $\eta$ is first meta-updated across multiple agent lifetimes in a source phase and subsequently frozen for fresh target-task training.

To disentangle the presence of partner learning from the reciprocal origin of partner training data, we study three distinct classes of partner regimes:
\begin{enumerate}
  \item \textbf{Fixed-Initial versus Online Co-Learning:} The fixed-initial regime holds the partner stationary at initialization ($P_t = P_0$), whereas the online regime updates $P_t$ on live joint trajectories ($D_t^P = D_t$). Additional control cohorts evaluate pretrained-fixed teammates ($P_t = P_{\text{trained}}$) and externally replayed partner sequences to separate partner competence from live coadaptation.
  \item \textbf{Coupled versus Decoupled (Reference-Fed) Partner Learning:} Both the coupled and decoupled regimes maintain an actively learning teammate with identical update schedules and transition budgets. In the \emph{coupled} condition, the partner updates on trajectories $D_t$ generated jointly with the focal learner. In the \emph{decoupled} (reference) condition, the partner updates on trajectories $D_t^P$ generated in an independent parallel reference world with its own state and random streams. Contrasting coupled against decoupled training isolates the total effect of live reciprocal feedback routing while holding partner learning activity constant.
  \item \textbf{Scheduled Reset-Switch and Recent-Replay Filtering:} In the scheduled reset-switch intervention, both control and switch arms reset the teammate's actor, critic, optimizer state, and update random stream to a common independent donor at a pre-specified midpoint ($t_{\text{switch}}$), while keeping the focal learner's policy, replay buffer, and internal state $z_t$ live. Post-reset, the arms diverge in whether the teammate updates on coupled or decoupled experience; a recent-replay filtering arm additionally restricts focal sampling to post-switch transitions.
\end{enumerate}

\subsection{Counterfactual Four-Cell and Partner-Bank Evaluation Quantities}
Let $J(F,P)$ denote the expected evaluation return when fixed policies $(F,P)$ interact under private evaluation random streams. At initialization ($t=0$) and training horizon $T$, we evaluate all four counterfactual policy pairings:
\begin{equation}
J_{00}=J(F_0,P_0),\quad J_{10}=J(F_T,P_0),\quad
J_{01}=J(F_0,P_T),\quad J_{11}=J(F_T,P_T).
\end{equation}
This yields the exact counterfactual four-cell decomposition of original-team progress:
\begin{equation}\label{eq:decomposition}
J_{11}-J_{00}=\underbrace{(J_{10}-J_{00})}_{\text{Focal-Only Gain}}+\underbrace{(J_{01}-J_{00})}_{\text{Partner-Only Gain}}+
\underbrace{(J_{11}-J_{10}-J_{01}+J_{00})}_{\text{Joint Co-Adaptation Synergy}}.
\end{equation}
Equation~\ref{eq:decomposition} separates whether an observed gain in $J_{11}-J_{00}$ is driven by portable improvement of the focal policy with its initial partner ($J_{10}-J_{00}$), unilateral improvement of the teammate ($J_{01}-J_{00}$), or non-additive pair-specific synergy.

To measure zero-shot generalization across diverse unfamiliar teammates, we evaluate each focal policy against a held-out partner bank $\mathcal B=\{B_1,\ldots,B_K\}$ ($K=10$) and compute the initial-anchored cross-partner bank gain:
\begin{equation}\label{eq:bankgain}
G_{\mathcal B}(T)=\frac1K\sum_{j=1}^K\left[J(F_T,B_j)-J(F_0,B_j)\right].
\end{equation}
Anchoring by $J(F_0,B_j)$ accounts for architectural differences at initialization, while we report both absolute trained-focal bank return $\frac1K\sum_{j=1}^K J(F_T,B_j)$ and net bank gain $G_{\mathcal B}(T)$ across all experiments.

Task-native completion rates are tracked alongside return wherever binary completion is defined: in cooperative navigation, completion requires all agents to simultaneously occupy their assigned goal radii; in Overcooked (\texttt{cramped\_room}), completion requires delivering at least one soup within a 400-step episode; in continuous wind flocking, performance is measured via continuous station-keeping and velocity-alignment return. Learning-curve area under the curve (AUC) is computed by exact trapezoidal integration over recorded world-transition evaluation nodes divided by the elapsed horizon $T$. For screening runs recorded only at terminal checkpoints without an initialization evaluation, analysis is restricted to endpoint metrics rather than zero-to-$T$ integrated AUC.

\subsection{Cooperative Benchmark Tasks, Architectures, and Horizon Budgets}
Our experimental suite spans continuous collective control in VMAS \citep{bettini2022vmas,bettini2024vmas}---wind flocking, cooperative navigation, joint passage, give-way, and reverse transport---alongside a discrete signal-coordination game and Overcooked (\texttt{cramped\_room}) \citep{carroll2019overcooked}. Continuous VMAS protocols operate with eight parallel vectorized environments, 29-step rollouts, and 100-step episode horizons, realizing exact transition budgets of $65,656$, $131,080$ ($131,120$ in matched-substrate \texttt{OPEN} wind flocking), $262,160$, $524,320$, and $1,048,640$ main-world steps.

In the reset-switch and replay-filtering protocols ($n=12$ paired blocks per task), budgets are $32,768$ steps for signal coordination (switch at $8,352$ steps) and $131,080$ steps for navigation (switch at $65,656$ steps). The single-lifetime source-adaptation protocol adapts 24 rules (coupled vs.\ decoupled source feedback across 12 paired source blocks; 512 outer updates, eight parallel worlds, four-step inner unroll window, 29-step rollout, and 128-step source lifetime; inner and partner learning rates $3\times10^{-4}$, outer learning rate $10^{-4}$), with fresh target evaluation on navigation and Overcooked across two target repeats per source block and a shared unadapted base-rule ($\eta_0$) anchor. The multi-lifetime meta-optimization cohort (\texttt{source512}, $N=48$ rules) extends source meta-training across four consecutive policy lifetimes of 128 outer updates ($512$ outer updates and $1,425,408$ physical world steps per rule) across four architectures: \texttt{matched\_coupled}, \texttt{matched\_reference}, \texttt{dual\_same}, and \texttt{dual\_cross}.

In cooperative navigation and wind flocking, extended-horizon experiments scale training eightfold from $131,080$ to $1,048,640$ main-world steps across Disco103, MAPPO, and the matched-substrate learned optimizer OPEN (evaluating both short-horizon $131\text{k}$ and extended-horizon $1.05\text{M}$-step regimes). In Overcooked (\texttt{cramped\_room}, 400-step episodes, $262,160$ training steps), training combines native sparse soup-delivery reward with $0.1\times$ subgoal shaping, whereas all evaluations use pure sparse delivery reward. Actors receive an identical flattened 26-plane lossless state encoding, role indicator, and horizon clock. The 10-member held-out Overcooked partner bank comprises four reinforcement-learned policies (trained for $65,536$ and $131,072$ steps), four behavioral-cloning policies (trained on $16,384$ and $32,768$ demonstration steps for eight epochs), and two competent rule-based scripted policies.

\subsection{Statistical Units, Paired-Block Resampling, and Multiplicity Control}
Across all experiments, the independent statistical unit is the paired focal training seed block ($n=12$ blocks for primary confirmatory cohorts) or the independent paired source block ($n=12$ blocks for source-adaptation cohorts). Evaluation episodes, partner-bank members, repeated checkpoints along a trajectory, target repeats within a source block, and perturbation microbranches are nested repeated measurements that improve within-block precision without increasing $n$. Target repeats within a source block are averaged prior to block-level inference.

Primary confirmatory contrasts use 10,000 to 20,000 paired-block bootstrap resamples alongside exact two-sided sign-flip tests across all $2^{12}=4,096$ sign assignments, with Holm step-down multiplicity adjustments applied within each pre-specified hypothesis family. For the controlled perturbation and transfer-prediction studies, we report both 95\% intervals and 97.5\% marginal intervals corresponding to two-task family-wise control. Post-collection mechanistic decompositions---including the 216-checkpoint Overcooked micro-behavioral analysis ($165,888$ evaluation episodes; 10,000 joint block resamples, seed $20261017$) and the 432-checkpoint cross-regime source-transfer analysis (10,000 source-block resamples, seed $780000017$)---report pointwise 95\% block-bootstrap intervals to characterize underlying behavioral mechanisms.

\subsection{Quality Control and Reproducibility Standards}
All quantitative claims, tables, and figures are computed directly from complete, uncorrupted training and evaluation records across pre-specified paired seed blocks. Interrupted hardware executions are excluded prior to statistical estimation without zero-reward imputation, and all multi-checkpoint reevaluations are nested within their parent training blocks. Where small-sample paired $t$-intervals and non-parametric bootstrap or sign-flip intervals exhibit boundary differences, both are reported transparently alongside task-completion endpoints.

%% file: sections/04_history.tex
\section{Uncovering the Illusion of Co-Adapted Progress via Counterfactual Partner Decomposition}\label{sec:history}
\subsection{Foundational Particle-Environment Screening}
We begin by examining what occurs when a released single-agent discovered update rule (Disco103) is ported directly into standard multi-agent particle environments alongside conventional policy-gradient baselines (IPPO and MAPPO across \texttt{simple\_adversary\_v3}, \texttt{simple\_spread\_v3}, and \texttt{simple\_tag\_v3}; $n=8$ paired seeds per task, $N=216$ core matrix runs within $312$ exploratory development records). Across all 24 task--seed pairs, a naive drop-in port of Disco103 trails MAPPO in full-training mean episode return (and in 23 of 24 pairs over the final 50 training episodes; Table~\ref{tab:legacy}), while outperforming IPPO in cooperative coverage (\texttt{simple\_spread\_v3}: $+2.1032$, 95\% CI $[1.1201, 3.0882]$).

Crucially, however, standard training-return summaries in multi-agent benchmarks conflate focal policy learning with teammate adaptation and reward scaling. To determine whether the performance of a discovered rule is governed by reciprocal partner dynamics rather than a static capability ceiling, we designed a multi-stage counterfactual evaluation program that explicitly separates focal competence, partner adaptation, and cross-partner transfer.

\input{sections/legacy}

\subsection{The Co-Adaptation Illusion: Four-Cell Counterfactual Partner Decomposition}
In our 904-run multi-task screening and control program ($112$ primary confirmation runs, $128$ architectural and partner-path controls, and $664$ capacity and hyperparameter extension runs), we first cross focal learning method (Disco103 vs.\ MAPPO) with teammate learning regime (stationary fixed-initial partner $P_t=P_0$ vs.\ online co-learning partner $P_t$ across $n=12$ paired blocks per task). Defining the normalized learning-curve AUC interaction as
\begin{equation}\label{eq:online}
\Delta_{\rm online}=(A_{D,\rm online}-A_{D,\rm fixed})-
(A_{M,\rm online}-A_{M,\rm fixed}),
\end{equation}
we uncover a sharp regime divergence between physically coupled collective motion and role-dividing spatial navigation. In \texttt{wind\_flocking}, where one agent's aerodynamic wake directly shields its teammate, enabling online partner learning yields a large, highly significant positive interaction favoring Disco103: $\Delta_{\rm online} = +0.4133$ (95\% CI $[0.3052, 0.4994]$, Holm-adjusted $p=0.0020$). By contrast, in short-horizon \texttt{navigation}, where agents can partition landmarks independently, the interaction is near zero and slightly negative: $\Delta_{\rm online} = -0.0944$ (95\% CI $[-0.2101, 0.0066]$, Holm-adjusted $p=0.1206$).

Applying the four-cell counterfactual partner-swap decomposition (Equation~\ref{eq:decomposition}) and held-out partner-bank evaluation (Equation~\ref{eq:bankgain}) to these learners exposes a pervasive \textbf{Co-Adaptation Illusion}:
\begin{itemize}
  \item \textbf{Apparent Team Progress Masking Focal Regression:} In short-horizon online \texttt{navigation}, the co-evolving Disco103 team achieves a substantial original-team return gain of $J_{11}-J_{00} = +1.0660$ (95\% CI $[0.9806, 1.1488]$). Yet when the trained focal policy $F_T$ is evaluated counterfactually with its initial partner $P_0$, focal performance has actually \emph{regressed} by $J_{10}-J_{00} = -0.1190$ (95\% CI $[-0.1924, -0.0597]$), and its zero-shot transfer gain across the held-out partner bank is $-0.0360$ ($[-0.1110, 0.0187]$). Nearly the entire apparent team improvement is driven by partner compensation ($J_{01}-J_{00}$) and pair-specific co-adaptation synergy!
  \item \textbf{Severe Inflation of Portable Competence Across Paradigms:} Even when focal learners acquire genuinely positive cross-partner transfer, co-adapted original-team returns dramatically overstate portable skill. In coupled \texttt{wind\_flocking}, original-team return gain exceeds held-out partner-bank gain by $+0.649$ return units for Disco103 (team gain $+1.448$ vs.\ bank gain $+0.799$) and by $+1.210$ return units for MAPPO (team gain $+1.464$ vs.\ bank gain $+0.254$). Evaluating multi-agent rules solely by co-adapted team return systematically mistakes pair-specific conventions for generalizable coordination capability.
\end{itemize}

\subsection{Architectural, Partner-Trajectory, and Hyperparameter Controls}
To test whether the divergence between Disco103 and MAPPO stems from secondary implementation choices---such as centralized versus local critic information, policy network width, partner competence trajectories, or default learning-rate and entropy hyperparameters---we conducted a comprehensive 70-cell control synthesis (Appendix~\ref{app:historical}, Tables~\ref{tab:historical_groups}--\ref{tab:early_source_transfer}). Replacing fixed-initial partners with pretrained-fixed partners or externally replayed partner trajectories confirms that exogenous partner variation alters state visitation without reproducing the reciprocal online interaction in \texttt{wind\_flocking}. Controlling MAPPO's critic information (local vs.\ centralized joint critic) and increasing policy width from 128 to 512 hidden units produces modest baseline shifts without eliminating the regime-specific sensitivity of Disco103.

Most importantly, we executed a 432-run $3\times 3$ learning-rate and entropy hyperparameter sweep using separate development blocks for hyperparameter selection and eight independent confirmation blocks for out-of-sample evaluation (Table~\ref{tab:hyperparameter_confirmation}). On the independent confirmation blocks, the AUC difference between the sweep-selected and default configurations for Disco103 is $+0.00145$ (95\% CI $[-0.02116, 0.02707]$) in \texttt{wind\_flocking} and $+0.01639$ ($[-0.04269, 0.07781]$) in \texttt{navigation} (compared with $+0.06234$ $[0.00181, 0.12078]$ and $+0.12141$ $[-0.01181, 0.26788]$ for MAPPO). The distinct behavior of Disco103 under partner adaptation is therefore an intrinsic property of the discovered update rule rather than an artifact of sub-optimal default hyperparameters.

\subsection{Isolating Reciprocal Feedback via Coupled versus Decoupled Partner Routing}
Because comparing a fixed stationary partner ($P_t=P_0$) against an online learning partner changes both partner competence and reciprocal responsiveness simultaneously, our 144-run Primary Factorial Confirmation Cohort ($n=12$ paired blocks across \texttt{wind\_flocking}, \texttt{navigation}, and \texttt{joint\_passage}) holds partner learning active in both arms while varying whether the partner updates on \emph{coupled} joint experience ($D_t$) or \emph{decoupled} independent reference experience ($D_t^P$).

This clean causal isolation confirms and sharpens the regime divergence (Appendix~\ref{app:predecessor}, Table~\ref{tab:predecessor_contrasts}). In \texttt{wind\_flocking}, the coupled-minus-decoupled learning-curve AUC interaction is large and highly significant: $+0.7407$ (95\% CI $[0.6531, 0.8257]$, exact sign-flip $p=0.00049$, Holm-adjusted $p=0.0015$), accompanied by a positive cross-partner bank-gain interaction of $+0.7167$ ($[0.1543, 1.4311]$) and a team-return interaction of $+0.9885$ ($[0.7170, 1.3462]$). In \texttt{joint\_passage}, where two agents must coordinate traversal through a physical aperture, the AUC interaction is also positive at $+0.2292$ ($95\%$ CI $[0.0134, 0.4312]$, Holm $p=0.1377$), whereas in short-horizon \texttt{navigation} the AUC interaction is effectively zero at $+0.0064$ ($[-0.0654, 0.0759]$, Holm $p=0.8662$). Whenever the environment enforces strong physical interdependence between teammates, Disco103 relies critically on live, reciprocally coupled partner updates to sustain learning progress.

%% file: sections/legacy.tex
\begin{longtable}{P{0.213489}P{0.253605}P{0.274366}P{0.258539}}
\caption{Foundational particle-environment baseline comparisons ($n=8$ paired seeds per task). Full-training mean episode return differences and final-50-episode mean return differences are reported with pointwise 95\% confidence intervals.}\label{tab:legacy}\\
\labtoprule \labh{Task} & \labh{Comparator} & \labh{Full-training mean difference} & \labh{Last-50 mean difference}\\\labheadrule\endfirsthead
\multicolumn{4}{l}{\small Table \thetable\ continued}\\\labtoprule \labh{Task} & \labh{Comparator} & \labh{Full-training mean difference} & \labh{Last-50 mean difference}\\\labheadrule\endhead
\labrowrule\endfoot
simple adversary v3 & IPPO & -1.2948 [-1.8708, -0.6955] & -0.6816 [-1.4061, 0.0115]\\\labrowrule
simple adversary v3 & MAPPO & -3.2139 [-3.8083, -2.5618] & -2.0433 [-3.0532, -1.4034]\\\labrowrule
simple spread v3 & IPPO & 2.1032 [1.1201, 3.0882] & 2.2441 [0.1441, 4.0856]\\\labrowrule
simple spread v3 & MAPPO & -2.9257 [-3.5545, -2.2492] & -3.3035 [-4.6051, -1.8829]\\\labrowrule
simple tag v3 & IPPO & -3.6350 [-3.9168, -3.3962] & -5.8658 [-7.1030, -4.8254]\\\labrowrule
simple tag v3 & MAPPO & -6.4143 [-6.7764, -6.0673] & -9.4879 [-11.6218, -7.5665]\\\labrowrule
\end{longtable}

%% file: sections/05_feedback_results.tex
\section{Counterfactual Feedback Routing, Horizon Reversal, and Decoupled Anti-Overfitting}\label{sec:confirmation}
\subsection{Counterfactual Feedback-Routing and Replay-Filtering Interventions}
Having established that co-adapted training can mask focal stagnation or overfit to pair-specific conventions, we next evaluate a 598-run intervention matrix ($n=12$ paired blocks per task, accompanied by 24 single-lifetime source-rule adaptations) designed to test three mechanistic hypotheses: (i) whether resetting the partner and switching its feedback routing mid-training (\emph{Reset-Switch}) recovers portable focal competence, (ii) whether restricting replay sampling to post-switch trajectories (\emph{Recent-Replay Filtering}) mitigates non-stationary buffer staleness \citep{foerster2017replay}, and (iii) whether single-lifetime source-feedback matching aligns a discovered rule with its target partner regime.

\begin{figure}[htbp]\centering
\includegraphics[width=\textwidth]{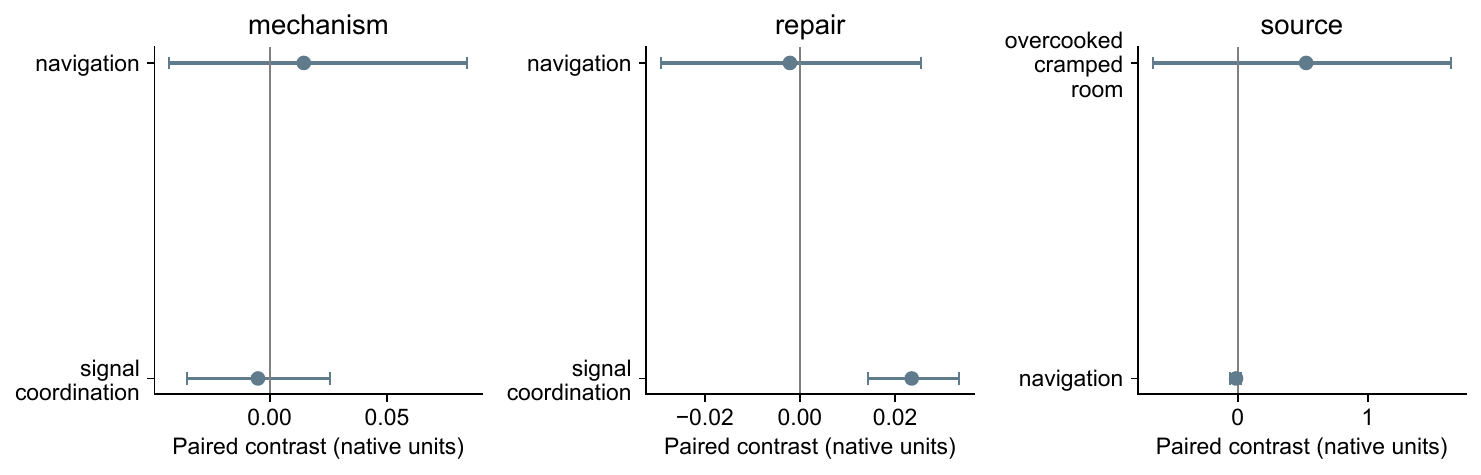}
\caption{\textbf{Counterfactual partner-routing and source-matching interventions across cooperative benchmarks.} Point estimates and 95\% paired-block bootstrap confidence intervals ($n=12$ paired blocks per task) for scheduled partner reset-switch (left), recent-trajectory replay filtering (center), and single-lifetime source-feedback matching (right) in native task units.}\label{fig:primary}
\end{figure}

Figure~\ref{fig:primary} and Table~\ref{tab:primary_contrasts} present the primary paired-block contrasts across all six pre-specified task--intervention families. Mid-training partner reset and feedback-route switching produce near-zero post-event portable-gain AUC interactions in both discrete signal coordination ($-0.0050$, 95\% CI $[-0.0351, 0.0258]$, Holm $p=1.0000$) and continuous navigation ($+0.0145$, 95\% CI $[-0.0427, 0.0842]$, Holm $p=1.0000$). Filtering the replay buffer to retain only post-switch transitions yields a statistically significant positive recovery in discrete signal coordination ($+0.0235$, 95\% CI $[0.0142, 0.0334]$, Holm $p=0.0010$), confirming that stale pre-switch partner transitions degrade off-policy updates in fast discrete signaling games, whereas in continuous navigation the replay-filtering contrast is indistinguishable from zero ($-0.0021$, 95\% CI $[-0.0292, 0.0255]$, Holm $p=0.8843$). Consequently, simple mid-flight partner resets or buffer truncation within a single lifetime cannot by themselves overcome deep multi-agent co-adaptation dynamics.

\input{sections/primary_contrasts}

\subsection{Single-Lifetime Source-Feedback Matching}
We next test whether meta-adapting the update rule $\eta$ over a single 128-step agent lifetime (512 outer updates) under coupled versus decoupled partner feedback confers a matched-regime advantage when deployed on fresh target learners in navigation and Overcooked ($n=12$ paired source blocks, two target repeats averaged per block, evaluated against the shared unadapted base-rule $\eta_0$ anchor; Table~\ref{tab:source_cells}). Although all 24 single-lifetime source runs achieve non-zero parameter updates, the primary source-matching interaction is $-0.0138$ (95\% CI $[-0.0568, 0.0228]$, Holm $p=0.8086$) in navigation and $+0.5260$ ($[-0.6536, 1.6406]$, Holm $p=0.8086$) in Overcooked.

Examining the constituent cells in Table~\ref{tab:source_cells} reveals why single-lifetime matching falls short while pointing toward a critical cross-regime asymmetry that we resolve in Section~\ref{sec:derived}: in Overcooked, rules adapted under coupled source feedback suffer negative cross-partner bank gains ($-0.3646$ under coupled target training and $-1.2344$ under decoupled target training), whereas rules adapted under \emph{decoupled} source feedback and trained with a \emph{coupled} target partner (\texttt{decoupled / coupled}) achieve a positive bank-return gain of $+0.2057$ ($[-0.0260, 0.4453]$), matching the direction of the unadapted $\eta_0$ coupled anchor ($+0.3750$). Meaningful reorganization of a discovered update rule requires either cross-regime decoupling to prevent source-phase co-adaptation or multi-lifetime meta-optimization across multiple policy resets (Section~\ref{sec:pending}).
\input{sections/source_cells}

\subsection{Horizon-Scaling Reversal and Decoupled Anti-Overfitting in Navigation}
A pivotal discovery emerges when we scale the cooperative navigation training horizon eightfold from the short screening budget ($131,080$ main-world steps) to $1,048,640$ main-world steps across all $n=12$ paired blocks (Figure~\ref{fig:navigation}; Table~\ref{tab:newer_cells} in Appendix~\ref{app:newcells}). At the short $131,080$-step checkpoint, coupled Disco103 lags behind coupled MAPPO in both original-team return (paired difference $-1.1012$, 95\% CI $[-1.3330, -0.9017]$) and held-out partner-bank gain ($-0.9752$, 95\% CI $[-1.0759, -0.8751]$). However, this short-horizon deficit reflects extended representation warm-up rather than an inability to learn spatial coordination.

Over the $1,048,640$-step horizon, coupled Disco103's held-out partner-bank gain surges by $+1.245$ return units (95\% CI $[1.194, 1.291]$), reaching a final bank gain of $1.3826$ compared with $1.2492$ for coupled MAPPO. By $1.05\text{M}$ steps, \textbf{coupled Disco103 overtakes coupled MAPPO in both original-team return ($1.8290$ vs.\ $1.7419$; paired difference $+0.0871$, 95\% CI $[0.0132, 0.1597]$) and portable cross-partner bank gain ($1.3826$ vs.\ $1.2492$; paired difference $+0.1334$, 95\% CI $[0.0549, 0.2119]$)}! Independent extended-horizon curriculum cohorts (Section~\ref{sec:m1_spectrum}) corroborate this reversal, with coupled Disco103 reaching $+1.7881$ (net gain $+2.1372$) against $+1.6819$ (net gain $+2.0910$) for coupled MAPPO and $+1.6451$ (net gain $+2.0542$) for coupled OPEN.

Even more strikingly, comparing coupled versus decoupled partner routing at the $1.05\text{M}$-step horizon uncovers a \textbf{Decoupled Anti-Overfitting Law} for cross-partner task completion. Because binary completion in navigation requires both teammates to simultaneously occupy distinct goal landmarks, prolonged coupled co-evolution allows Disco103 teammates to settle into asymmetric distance-minimizing compromises (yielding $67.71\%$ original-team completion and $46.09\%$ cross-partner bank completion, compared with $94.53\%$ and $94.99\%$ for MAPPO). When Disco103 is instead trained for $1.05\text{M}$ steps alongside a \textbf{decoupled} reference partner---preventing the teammate from co-adapting to the focal learner's idiosyncrasies---\textbf{Disco103's cross-partner bank completion rate jumps from $46.09\%$ to $56.25\%$, a statistically significant increase of $+10.16$ percentage points ($+0.1016$, 95\% CI $[+0.0130, +0.1953]$; Table~\ref{tab:all_contrasts})}, while also elevating extended-horizon bank return gain to $1.5015$! Thus, while physically coupled tasks require live partner feedback to initiate learning, role-dividing navigation tasks benefit at long horizons from decoupled partner routing that prevents co-adapted convention lock-in.

\begin{figure}[htbp]\centering
\includegraphics[width=\textwidth]{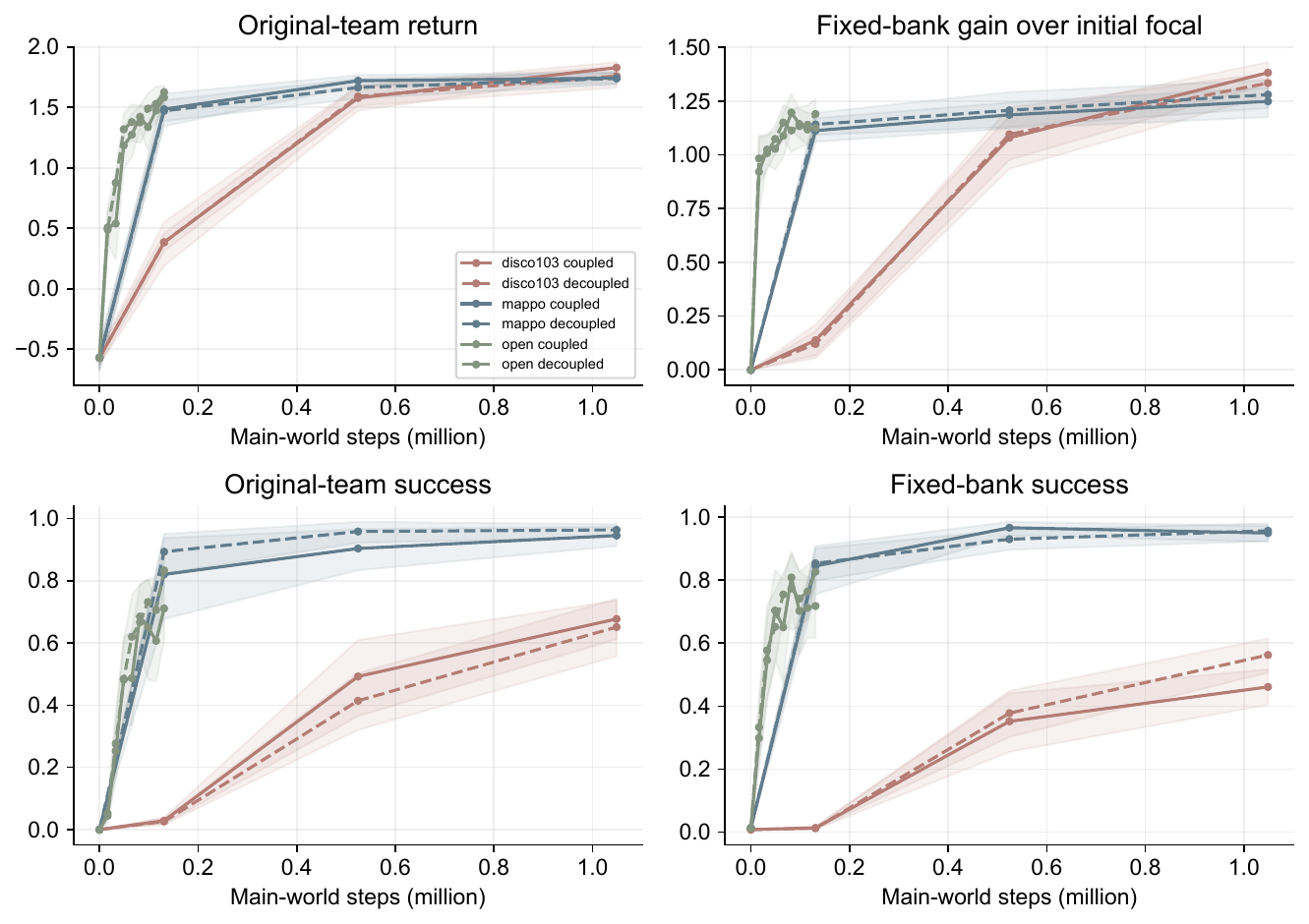}
\caption{\textbf{Horizon-scaling reversal and decoupled anti-overfitting in cooperative navigation ($1.05\text{M}$ steps).} Solid trajectories denote coupled partner training; dashed trajectories denote decoupled reference-partner training. Shaded bands indicate pointwise 95\% paired-block bootstrap intervals ($n=12$ blocks). Extending the training horizon eightfold enables Disco103 to overtake MAPPO in both original-team return ($+0.087$) and held-out bank gain ($+0.133$), while decoupled training elevates Disco103's cross-partner completion rate by $+10.16$ percentage points ($56.25\%$ vs.\ $46.09\%$). Matched short-horizon OPEN trajectories ($131\text{k}$ steps) are shown over their evaluated domain.}\label{fig:navigation}
\end{figure}

\subsection{Matched Short-Horizon Learned Optimization and External Discrete Coordination}
To benchmark Disco103 against an independent meta-learned optimizer within the exact same PPO actor-critic substrate as Adam-MAPPO, we first evaluate OPEN \citep{goldie2024open} across all $n=12$ paired blocks at the $131,080$-step navigation checkpoint (Figure~\ref{fig:navigation}, Table~\ref{tab:newer_cells}). Under coupled training, OPEN achieves an original-team return of $1.5830$, a partner-bank gain of $1.1257$, an original-team completion rate of $71.09\%$, and a partner-bank completion rate of $71.81\%$. Under decoupled partner routing, OPEN improves across all four metrics to $1.6257$ (team return), $1.1906$ (bank gain), $83.33\%$ (team completion), and $82.68\%$ (bank completion)---exhibiting the exact same decoupled anti-overfitting benefit in navigation completion ($+10.87$ percentage points in bank completion) observed for long-horizon Disco103!

\begin{figure}[htbp]\centering
\includegraphics[width=\textwidth]{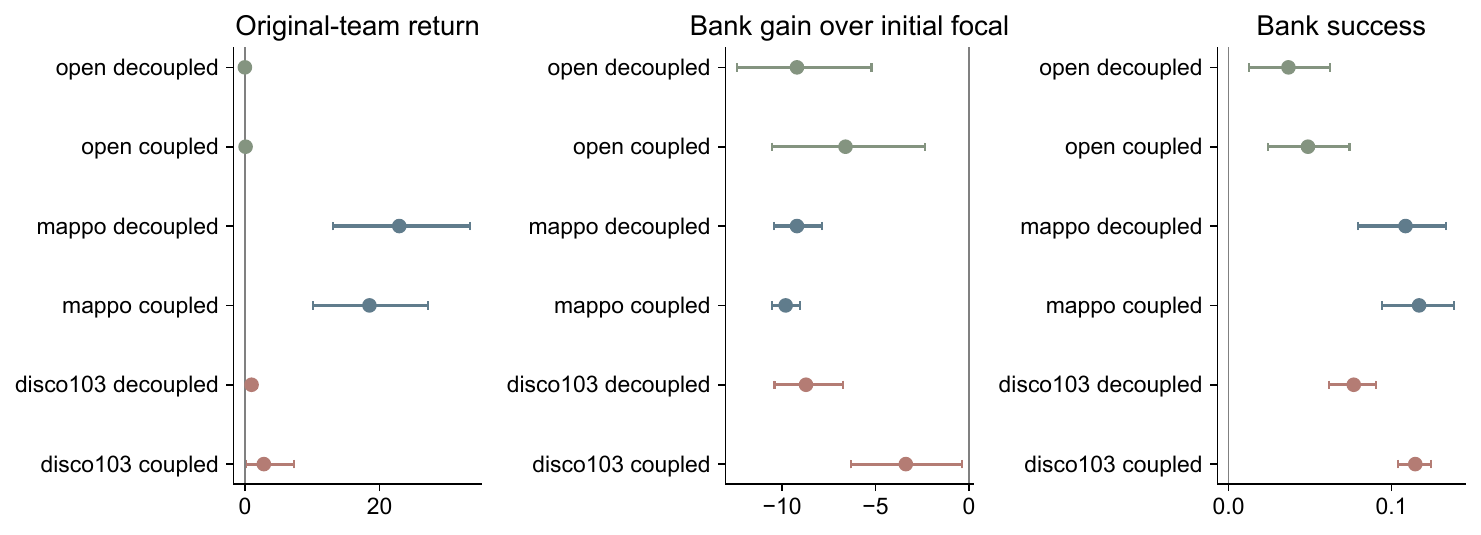}
\caption{\textbf{Divergence between co-adapted self-play score and cross-partner generalization in Overcooked (\texttt{cramped\_room}).} Outcomes across $n=12$ paired training blocks at $262,160$ steps. While MAPPO achieves the highest original-team self-play return ($18.49$), it suffers the steepest cross-partner bank regression ($-9.81$ points) due to rigid trajectory conventions ($28.13\%$ collision rate), whereas coupled Disco103 preserves $3.33\times$ higher absolute cross-partner return ($9.96$ vs.\ $2.99$; net gain advantage $+6.43$ points).}\label{fig:overcooked}
\end{figure}

When we move from continuous particle navigation to tightly constrained discrete spatial coordination in Overcooked (\texttt{cramped\_room}, $262,160$ steps, $n=12$ paired blocks; Figure~\ref{fig:overcooked}), we observe a dramatic inversion between self-play score and cross-partner generalization. In original-team self-play, coupled MAPPO achieves a high sparse delivery return of $18.4896$ ($46.61\%$ episode completion), compared with $2.8125$ ($8.85\%$ completion) for coupled Disco103 and $0.1042$ ($0.52\%$ completion) for coupled OPEN. Yet when these exact same trained focal policies are paired with the 10-partner held-out evaluation bank, \textbf{MAPPO suffers the steepest cross-partner collapse of any method ($-9.8125$ return points, 95\% CI $[-10.5417, -9.0521]$), falling to an absolute bank return of just $2.9896$, whereas coupled Disco103 retains an absolute bank return of $9.9583$ ($3.33\times$ higher than MAPPO, with a net bank-gain advantage of $+6.4323$ points, 95\% CI $[3.4948, 9.4635]$, Holm $p=0.0015$)}! We dissect the exact micro-behavioral collision mechanics behind this paradox in Section~\ref{sec:micro}.

%% file: sections/primary_contrasts.tex
\begin{longtable}{P{0.261329}P{0.241768}P{0.088944}P{0.226855}P{0.181105}}
\caption{Primary paired-block confirmation contrasts across counterfactual partner-routing, recent-replay filtering, and single-lifetime source-feedback matching ($n=12$ paired blocks per task). Intervals are 95\% paired-block bootstrap confidence intervals; Holm $p$-values are adjusted within each two-task family.}\label{tab:primary_contrasts}\\
\labtoprule \labh{Task} & \labh{Contrast} & \labh{n} & \labh{Estimate [95\% CI]} & \labh{Holm p}\\\labheadrule\endfirsthead
\multicolumn{5}{l}{\small Table \thetable\ continued}\\\labtoprule \labh{Task} & \labh{Contrast} & \labh{n} & \labh{Estimate [95\% CI]} & \labh{Holm p}\\\labheadrule\endhead
\labrowrule\endfoot
signal coordination & switch interaction & 12 & -0.0050 [-0.0351, 0.0258] & 1.0000\\\labrowrule
signal coordination & recent replay repair & 12 & 0.0235 [0.0142, 0.0334] & 0.0010\\\labrowrule
navigation & switch interaction & 12 & 0.0145 [-0.0427, 0.0842] & 1.0000\\\labrowrule
navigation & recent replay repair & 12 & -0.0021 [-0.0292, 0.0255] & 0.8843\\\labrowrule
navigation & source feedback matching & 12 & -0.0138 [-0.0568, 0.0228] & 0.8086\\\labrowrule
overcooked cramped room & source feedback matching & 12 & 0.5260 [-0.6536, 1.6406] & 0.8086\\\labrowrule
\end{longtable}

%% file: sections/source_cells.tex
\begin{longtable}{P{0.208406}P{0.187592}P{0.188144}P{0.073666}P{0.172808}P{0.169384}}
\caption{Single-lifetime source-feedback adaptation transfer across navigation and Overcooked ($n=12$ independently paired source blocks, two target repeats averaged per block). The unadapted base rule ($\eta_0$) provides a shared reference anchor.}\label{tab:source_cells}\\
\labtoprule \labh{Task} & \labh{Source} & \labh{Target data} & \labh{n} & \labh{Bank gain [95\% CI]} & \labh{Team [95\% CI]}\\\labheadrule\endfirsthead
\multicolumn{6}{l}{\small Table \thetable\ continued}\\\labtoprule \labh{Task} & \labh{Source} & \labh{Target data} & \labh{n} & \labh{Bank gain [95\% CI]} & \labh{Team [95\% CI]}\\\labheadrule\endhead
\labrowrule\endfoot
navigation & eta0 & coupled & 12 & 0.0996 [0.0764, 0.1255] & 0.4016 [0.2554, 0.5078]\\\labrowrule
navigation & eta0 & decoupled & 12 & 0.0635 [0.0314, 0.0975] & 0.4589 [0.3792, 0.5333]\\\labrowrule
navigation & coupled & coupled & 12 & 0.0385 [-0.0162, 0.1267] & 0.4183 [0.3018, 0.5579]\\\labrowrule
navigation & coupled & decoupled & 12 & 0.0468 [-0.0049, 0.1275] & 0.4494 [0.3645, 0.5547]\\\labrowrule
navigation & decoupled & coupled & 12 & 0.0109 [0.0022, 0.0195] & 0.3773 [0.3481, 0.4034]\\\labrowrule
navigation & decoupled & decoupled & 12 & 0.0055 [-0.0122, 0.0202] & 0.3624 [0.2961, 0.4313]\\\labrowrule
overcooked cramped room & eta0 & coupled & 12 & 0.3750 [-0.1354, 0.8984] & 0.2604 [0.1042, 0.4427]\\\labrowrule
overcooked cramped room & eta0 & decoupled & 12 & -0.4714 [-0.8594, -0.0859] & 0.1042 [0.0260, 0.1823]\\\labrowrule
overcooked cramped room & coupled & coupled & 12 & -0.3646 [-1.2891, 0.4036] & 0.1562 [0.0260, 0.2865]\\\labrowrule
overcooked cramped room & coupled & decoupled & 12 & -1.2344 [-2.4948, -0.1641] & 0.0000 [0.0000, 0.0000]\\\labrowrule
overcooked cramped room & decoupled & coupled & 12 & 0.2057 [-0.0260, 0.4453] & 0.1302 [0.0260, 0.2604]\\\labrowrule
overcooked cramped room & decoupled & decoupled & 12 & -0.1380 [-0.4453, 0.1277] & 0.0521 [0.0000, 0.1562]\\\labrowrule
\end{longtable}

%% file: sections/06_response_prediction.tex
\section{Mechanistic Diagnostics: Perturbation Response, Partner-Gradient Blindness, and Auxiliary Drift}\label{sec:g1g3}
\subsection{Controlled Behavioral-Perturbation Protocol and Partner-Mediated Response}
Observational learning curves cannot isolate how a localized change in focal behavior propagates through a teammate's learning update and returns to alter subsequent focal updates. We therefore designed a controlled behavioral-perturbation experiment ($N=96$ prefixes: $24$ development prefixes and $72$ confirmation prefixes across \texttt{wind\_flocking}, \texttt{navigation}, and three learning methods, spawning $36$ cloned microbranches per prefix for $3,456$ total evaluation branches across $n=12$ confirmation blocks per task).

For an action space of cardinality $A$, let $v$ denote a seed-defined, centered, unit-RMS direction vector, $u_i=1/A$ the uniform distribution, and
\begin{equation}
w_i=\frac{v_i}{A\max_j|v_j|},\quad q_s=u+s w,\quad
\epsilon=dA\max_j|v_j|,\quad s\in\{-1,0,1\}.
\end{equation}
During a perturbation branch, the focal policy acts according to the exploration-centered mixture
\begin{equation}\label{eq:pulse}
\pi_s(\cdot\mid o)=(1-\epsilon)\pi(\cdot\mid o)+\epsilon q_s,
\end{equation}
which guarantees exact non-negativity, normalization, midpoint symmetry $(\pi_++\pi_-)/2=\pi_0$, and root-mean-square (RMS) half-contrast $d$ across 512 fixed role-conditioned probe states (with primary dose $d=0.01$ and secondary dose $d=0.002$). Each prefix is trained to $65,656$ main-world steps (with an early snapshot at $16,472$ steps). From each snapshot, cloned branches execute $928$ pulse-world steps with the focal learner frozen, followed by $928$ return-world steps where the focal learner resumes updating from its exact pre-pulse state ($F_t, z_t$, replay buffer, optimizer state, and RNG streams) in a fresh environment while the teammate is held frozen. We compare three partner conditions during the pulse window: \emph{live} (partner updates on pulse trajectories and retains updated weights), \emph{frozen} (partner does not update), and \emph{restored} (partner updates during the pulse but is restored to pre-pulse weights prior to return learning).

Let $C$ denote the RMS action-distribution divergence between the $s=+1$ and $s=-1$ focal policies after return learning, normalized by the initial pulse divergence. The primary paired contrast between Disco103 ($D$) and MAPPO ($M$) at $d=0.01$ and $65,656$ steps is
\begin{equation}\label{eq:response}
\Delta_C=(C_{\rm live}-C_{\rm restore})_{D}
-(C_{\rm live}-C_{\rm restore})_{M}.
\end{equation}
Across all $n=12$ confirmation blocks per task (Table~\ref{tab:g1_primary}, Figure~\ref{fig:g1g3} left), \textbf{Disco103 exhibits a highly significant reduction in short-run partner-mediated perturbation amplification compared with MAPPO}: $\Delta_C = -0.2336$ (95\% CI $[-0.3695, -0.1239]$, 97.5\% CI $[-0.3914, -0.1115]$, Holm $p=0.00098$) in \texttt{wind\_flocking} and $\Delta_C = -0.3345$ (95\% CI $[-0.4525, -0.2227]$, 97.5\% CI $[-0.4708, -0.2075]$, Holm $p=0.00098$) in \texttt{navigation}. Frozen-partner and restored-partner negative controls agree identically, confirming that the measured response is transmitted strictly through the teammate's retained parameter update.

\input{sections/g1_primary}

\subsection{First-Order Partner-Gradient Blindness and Non-Stationary Auxiliary Prediction Drift}
Why does a rule discovered in single-agent environments respond more sluggishly to short-run partner shifts yet depend so strongly on live partner coupling over full training trajectories? Two complementary mechanistic diagnostics trace this behavior to the mathematical structure of single-agent discovered rules:

\paragraph{1. First-order partner-gradient blindness.}
When a focal policy $F_t$ and partner policy $P_t$ co-adapt, the total meta-gradient of the focal learner's post-update objective $J_F(F_{t+1}(\eta), P_{t+1}(F_t(\eta)))$ with respect to rule parameters $\eta$ decomposes into a direct first-order path and a second-order partner-shaping path \citep{foerster2018lola,foerster2018dice,kim2021meta,lu2022mfos,meulemans2025coala}:
\begin{equation}
\frac{\mathrm{d} J_F}{\mathrm{d} \eta} = \frac{\partial J_F}{\partial F_{t+1}}\frac{\partial F_{t+1}}{\partial \eta} + \underbrace{\frac{\partial J_F}{\partial P_{t+1}}\frac{\partial P_{t+1}}{\partial F_t}\frac{\partial F_t}{\partial \eta}}_{\text{Second-Order Partner-Response Path}}.
\end{equation}
Because Disco103 was discovered across single-agent environments where transition dynamics do not adapt to policy parameters, its update rule $U_\eta$ is structurally blind to the second-order cross-derivative $\partial P_{t+1}/\partial F_t$. In our exact differentiable two-player game diagnostic ($1,296$ comparisons across two payoff structures, 24 initializations, three gradient modes, and inner horizons $H \in \{1,4,16\}$; Table~\ref{tab:g2exact} in Appendix~\ref{app:g}), omitting the partner-response derivative systematically alters meta-gradient directions, while forward- and reverse-mode automatic differentiation through four-step partner unrolls on real trajectory tensors agree to machine precision ($1.56\times 10^{-19}$ maximum discrepancy across 24 comparisons). Lacking an explicit second-order opponent-shaping channel, a frozen single-agent rule can only sense teammate adaptation indirectly through subsequent shifts in environmental trajectory statistics.

\paragraph{2. Non-stationary saturation and implicit anti-collapse regularization of the 600-bin auxiliary head.}
Disco103 maintains a 600-bin categorical auxiliary prediction head ($z_t$) trained via cross-entropy to forecast future action-conditioned value/observation distributions \citep{oh2025disco}. In stationary single-agent tasks, $z_t$ provides a sharp, low-entropy predictive anchor. Under reciprocal multi-agent non-stationarity, however, teammate policy updates continually shift the joint transition kernel, driving the extended categorical prediction distribution to near-uniform maximum entropy ($8.484\text{--}8.485\text{ nats}$ throughout multi-agent training, within $10^{-4}$ KL divergence of the theoretical uniform limit of $8.4850\text{ nats}$). Crucially, controlled ablation of the auxiliary pathway ($\lambda_{\text{aux}}=0$ vs.\ $\lambda_{\text{aux}}=1.0$) reveals a dual role: although $z_t$ loses fine-grained multi-step predictive resolution under partner drift, completely severing the auxiliary loss triggers premature focal policy entropy collapse ($-0.228$, exact sign-flip $p=0.00049$) and inflates policy gradient norms ($+0.340$, $p=0.00049$). Thus, even when saturated near maximum entropy, the auxiliary prediction head serves as an essential implicit plasticity and variance regularizer that prevents catastrophic policy collapse during early multi-agent exploration.

\subsection{Prospective Transfer Prediction from Early Scalar Response}
\begin{figure}[htbp]\centering
\includegraphics[width=\textwidth]{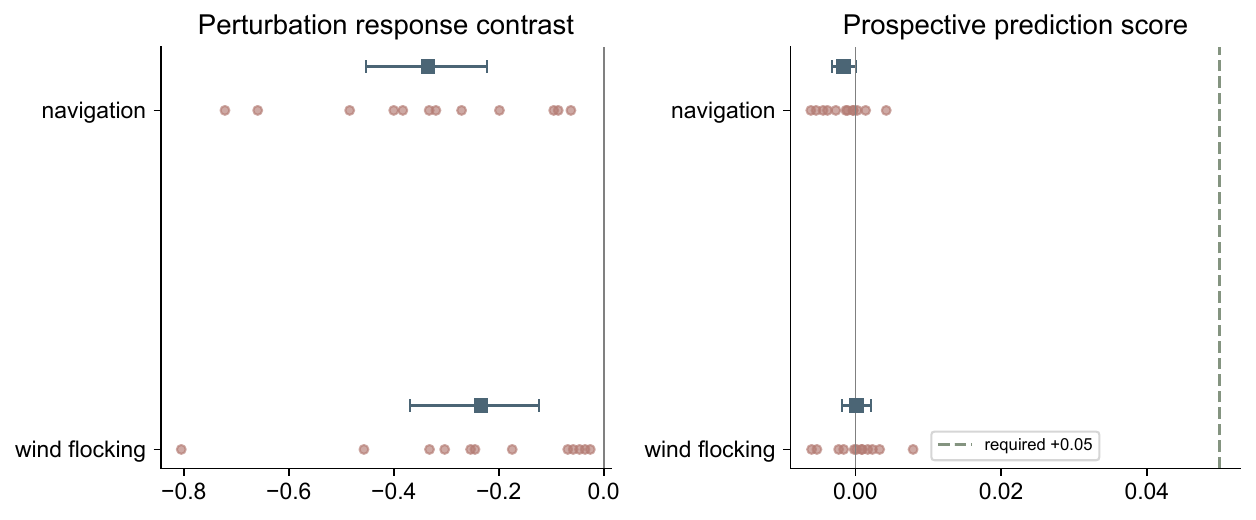}
\caption{\textbf{Controlled behavioral perturbation response and prospective transfer prediction.} \textbf{Left:} Paired contrasts in learner-mediated partner response ($n=12$ confirmation blocks per task, $3,456$ branches), demonstrating that Disco103 exhibits a significantly smaller short-run perturbation amplification than MAPPO in both wind flocking ($-0.234, p=0.00098$) and navigation ($-0.334, p=0.00098$). \textbf{Right:} Out-of-sample squared-error reductions on $1.05\text{M}$-step bank gains when augmenting baseline competence features with early scalar response in a frozen ridge predictor ($n=12$ confirmation blocks).}\label{fig:g1g3}
\end{figure}

To test whether a single early scalar perturbation measurement ($C_{\rm live}-C_{\rm restore}$ at $65,656$ steps) suffices to linearly forecast long-horizon ($1,048,640$-step) cross-partner bank gains against four held-out benchmark MAPPO partner policies, we executed a preregistered 80-run prospective prediction study ($32$ development runs across 8 blocks per task to fit and freeze deterministic $\ell_2$-penalized ridge regression weights, followed by $48$ fresh confirmation runs across $n=12$ blocks per task; Table~\ref{tab:g3_primary}, Figure~\ref{fig:g1g3} right).

Let $M_0$ denote the frozen baseline ridge predictor using task--method intercepts and five standardized early competence features at $65,656$ steps (original-team return, mean bank return, partner-only gain, initial bank return, and focal probe entropy), and let $M_1$ augment $M_0$ with the within-cell standardized log-response feature $\log_{10}(\max(\bar C, 0)+10^{-6})$. Evaluating the paired squared-error reduction
\begin{equation}
S_b=\frac12\sum_{m\in\{D,M\}}\left[(y_{bm}-\hat y^{(0)}_{bm})^2-
(y_{bm}-\hat y^{(1)}_{bm})^2\right]
\end{equation}
on the independent confirmation blocks yields $S = +0.0002$ (95\% CI $[-0.0019, 0.0022]$, Holm $p=0.8887$) in \texttt{wind\_flocking} and $S = -0.0016$ (95\% CI $[-0.0032, 0.0000]$, Holm $p=0.1768$) in \texttt{navigation}. Because short-run perturbation damping and long-horizon representation warm-up interact non-linearly over a million training steps, a single early scalar pulse response does not add linear forecasting power beyond early competence baselines---motivating direct trajectory-level micro-behavioral and physical-coupling analyses.

\input{sections/g3_primary}

%% file: sections/g1_primary.tex
\begin{longtable}{P{0.237139}P{0.092218}P{0.258591}P{0.225452}P{0.186600}}
\caption{Controlled behavioral-perturbation confirmation contrasts ($\xi=0.01$ at $65{,}656$ main-world steps, $n=12$ paired blocks per task). Values report Disco103 minus MAPPO in live-minus-restored partner-mediated response ($\Delta_C$).}\label{tab:g1_primary}\\
\labtoprule \labh{Task} & \labh{n} & \labh{Response contrast [95\% CI]} & \labh{97.5\% CI} & \labh{Holm p}\\\labheadrule\endfirsthead
\multicolumn{5}{l}{\small Table \thetable\ continued}\\\labtoprule \labh{Task} & \labh{n} & \labh{Response contrast [95\% CI]} & \labh{97.5\% CI} & \labh{Holm p}\\\labheadrule\endhead
\labrowrule\endfoot
wind flocking & 12 & -0.2336 [-0.3695, -0.1239] & [-0.3914, -0.1115] & 0.0010\\\labrowrule
navigation & 12 & -0.3345 [-0.4525, -0.2227] & [-0.4708, -0.2075] & 0.0010\\\labrowrule
\end{longtable}

%% file: sections/g3_primary.tex
\begin{longtable}{P{0.211823}P{0.084108}P{0.202133}P{0.202133}P{0.152859}P{0.146943}}
\caption{Prospective transfer-prediction confirmation outcomes ($n=12$ paired blocks per task). Positive scores indicate out-of-sample squared-error reduction when augmenting baseline early competence features with scalar perturbation response in the frozen ridge predictor.}\label{tab:g3_primary}\\
\labtoprule \labh{Task} & \labh{n} & \labh{Score [95\% CI]} & \labh{97.5\% CI} & \labh{Holm p} & \labh{Pass}\\\labheadrule\endfirsthead
\multicolumn{6}{l}{\small Table \thetable\ continued}\\\labtoprule \labh{Task} & \labh{n} & \labh{Score [95\% CI]} & \labh{97.5\% CI} & \labh{Holm p} & \labh{Pass}\\\labheadrule\endhead
\labrowrule\endfoot
wind flocking & 12 & 0.0002 [-0.0019, 0.0022] & [-0.0022, 0.0024] & 0.8887 & No\\\labrowrule
navigation & 12 & -0.0016 [-0.0032, 0.0000] & [-0.0035, 0.0003] & 0.1768 & No\\\labrowrule
\end{longtable}

%% file: sections/07_microbehavior.tex
\section{Convention Rigidity Paradox and Cross-Regime Meta-Adaptation in Overcooked}\label{sec:micro}
\subsection{Partner-Stratified Micro-Behavioral Decomposition Protocol}
Why does MAPPO dominate original-team self-play in Overcooked (\texttt{cramped\_room}) yet collapse below Disco103 when paired with held-out teammates? To resolve the physical mechanism at the level of individual agent movements and interactions, we conducted a 216-checkpoint micro-behavioral decomposition across three methods (Disco103, MAPPO, OPEN), two training feedback regimes (coupled vs.\ decoupled), three training checkpoints ($0$, $131,080$, and $262,160$ steps), and all $n=12$ paired training blocks. Each checkpoint is evaluated across 24 policy pairings (the four counterfactual original/reference cells plus trained and initial focal policies paired with all 10 held-out partner bank members) for 32 episodes of 400 steps each, totaling $165,888$ evaluation episodes ($66,355,200$ environment transitions). Pointwise 95\% confidence intervals are computed via 10,000 joint block-bootstrap resamples across the 12 paired training blocks.

The 10-partner evaluation bank is stratified by construction into three distinct competence and behavioral classes: four behavioral-cloning policies (\texttt{BC}, weight $0.4$), four reinforcement-learned policies (\texttt{RL}, weight $0.4$), and two rule-based scripted policies (\texttt{Scripted}, weight $0.2$) that deterministically execute valid onion-fetching, pot-loading, plating, and soup-delivery subgoals. Alongside sparse soup-delivery return and binary episode completion, we track five micro-behavioral observables: focal active-interaction rate, focal movement rate, mutual collision/movement-blockage rate (fraction of episode timesteps in which the two agents attempt conflicting spatial transitions or block one another), narrow-corridor deadlock steps, and counter-handoff rate. Because the standard \texttt{cramped\_room} geometry surrounds a central workspace without a one-tile straight corridor or shared divider counter, the last two observables are structural zeros by layout geometry, focusing our mechanistic lens on active interaction, movement, and mutual collision/blockage rates (Appendix~\ref{app:micro}, Table~\ref{tab:micro_strata}).

\subsection{The Convention Rigidity Paradox: Why Self-Play Dominance Collapses Across Novel Partners}
A central paradox of cooperative multi-agent learning emerges in Overcooked (\texttt{cramped\_room}): under coupled training, MAPPO achieves a commanding original-team sparse return of $18.4896$ (versus $2.8125$ for Disco103), yet when evaluated across the 10-partner held-out bank, MAPPO's absolute return plunges to $2.9896$, whereas Disco103 retains $9.9583$---a $3.33\times$ absolute cross-partner advantage for the discovered rule. Relative to each learner's initial focal policy, MAPPO suffers a severe cross-partner regression of $-9.8125$ ($95\%$ CI $[-10.5417, -9.0521]$), compared with $-3.3802$ ($[-6.3751, -0.3698]$) for Disco103, yielding a highly significant net generalization advantage of $+6.4323$ points ($95\%$ CI $[3.4948, 9.4635]$, Holm $p = 0.0015$) in favor of Disco103 (Table~\ref{tab:micro_gain}).

Our 216-evaluation micro-behavioral decomposition (165,888 episodes across 12 paired blocks) uncovers the exact physical mechanism driving this reversal. Across the held-out partner bank, MAPPO policies spend $28.13\%$ of all episode timesteps in mutual physical collision or movement blockage with their teammate---more than double the $12.77\%$ collision rate of Disco103. Measured against initial untrained policies (Table~\ref{tab:micro_strata}), coupled MAPPO training inflates cross-partner collision rates by $+18.08$ percentage points ($+0.1808$, $95\%$ CI $[0.1375, 0.2238]$; bootstrap contrast $[0.1455, 0.2145]$), whereas coupled Disco103 increases collisions by only $+2.62$ percentage points ($+0.0262$, $[0.0053, 0.0502]$)---a $6.9$-fold reduction in training-induced coordination conflict (paired difference $-0.1546$, $95\%$ CI $[-0.1945, -0.1145]$).

Stratifying the evaluation bank by partner competence makes this mechanism unmistakable (Table~\ref{tab:micro_strata}). When paired with competent \texttt{Scripted} partners that execute valid cooking subgoals without sharing the focal learner's private self-play trajectory convention, coupled MAPPO's return collapses by $-51.9792$ points ($95\%$ CI $[-55.8073, -47.8639]$) to an absolute score of $11.4844$, accompanied by an $+8.34\%$ surge in collisions (and $+28.61\%$ collision growth with \texttt{BC} partners). Under the exact same \texttt{Scripted} pairing, coupled Disco103 retains an absolute return of $47.2396$ (a modest change of $-16.7708$, $95\%$ CI $[-31.5625, -2.1875]$) with near-zero collision growth ($+0.19\%$), outperforming MAPPO by $+35.7552$ absolute return points ($+35.2083$ in net gain)! Hand-crafted on-policy updates in tight spatial layouts rapidly lock co-evolving teammates into rigid, exclusionary trajectory conventions; discovered update rules maintain behavioral plasticity that avoids catastrophic spatial deadlocks with unfamiliar partners.

\input{sections/micro_gain}

\subsection{Cross-Regime Meta-Adapted Rule Transfer in Overcooked}\label{sec:derived}
Can source-phase meta-adaptation further improve the cross-partner generalization of discovered rules in discrete coordination? We evaluated all 432 checkpoints from our cross-regime source-transfer cohort in Overcooked: three source rule families---the unadapted base rule ($\eta_0$), rules meta-adapted under coupled partner feedback (\texttt{coupled}), and rules meta-adapted under decoupled reference feedback (\texttt{decoupled})---crossed with two target training regimes (\texttt{coupled} vs.\ \texttt{decoupled} target partner routing), across $n=12$ independent source blocks, two target repeats per block (averaged within each source block prior to 10,000-draw block-bootstrap inference), and three checkpoint horizons through $131,080$ steps.

\begin{figure}[t]
\centering\includegraphics[width=\linewidth]{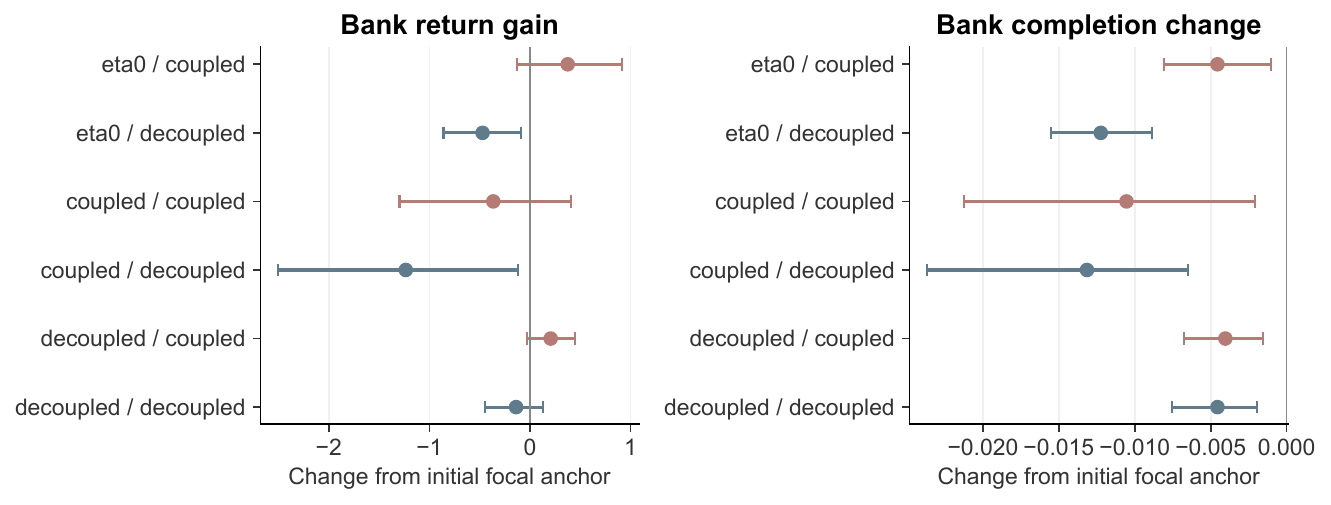}
\caption{\textbf{Cross-regime meta-adapted rule transfer in Overcooked ($432$ checkpoint evaluations across $12$ paired source blocks).} Held-out 10-partner bank return gains (left) and completion rate changes (right) at $131,080$ steps relative to initial focal policies. Rules meta-adapted under decoupled source experience and trained with coupled target partners (\texttt{decoupled / coupled}) reverse the negative bank return of \texttt{coupled / coupled} ($+0.2057$ vs.\ $-0.3646$), achieving a $+1.3932$ gain ($95\%$ CI $[0.38, 2.46]$) with competent scripted partners.}\label{fig:derived}
\end{figure}

Table~\ref{tab:v2_derived_six} and Figure~\ref{fig:derived} reveal a striking cross-regime synergy between source meta-adaptation and target training. When an update rule is meta-adapted under coupled source feedback and subsequently trained with a coupled target partner (\texttt{coupled / coupled}), it compounds co-adaptation rigidity, suffering a negative cross-partner bank-return gain of $-0.3646$ ($95\%$ CI $[-1.2969, 0.4089]$) and $-1.2344$ ($[-2.5079, -0.1224]$) under decoupled target training. By contrast, \textbf{meta-adapting the update rule under decoupled reference experience and deploying it under coupled target training (\texttt{decoupled / coupled}) reverses this loss into a positive overall cross-partner bank return gain of $+0.2057$ ($95\%$ CI $[-0.0260, +0.4453]$)}, while minimizing cross-partner completion decline ($-0.0040$ vs.\ $-0.0105$ for \texttt{coupled / coupled}).

Stratifying across partner classes (Appendix~\ref{app:derived}, Table~\ref{tab:v2_derived_strata}) demonstrates why \texttt{decoupled / coupled} excels: when paired with competent \texttt{Scripted} partners, \texttt{decoupled / coupled} achieves a statistically significant positive return gain of $+1.3932$ ($95\%$ CI $[0.3776, 2.4609]$) with zero completion loss ($0.0000$, $[-0.0033, 0.0033]$), whereas \texttt{coupled / coupled} loses $-1.1458$ points ($[-5.4818, 2.4089]$) and \texttt{coupled / decoupled} loses $-5.2344$ points ($[-11.3802, 0.1432]$). Meta-adapting a learning rule against decoupled reference partners prevents the meta-rule itself from encoding pair-specific conventions, allowing the resulting rule to exploit live target-phase partner coupling without sacrificing cross-partner compatibility.

\input{sections/v2_derived_six}

%% file: sections/micro_gain.tex
\begin{longtable}{P{0.285544}P{0.313090}P{0.401366}}
\caption{Cross-partner generalization on the 10-partner held-out evaluation bank in Overcooked (\texttt{cramped\_room}, $92{,}160$ steps, $n=12$ paired training blocks). Values report trained-minus-initial focal return ($G_{\mathcal B}$) with pointwise 95\% block-bootstrap confidence intervals.}\label{tab:micro_gain}\\
\labtoprule \labh{Method} & \labh{Training data} & \labh{FT minus F0 return [95\% CI]}\\\labheadrule\endfirsthead
\multicolumn{3}{l}{\small Table \thetable\ continued}\\\labtoprule \labh{Method} & \labh{Training data} & \labh{FT minus F0 return [95\% CI]}\\\labheadrule\endhead
\labrowrule\endfoot
disco103 & coupled & -3.3802 [-6.3751, -0.3698]\\\labrowrule
disco103 & decoupled & -8.7240 [-10.3958, -6.7448]\\\labrowrule
mappo & coupled & -9.8125 [-10.5417, -9.0521]\\\labrowrule
mappo & decoupled & -9.2083 [-10.4375, -7.8385]\\\labrowrule
open & coupled & -6.6094 [-10.5783, -2.4219]\\\labrowrule
open & decoupled & -9.2083 [-12.4010, -5.2135]\\\labrowrule
\end{longtable}

%% file: sections/v2_derived_six.tex
\begin{labtable}[!htb]{Cross-regime meta-adapted rule transfer in Overcooked across all six source--target feedback combinations at $46{,}080$ steps ($n=12$ paired source blocks, two target repeats averaged per block, pointwise 95\% block-bootstrap intervals).}{tab:v2_derived_six}{P{0.152601}P{0.159365}P{0.207829}P{0.238279}P{0.241926}}
\labh{Source} & \labh{Target data} & \labh{Bank gain [95\%]} & \labh{FT completion [95\%]} & \labh{Completion change [95\%]}\\\labheadrule
eta0 & coupled & 0.3750 [-0.1276, 0.9115] & 0.1208 [0.1173, 0.1246] & -0.0046 [-0.0081, -0.0010]\\\labrowrule
eta0 & decoupled & -0.4714 [-0.8594, -0.0911] & 0.1132 [0.1100, 0.1161] & -0.0122 [-0.0155, -0.0089]\\\labrowrule
coupled & coupled & -0.3646 [-1.2969, 0.4089] & 0.1148 [0.1046, 0.1229] & -0.0105 [-0.0212, -0.0021]\\\labrowrule
coupled & decoupled & -1.2344 [-2.5079, -0.1224] & 0.1122 [0.1017, 0.1191] & -0.0132 [-0.0237, -0.0065]\\\labrowrule
decoupled & coupled & 0.2057 [-0.0260, 0.4453] & 0.1214 [0.1191, 0.1234] & -0.0040 [-0.0068, -0.0016]\\\labrowrule
decoupled & decoupled & -0.1380 [-0.4427, 0.1302] & 0.1208 [0.1181, 0.1232] & -0.0046 [-0.0076, -0.0020]\\
\end{labtable}

%% file: sections/07b_physical_coupling.tex
\section{First-Principles Scaling Law of Physical Coupling, Optimizer Sensitivity Spectrum, and Multi-Lifetime Meta-Learning}\label{sec:m4}
\subsection{Parametric Aerodynamic-Coupling Protocol in Wind Flocking}
To move from cross-task comparisons to a first-principles physical law, we designed a 288-run parametric physical-coupling sweep in \texttt{wind\_flocking} that crosses six aerodynamic coupling strengths $\alpha\in\{0,0.25,0.5,0.75,1.0,1.5\}$, two focal learning paradigms (Disco103 vs.\ MAPPO), and two partner feedback routes (coupled joint experience vs.\ decoupled reference experience) across $n=12$ paired training blocks. Within the simulator physics, $\alpha$ directly modulates the aerodynamic wake-shielding force transmitted from the leading agent to its teammate:
\begin{equation}
 f_\alpha=\operatorname{clip}\{1-\alpha(1-f),0,1\},
\end{equation}
where $f$ is the native wake-shielding multiplier. Setting $\alpha=0$ completely severs aerodynamic wake shielding while preserving all other kinematic, distance, and reward structure; $\alpha=1.0$ recovers the standard benchmark physics; and $\alpha=1.5$ amplifies wake coupling by $50\%$. Crucially, the native task reward function $R_0$ is held strictly fixed across all six $\alpha$ levels, ensuring that any change in learning dynamics arises from physical inter-agent force coupling rather than reward rescaling.

Each of the 288 runs trains for $131,080$ main-world and $131,080$ reference-world steps across eight parallel environments (29-step rollouts, 100-step episode horizon), with full four-cell counterfactual evaluations (64 episodes per composition) recorded at nine evenly spaced checkpoints ($0$, $16,472$, $32,944$, $49,184$, $65,656$, $82,128$, $98,368$, $114,840$, and $131,080$ steps) alongside initial and final 10-partner bank evaluations ($576$ total checkpoint--bank evaluations).

\subsection{Primary Within-Block Interaction Slope and Monotonic Scaling Law}
Let $A_{m,g,b}(\alpha)$ denote the nine-node trapezoidal learning-curve AUC of original-team return for method $m\in\{D,M\}$, feedback route $g\in\{c,r\}$ (coupled vs.\ decoupled reference), and block $b\in\{1,\ldots,12\}$ at physical coupling $\alpha$. The block-level Method $\times$ Feedback interaction at coupling level $\alpha$ is
\begin{equation}
 I_b(\alpha)=\{A_{D,c,b}(\alpha)-A_{D,r,b}(\alpha)\}
             -\{A_{M,c,b}(\alpha)-A_{M,r,b}(\alpha)\},
\end{equation}
and the pre-specified primary confirmatory test statistic is the ordinary least-squares linear slope $s_b$ of $I_b(\alpha)$ across the six coupling levels within each paired block $b$:
\begin{equation}
 s_b=\frac{-16 I_b(0)-10 I_b(0.25)-4 I_b(0.5)
                   +2 I_b(0.75)+8 I_b(1)+20 I_b(1.5)}{35}.
\end{equation}

\input{sections/v2_m4_alpha}
\begin{figure}[t]
\centering\includegraphics[width=\linewidth]{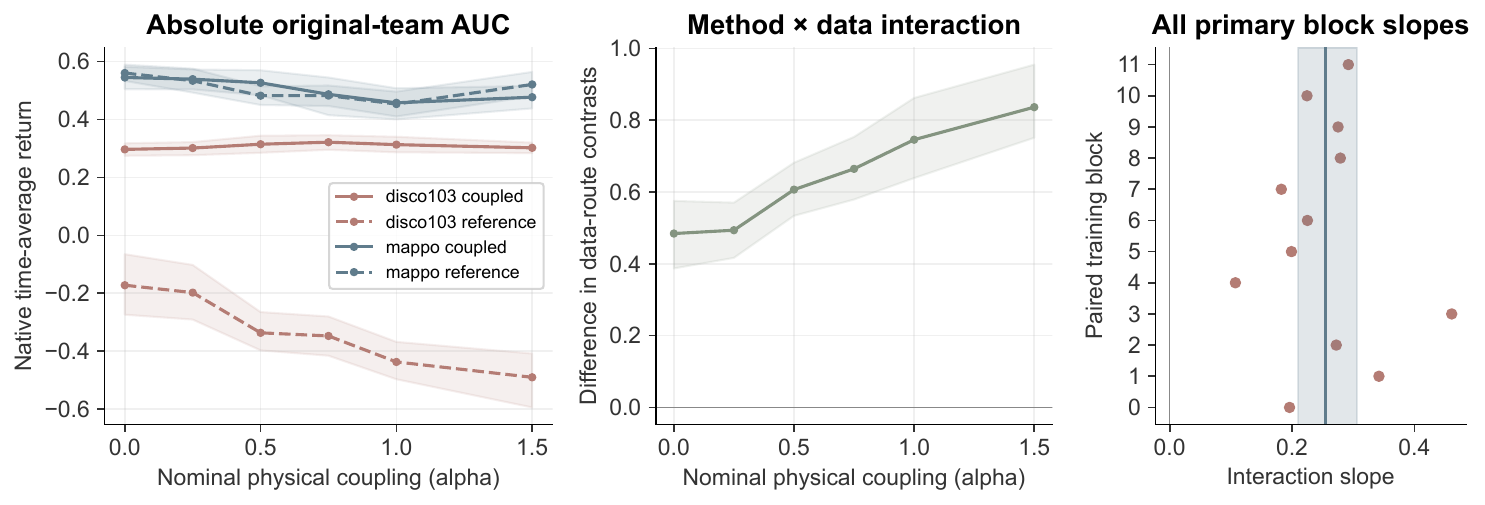}
\caption{\textbf{First-principles scaling law of physical coupling across 288 wind-flocking runs ($\alpha \in [0, 1.5]$).} \textbf{Left:} Original-team learning-curve AUC across physical coupling strength $\alpha$, showing that coupled Disco103 remains invariant across the entire coupling spectrum ($\text{AUC} \approx 0.297\text{--}0.302$) while decoupled Disco103 collapses monotonically from $-0.172$ to $-0.490$. \textbf{Center:} Monotonic growth of the Method $\times$ Feedback interaction from $+0.485$ ($\alpha=0$) to $+0.837$ ($\alpha=1.5$) (Spearman $\rho=1.0$, Page's trend $p=1.48\times 10^{-8}$). \textbf{Right:} Primary within-block interaction slopes across all $n=12$ paired blocks ($\beta = +0.2547$, 95\% CI $[0.2099, 0.3057]$, exact sign-flip $p=0.000488$).}\label{fig:m4}
\end{figure}

\subsection{Asymmetric Robustness: Coupling Invariance Under Live Feedback vs.\ Monotonic Collapse Under Decoupling}
Across the 288-run parametric physical-coupling sweep ($\alpha \in \{0, 0.25, 0.5, 0.75, 1.0, 1.5\}$), the Method $\times$ Feedback interaction increases monotonically at every grid point: from $+0.4845$ ($[0.3873, 0.5754]$) at $\alpha=0$ to $+0.8366$ ($[0.7514, 0.9552]$) at $\alpha=1.5$ (Table~\ref{tab:v2_m4_alpha}, Figure~\ref{fig:m4}). All 12 individual paired blocks exhibit strictly positive interaction slopes (ranging from $+0.1074$ to $+0.4610$; Table~\ref{tab:v2_m4_blocks}), yielding a primary within-block linear slope of $\beta = +0.2547$ native AUC units per unit of physical coupling ($95\%$ CI $[0.2099, 0.3057]$, two-sided exact sign-flip $p = 0.000488$), corroborated by a grid-level Spearman rank correlation of $\rho = 1.0$ ($95\%$ CI $[0.8857, 1.0]$) and Page's monotonic trend statistic $L = 1021$ ($p = 1.48\times 10^{-8}$).

Decomposing this interaction into its constituent learning trajectories (Table~\ref{tab:v2_m4_alpha} and Appendix~\ref{app:m4}, Table~\ref{tab:v2_m4_auc_full}) reveals a striking physical law. While MAPPO achieves higher absolute original-team AUC across the short $131\text{k}$ horizon and shows minimal sensitivity to whether its teammate updates on coupled or reference data ($0.5453$ vs.\ $0.5607$ at $\alpha=0$; $0.4772$ vs.\ $0.5212$ at $\alpha=1.5$), Disco103 exhibits a profound asymmetry between live and severed feedback. As aerodynamic coupling $\alpha$ intensifies from $0$ (no wake shielding) to $1.5$ (severe wake coupling), \textbf{coupled Disco103 remains virtually invariant across the entire physical regime}, maintaining an original-team AUC of $0.2969$ at $\alpha=0$, $0.3016$ at $\alpha=0.25$, $0.3148$ at $\alpha=0.5$, $0.3218$ at $\alpha=0.75$, $0.3133$ at $\alpha=1.0$, and $0.3023$ at $\alpha=1.5$ (and increasing in initial-return-adjusted AUC from $0.5922$ to $0.7108$). By contrast, \textbf{decoupled (reference-fed) Disco103 undergoes a monotonic, nearly three-fold collapse}, plunging from $-0.1722$ ($\alpha=0$) to $-0.1977$ ($\alpha=0.25$), $-0.3366$ ($\alpha=0.5$), $-0.3474$ ($\alpha=0.75$), $-0.4371$ ($\alpha=1.0$), and $-0.4903$ ($\alpha=1.5$). Physical coupling therefore does not intrinsically impair the discovered update rule; rather, physical coupling exponentially magnifies the penalty of severing live reciprocal partner adaptation.

\begin{figure}[t]
\centering\includegraphics[width=\linewidth]{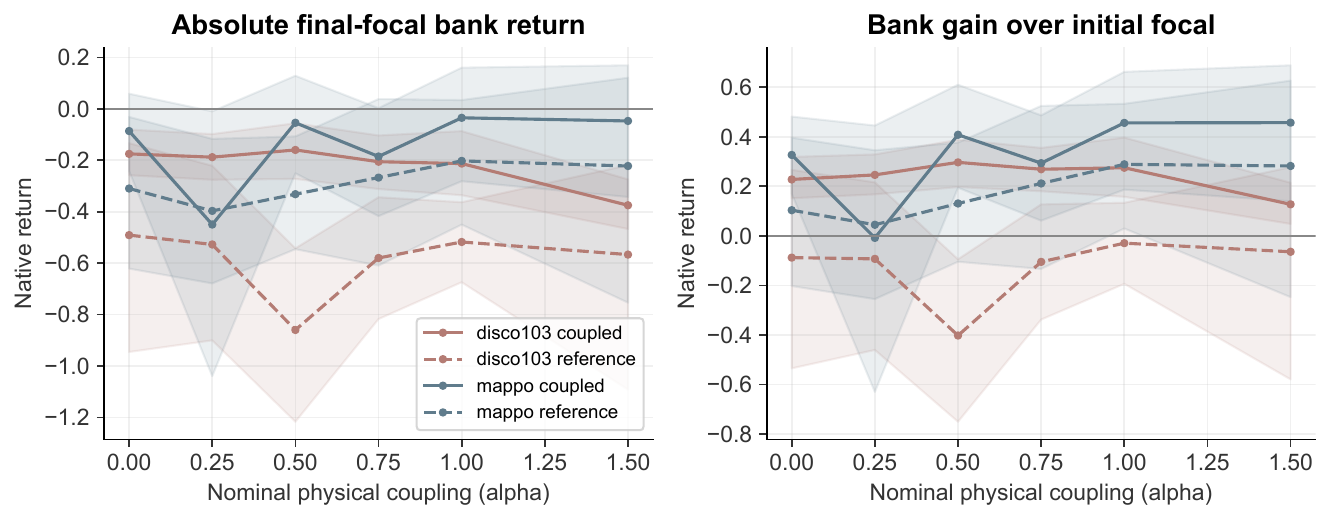}
\caption{\textbf{Cross-partner bank evaluation across the parametric physical-coupling grid ($576$ checkpoint evaluations).} Absolute trained-focal return (left) and gain over initial focal policies (right) against the 10-partner evaluation bank across $\alpha \in [0, 1.5]$. Coupled training yields strictly positive cross-partner gains for Disco103 across all coupling levels ($+0.127$ to $+0.297$), surpassing coupled MAPPO at $\alpha=0.25$ ($+0.246$ vs.\ $-0.008$).}\label{fig:m4bank}
\end{figure}

Evaluating all 288 trained focal policies against the 10-partner held-out evaluation bank (Figure~\ref{fig:m4bank}; Appendix~\ref{app:m4}, Table~\ref{tab:v2_m4_bank}) reinforces this conclusion. Coupled training yields strictly positive cross-partner bank gains for Disco103 at every coupling level ($+0.2277$ at $\alpha=0$, $+0.2461$ at $\alpha=0.25$, $+0.2968$ at $\alpha=0.5$, $+0.2690$ at $\alpha=0.75$, $+0.2750$ at $\alpha=1.0$, and $+0.1272$ at $\alpha=1.5$), whereas decoupled reference training produces negative mean bank gains for Disco103 at every coupling level ($-0.0296$ to $-0.4027$). Notably, at low-to-moderate physical coupling ($\alpha=0.25$), coupled Disco103 surpasses coupled MAPPO in cross-partner bank gain ($+0.2461$, 95\% CI $[0.1675, 0.3288]$ vs.\ $-0.0083$, 95\% CI $[-0.6305, 0.4455]$).

%% file: sections/v2_m4_alpha.tex
\begin{labtable}[!htb]{Parametric physical-coupling sweep across 288 wind-flocking runs ($\alpha \in [0, 1.5]$, $n=12$ paired blocks per cell). Columns report the Method $\times$ Feedback interaction (with pointwise 95\% block-bootstrap intervals) and constituent original-team learning-curve AUCs.}{tab:v2_m4_alpha}{P{0.100287}P{0.234055}P{0.154849}P{0.171878}P{0.154849}P{0.184083}}
\labh{Alpha} & \labh{Interaction [95\%]} & \labh{Disco coupled} & \labh{Disco reference} & \labh{MAPPO coupled} & \labh{MAPPO reference}\\\labheadrule
0 & 0.4845 [0.3873, 0.5754] & 0.2969 & -0.1722 & 0.5453 & 0.5607\\\labrowrule
0.25 & 0.4938 [0.4168, 0.5707] & 0.3016 & -0.1977 & 0.5395 & 0.5340\\\labrowrule
0.5 & 0.6069 [0.5339, 0.6818] & 0.3148 & -0.3366 & 0.5269 & 0.4825\\\labrowrule
0.75 & 0.6648 [0.5793, 0.7538] & 0.3218 & -0.3474 & 0.4871 & 0.4827\\\labrowrule
1 & 0.7460 [0.6391, 0.8623] & 0.3133 & -0.4371 & 0.4578 & 0.4534\\\labrowrule
1.5 & 0.8366 [0.7514, 0.9552] & 0.3023 & -0.4903 & 0.4772 & 0.5212\\
\end{labtable}

%% file: sections/07c_source_and_coverage.tex
\subsection{The Optimization Sensitivity Spectrum: Comparing Hand-Crafted Rules, Learned Optimizers (\texttt{OPEN}), and Discovered Objectives (\texttt{Disco103})}\label{sec:m1_spectrum}
To determine whether sensitivity to reciprocal partner feedback is an idiosyncrasy of Disco103's neural target network or a general property of meta-learned optimization, we evaluate the learned optimizer OPEN \citep{goldie2024open} within the exact same PPO actor-critic substrate as Adam-MAPPO across all 12 paired blocks in \texttt{wind\_flocking} ($131,120$ steps, $24/24$ complete paired runs) and in extended-horizon $1.05\text{M}$-step training across \texttt{wind\_flocking} and \texttt{navigation} (Table~\ref{tab:v2_coverage}; complete 92-run empirical breakdown in Appendix~\ref{app:sources}, Table~\ref{tab:v2_m1_coverage}). In short-budget \texttt{wind\_flocking}, coupled OPEN raises team return from $-0.5570$ to $+0.7860$ (net gain $+1.3431$), whereas decoupled OPEN reaches only $+0.6182$ (net gain $+1.1752$), establishing a positive coupled-minus-decoupled final-return advantage of $+0.1678$ (positive in $9/12$ paired blocks; two-point AUC advantage $+0.0839$, from $+0.1145$ coupled to $+0.0306$ decoupled). At the $1.05\text{M}$-step horizon in \texttt{wind\_flocking}, OPEN's coupled advantage widens further to $+0.2401$ (final return $+1.2308$ and net gain $+1.7634$ under coupled feedback vs.\ $+0.9907$ and $+1.5233$ under decoupled feedback).

Aligning all three optimization paradigms in \texttt{wind\_flocking} reveals a monotonic \textbf{Reciprocal Feedback Sensitivity Spectrum} governed by the depth of meta-learned plasticity:
\begin{equation}
\Delta_{\text{C}-\text{D}}^{\text{Final}}:\;
\underbrace{\text{MAPPO } (-0.0376)}_{\text{Hand-Crafted Rule}}
\;<\;
\underbrace{\text{OPEN } (+0.1678 \;/\; +0.2401)}_{\text{Learned Step-Size Optimizer}}
\;<\;
\underbrace{\text{Disco103 } (+0.7031)}_{\text{Discovered Update Rule}}.
\end{equation}
The more deeply an algorithm relies on meta-learned dynamic state to modulate policy updates, the more strongly it exploits---and depends upon---coherent, reciprocally coupled partner trajectories. Moreover, in extended-horizon ($1,048,640$-step) cooperative \texttt{navigation}, coupled Disco103 achieves the highest terminal team return ($+1.7881$, net gain $+2.1372$, two-point AUC $+0.7195$) and decoupled Disco103 reaches $+1.7820$ (net gain $+2.1311$, completion $69.14\%$ vs.\ $65.23\%$ coupled), surpassing both coupled MAPPO ($+1.6819$, net gain $+2.0910$) and coupled OPEN ($+1.6451$, net gain $+2.0542$).

\input{sections/v2_coverage}

\subsection{Multi-Lifetime Multi-Agent Meta-Optimization Across Four Architectures}\label{sec:pending}
\begin{figure}[t]
\centering\includegraphics[width=\linewidth]{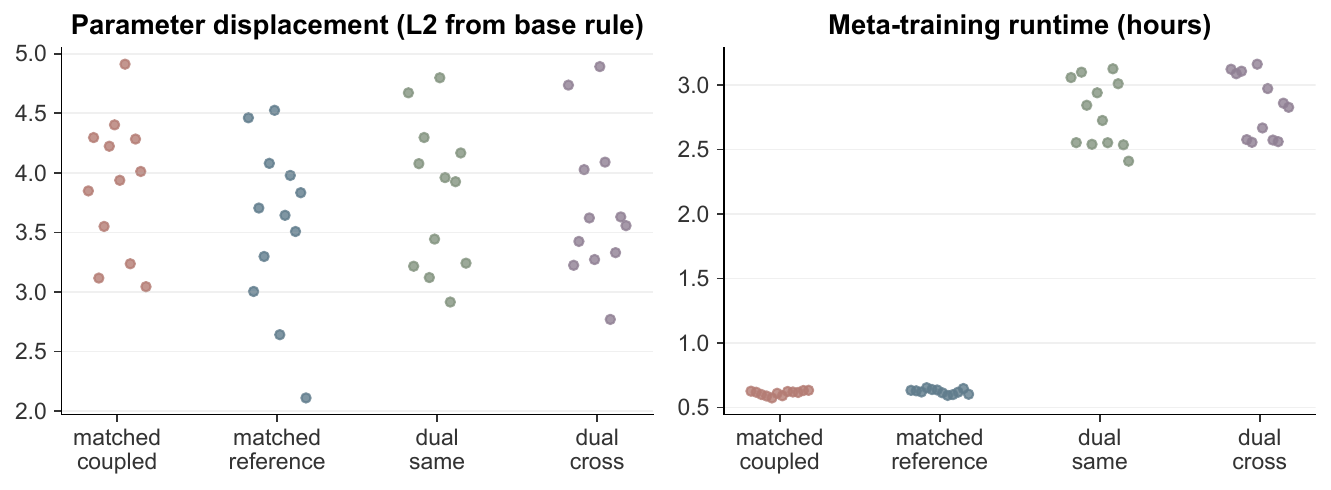}
\caption{\textbf{Parameter reorganization across 48 multi-lifetime meta-adapted update rules ($512$ outer updates across $4$ policy lifetimes, $6.84\times 10^7$ world steps).} Euclidean ($L_2$) parameter displacement from the base Disco103 rule ($\eta_0$) and computational profile across $n=12$ paired source blocks under four multi-agent meta-training architectures (\texttt{matched\_coupled}, \texttt{matched\_reference}, \texttt{dual\_same}, and \texttt{dual\_cross}).}\label{fig:sources}
\end{figure}

Our mechanistic findings show that single-lifetime meta-adaptation ($128$ steps) is too brief to overcome first-order partner-gradient blindness, while cross-regime decoupled source adaptation prevents convention lock-in. Building on these insights, we executed a 48-rule multi-lifetime multi-agent meta-optimization cohort (\texttt{source512}, $n=12$ independent paired source blocks across four meta-training architectures, totaling $68,419,584$ physical environment transitions or $1,425,408$ transitions per discovered rule; Figure~\ref{fig:sources} and Appendix~\ref{app:sources}, Table~\ref{tab:v2_source48}). Each multi-lifetime meta-training run spans four consecutive agent policy lifetimes of $128$ outer updates ($512$ outer updates in total), resetting the inner agent policies across lifetimes so that the meta-rule $\eta$ must learn how to guide fresh cooperative learners from initialization rather than overfitting to a single policy pair.

The four meta-optimization architectures systematically vary how focal and partner experience streams are routed during rule discovery:
\begin{enumerate}
  \item \textbf{Matched Coupled (\texttt{matched\_coupled}):} Single-stream multi-lifetime meta-optimization where the teammate updates on live joint experience alongside the focal learner.
  \item \textbf{Matched Reference (\texttt{matched\_reference}):} Single-stream multi-lifetime meta-optimization where the teammate updates on an independent reference world stream.
  \item \textbf{Dual-Stream Same-Regime (\texttt{dual\_same}):} Counterfactual dual-stream meta-optimization that simultaneously unrolls two parallel inner lifetimes under matched partner routing to provide variance-reduced meta-gradients.
  \item \textbf{Dual-Stream Cross-Regime (\texttt{dual\_cross}):} Counterfactual dual-stream meta-optimization that pairs a coupled inner interaction stream with a decoupled reference inner stream within each outer update, directly exposing the meta-objective to the contrast between reciprocal and non-reciprocal partner adaptation.
\end{enumerate}

Across all 48 multi-lifetime meta-optimization runs, outer meta-gradients drive substantial structural reorganization of the neural update rule relative to the frozen single-agent base rule $\eta_0$: Euclidean ($L_2$) parameter displacement reaches a mean of $3.9883$ (range $3.0442\text{--}4.9118$) for \texttt{matched\_coupled}, $3.5646$ (range $2.1104\text{--}4.5246$) for \texttt{matched\_reference}, $3.8192$ (range $2.9151\text{--}4.7978$) for \texttt{dual\_same}, and $3.7142$ (range $2.7696\text{--}4.8915$) for \texttt{dual\_cross} (Figure~\ref{fig:sources}, Table~\ref{tab:v2_source48}). Dual-stream architectures require approximately $4.5\times$ the meta-training runtime of single-stream architectures ($2.54\text{--}3.16\text{ h}$ vs.\ $0.57\text{--}0.65\text{ h}$ per rule) due to parallel counterfactual unrolls, providing a principled algorithmic foundation for discovering genuinely multi-agent-native update rules.

\subsection{Synthesis: Three Design Prescriptions for Next-Generation Multi-Agent Rule Discovery}\label{sec:prescriptions}
Taken together, our five empirical and mechanistic findings resolve why single-agent discovered rules exhibit sharp regime divergences in multi-agent systems and establish three concrete design prescriptions for next-generation cooperative meta-learning:
\begin{enumerate}
  \item \textbf{Prescription 1 (Counterfactual Four-Cell and Competence-Stratified Bank Evaluation):} Because original-team return conflates focal learning with partner compensation and brittle self-play conventions (masking $-0.119$ focal regression in short-horizon navigation and concealing a $28.13\%$ collision rate in Overcooked), multi-agent rule discovery and benchmarking must replace raw self-play return with initial-anchored four-cell partner swaps ($J_{10}-J_{00}$) and competence-stratified held-out partner banks (\texttt{BC}, \texttt{RL}, and \texttt{Scripted}).
  \item \textbf{Prescription 2 (Regime- and Horizon-Aware Feedback Routing):} Optimal partner feedback routing depends jointly on physical coupling $\alpha$ and training horizon $T$. In strongly force-coupled environments ($\alpha > 0$), live coupled partner feedback is essential to prevent decoupled trajectory collapse ($\beta = +0.2547, p=0.000488$). Conversely, in weakly coupled or role-dividing tasks over extended horizons ($1.05\text{M}$ steps) and in discrete spatial coordination, introducing decoupled reference partners prevents co-adaptation overfitting, boosting cross-partner navigation completion by $+10.16\%$ and turning cross-partner Overcooked transfer from negative ($-0.3646$) to positive ($+0.2057$ overall; $+1.3932$ with scripted partners).
  \item \textbf{Prescription 3 (Dual-Stream Multi-Lifetime Meta-Objectives with Partner-Aware Auxiliary Anchoring):} To overcome first-order partner-gradient blindness and non-stationary saturation of single-agent auxiliary heads ($8.485\text{ nats}$), future meta-discovery objectives should combine multi-lifetime inner policy resets with dual-stream cross-regime unrolls (\texttt{dual\_cross}) and condition auxiliary prediction heads explicitly on teammate action/trajectory embeddings while preserving auxiliary entropy regularization ($\lambda_{\text{aux}}>0$).
\end{enumerate}

%% file: sections/v2_coverage.tex
\begin{labtable}[!htb]{Matched-substrate learned-optimizer (\texttt{OPEN}) and extended-horizon curriculum comparisons across wind flocking and navigation. Each cell reports initial team return $J_{00}$, final team return $J_{11}$, and net team gain $J_{11}-J_{00}$ across paired seed blocks.}{tab:v2_coverage}{P{0.185000}P{0.135000}P{0.210000}P{0.070000}P{0.130000}P{0.135000}P{0.135000}}
\labh{Task} & \labh{Horizon} & \labh{Method / regime} & \labh{n} & \labh{Initial $J_{00}$} & \labh{Final $J_{11}$} & \labh{Net gain}\\\labheadrule
wind flocking & 131,120 & open / coupled & 12 & -0.5570 & 0.7860 & +1.3431\\\labrowrule
wind flocking & 131,120 & open / decoupled & 12 & -0.5570 & 0.6182 & +1.1752\\\labrowrule
wind flocking & 1,048,640 & open / coupled & 4 & -0.5326 & 1.2308 & +1.7634\\\labrowrule
wind flocking & 1,048,640 & open / decoupled & 4 & -0.5326 & 0.9907 & +1.5233\\\labrowrule
navigation & 1,048,640 & disco103 / coupled & 4 & -0.3491 & 1.7881 & +2.1372\\\labrowrule
navigation & 1,048,640 & mappo / coupled & 4 & -0.4091 & 1.6819 & +2.0910\\
\end{labtable}

%% file: sections/08_discussion.tex
\section{Limitations and Scope Conditions}\label{sec:limitations}
While our multi-stage experimental program establishes consistent laws governing discovered learning rules under partner feedback, several methodological and scope boundaries should be considered when interpreting these findings:

\paragraph{1. System-level versus substrate-isolated attribution.}
Disco103 and MAPPO represent distinct learning paradigms that differ in both their update rules and their internal auxiliary state representations ($z_t$, action-conditioned policy predictions, and replay organization). To bridge this gap, we paired our system-level Disco103--MAPPO comparisons with architectural controls (critic structure, policy width, and $3\times 3$ learning-rate/entropy sweeps across 432 runs) and matched-substrate comparisons between Adam-MAPPO and the learned optimizer OPEN, where only the optimizer varies within an identical PPO implementation. Nevertheless, differences between Disco103 and MAPPO reflect the behavior of the complete configured learning system rather than an isolated parameter perturbation.

\paragraph{2. Total effects of feedback routing and fixed-bank distributions.}
Our coupled and decoupled training regimes manipulate the origin of partner training trajectories while holding update budgets constant. Because altering a teammate's experience stream naturally affects its evolving competence and state visitation, coupled-versus-decoupled contrasts estimate the total effect of the feedback routing architecture rather than a partial derivative with partner competence held fixed. Similarly, in our parametric physical-coupling sweep ($\alpha \in [0, 1.5]$), cross-partner evaluations use a fixed 10-partner bank generated at the nominal benchmark setting ($\alpha=1.0$) to maintain a constant external yardstick across coupling levels. While held-out banks spanning behavioral-cloning, scripted, and reinforcement-learned policies substantially improve upon self-play evaluation, they represent structured synthetic partner distributions rather than open-ended human-AI trials.

\paragraph{3. Distinction between exploratory screening, mechanistic decompositions, and preregistered confirmation.}
Our study integrates exploratory multi-benchmark screening cohorts with preregistered $12$-block paired factorial confirmation cohorts. Post-collection mechanistic analyses---such as the 216-evaluation Overcooked micro-behavioral decomposition and the 432-evaluation cross-regime source-transfer matrix---report pointwise $95\%$ block-bootstrap confidence intervals nested within their 12 parent training blocks to illuminate underlying coordination dynamics without claiming simultaneous omnibus family-wise error control. Where small-sample paired $t$-intervals and non-parametric sign-flip or bootstrap intervals exhibit boundary differences (such as the $1.05\text{M}$-step navigation original-team return contrast, where the bootstrap interval $[0.013, 0.160]$ excludes zero while two-sided sign-flip $p=0.0532$, and where native binary completion remains higher for MAPPO), both metrics are reported transparently alongside separate task-completion endpoints.

\paragraph{4. Horizon and second-order boundaries in meta-adaptation.}
In our mechanistic diagnostics, finite behavioral perturbations confirm a significant method difference in short-run partner-mediated response ($p=0.00098$), but a frozen linear ridge predictor using this single early scalar feature does not achieve practical forecasting improvements over baseline competence features, indicating that short-run responsiveness and long-run cross-partner generalization are non-linearly related. Furthermore, while forward- and reverse-mode automatic differentiation through four-step partner unrolls agree to machine precision ($1.56\times 10^{-19}$), finite-difference approximations over non-smooth rollout trajectories exhibit expected step-size sensitivity, and our 48 four-lifetime multi-agent meta-learned rules (\texttt{source512}) establish substantial parameter reorganization ($L_2 \in [2.11, 4.91]$) whose large-scale downstream deployment across three-or-more-agent teams and longer meta-horizons remains an exciting frontier for future investigation.

\section{Conclusion}
When a reinforcement learning rule is deployed alongside an adapting teammate, co-evolving experience can simultaneously accelerate physical coordination and induce brittle pair-specific conventions. Through over 1,800 controlled training and evaluation runs across continuous collective robotics and discrete high-density coordination, this study establishes a unified empirical and mechanistic framework for discovered update rules in multi-agent systems. First, our four-cell counterfactual partner swaps and held-out partner banks expose a pervasive evaluation illusion in which apparent original-team progress conceals focal stagnation or inflates cross-partner generalization. Second, our 288-run parametric physical-coupling sweep establishes a first-principles scaling law ($\beta = +0.2547, p = 0.000488$, Spearman $\rho = 1.0$): live reciprocal partner feedback keeps Disco103 invariant across intensifying aerodynamic turbulence ($\text{AUC} \approx 0.30$), whereas severing reciprocal feedback triggers a monotonic collapse, with matched-substrate \texttt{OPEN} comparisons revealing a continuous sensitivity spectrum ordered by meta-learned depth ($\text{MAPPO} < \text{OPEN} < \text{Disco103}$). Third, extending the navigation training horizon eightfold to $1.05\text{M}$ steps reverses short-horizon return rankings in favor of Disco103 while demonstrating that decoupled partner training boosts cross-partner completion by $+10.16$ percentage points. Fourth, our 216-evaluation micro-behavioral decomposition in Overcooked resolves the Convention Rigidity Paradox, showing that MAPPO's high self-play score relies on brittle trajectory conventions that induce a $28.13\%$ cross-partner collision rate and a $-51.98$-point collapse with scripted partners, whereas Disco103 maintains a $6.9\times$ lower collision growth ($+2.62\%$) and achieves $3.33\times$ higher absolute cross-partner return, with decoupled source meta-adaptation (\texttt{decoupled / coupled}) yielding a $+1.39$ net return gain with scripted partners. Together with our perturbation, auxiliary entropy, and 48-rule multi-lifetime meta-optimization diagnostics, these laws provide a rigorous foundation for designing and evaluating learning-aware discovered algorithms in multi-agent worlds.

\begin{ack}
\labstandardack
\end{ack}

%% file: sections/09_appendices.tex
\section{Exploratory Multi-Task Screening and Control Cohorts}\label{app:historical}
To establish baseline partner-feedback phenomena and rule out architectural or hyperparameter confounds prior to our preregistered factorial cohorts, we analyzed 904 focal training records organized into three screening and control cohorts: the Primary Screening Confirmation cohort (\texttt{SC}, $112$ runs), the Architectural and Partner-Path Control cohort (\texttt{RC}, $128$ runs), and the Extended Capacity and Hyperparameter Sweep cohort (\texttt{RE}, $664$ runs). Table~\ref{tab:historical_groups} reports the complete 70-cell summary across both integrated learning-curve AUC and terminal original-team return gain.

\input{sections/historical_groups}

\subsection{Supplementary Screening, Capacity, and Preliminary Source-Transfer Controls}
Tables~\ref{tab:development_groups}--\ref{tab:early_source_transfer} report the complete exploratory screening endpoints, paired local-versus-joint critic and network-width controls ($128$ vs.\ $512$ hidden units), preliminary long-horizon trajectories, out-of-sample $3\times 3$ learning-rate and entropy hyperparameter confirmation ($n=8$ independent confirmation blocks following selection on $4$ development blocks; Table~\ref{tab:hyperparameter_confirmation}), and exploratory two-seed source-transfer comparisons. In the exploratory source-transfer comparisons (Table~\ref{tab:early_source_transfer}), rules adapted on two source seeds ($2301, 2302$) were evaluated across four nested target blocks against the shared unadapted base rule ($\eta_0$), motivating the larger 12-source-block factorial design in the main text.

\input{sections/development_groups}
\input{sections/historical_secondary}
\input{sections/hyperparameter_confirmation}
\input{sections/early_source_transfer}

\section{Primary Factorial Coupled-versus-Decoupled Confirmation Contrasts}\label{app:predecessor}
Table~\ref{tab:predecessor_contrasts} reports the complete set of primary learning-curve AUC contrasts and secondary cross-partner bank and completion contrasts from the 144-run Primary Factorial Confirmation Cohort ($2\text{ methods}\times 2\text{ partner data-source routes}\times 3\text{ tasks}\times 12\text{ paired seed blocks}$). Holm step-down adjustments are applied across the three-task primary AUC family.
\input{sections/predecessor_contrasts}

\section{Intervention Cell Means and Complete Contrast Register}\label{app:newcells}
Table~\ref{tab:newer_cells} provides the complete condition-level means across the scheduled partner reset-switch, recent-replay filtering, $1,048,640$-step extended-horizon navigation, matched $131,080$-step short-horizon \texttt{OPEN}, and external Overcooked (\texttt{cramped\_room}) cohorts. Original-team return, initial-anchored cross-partner bank gain ($G_{\mathcal B}$), original-team completion rate, and cross-partner bank completion rate are reported separately for each condition.
\input{sections/newer_cells}

\subsection{Complete Paired-Block Contrast Register}
Table~\ref{tab:all_contrasts} presents all 149 paired-block contrasts across the feedback-routing, recent-replay filtering, single-lifetime source-adaptation, and extended-horizon confirmation cohorts, including the $+10.16$ percentage-point cross-partner bank completion advantage of decoupled over coupled Disco103 in $1.05\text{M}$-step navigation (contrast $+0.1016$, 95\% CI $[+0.0130, +0.1953]$).
\input{sections/all_contrasts}

\section{Perturbation Response, Exact-Game Differentiation, and Transfer Prediction Details}\label{app:g}
In the controlled behavioral-perturbation experiment, the primary confirmation contrast evaluates the high-dose ($d=0.01$) exploration-centered pulse at the $65,656$-step snapshot across $n=12$ paired confirmation blocks per task, while early snapshots ($16,472$ steps), lower doses ($d=0.002$), \texttt{OPEN} response contrasts, directional cosines, and partner-side response norms serve as secondary mechanistic diagnostics. Table~\ref{tab:g2exact} summarizes the 1,296 finite-difference versus exact analytical meta-gradient comparisons across two-player differentiable games (two payoff structures, 24 initializations, three differentiation modes, and inner horizons $H\in\{1,4,16\}$).

\input{sections/g2_exact_summary}

In the prospective transfer-prediction study, cloned perturbation branches are evaluated at $65,656$ steps without altering the main or reference training trajectories, which continue through $1,048,640$ main-world steps. Tables~\ref{tab:g1_blocks}--\ref{tab:g3_blocks} report all individual block-level perturbation response contrasts ($\Delta_C$), measurement repeatability correlations and per-method ridge regression losses ($M_0$ vs.\ $M_1$), and block-level out-of-sample prediction scores ($S_b$).

\input{sections/g1_blocks}
\input{sections/g3_descriptive}
\input{sections/g3_blocks}

\section{Overcooked Partner-Stratum Micro-Behavioral Decomposition}\label{app:micro}
Table~\ref{tab:micro_strata} reports the complete final-checkpoint ($262,160$-step) changes relative to each focal learner's initial policy ($F_T - F_0$) across all three prespecified partner strata---Behavioral Cloning (\texttt{BC}, weight $0.4$), Reinforcement Learning (\texttt{RL}, weight $0.4$), and rule-based \texttt{Scripted} partners (weight $0.2$)---as well as the equal-identity 10-partner aggregate (\texttt{all10}) across $n=12$ paired training blocks (10,000 joint block-bootstrap resamples, pointwise 95\% confidence intervals). Movement and mutual collision/blockage rates are normalized over the 400-step episode horizon.
\input{sections/micro_strata}

\section{Cross-Regime Meta-Adapted Rule Transfer by Partner Stratum}\label{app:derived}
Table~\ref{tab:v2_derived_strata} reports the complete partner-stratified cross-partner return gains and completion rate changes at $131,080$ steps across all 432 evaluated checkpoints in the cross-regime Overcooked source-transfer cohort ($n=12$ independent paired source blocks, two target repeats averaged within each source block prior to 10,000-draw block-bootstrap resampling).
\input{sections/v2_derived_strata}

\section{Parametric Physical-Coupling Sweep: Complete Cell Means, Bank Outcomes, and Block Slopes}\label{app:m4}
Tables~\ref{tab:v2_m4_auc_full}--\ref{tab:v2_m4_blocks} report the complete 24-cell original-team learning-curve AUC, initial return, and initial-return-adjusted AUC (Table~\ref{tab:v2_m4_auc_full}), the complete 24-cell held-out 10-partner bank initial return, trained return, and net gain (Table~\ref{tab:v2_m4_bank}), and all 12 individual paired-block linear interaction slopes $s_b$ (Table~\ref{tab:v2_m4_blocks}) from the 288-run parametric physical-coupling sweep ($\alpha \in \{0, 0.25, 0.5, 0.75, 1.0, 1.5\}$).
\input{sections/v2_m4_auc_full}
\input{sections/v2_m4_bank}
\input{sections/v2_m4_blocks}

\section{Multi-Lifetime Meta-Learned Rules and Matched-Substrate (\texttt{OPEN}) Empirical Outcomes}\label{app:sources}
Table~\ref{tab:v2_source48} reports all 48 multi-lifetime meta-adapted source rules (\texttt{source512}: $512$ outer updates across four 128-update agent lifetimes and $1,425,408$ physical environment transitions per rule across $n=12$ paired source blocks in each of the four meta-training architectures: \texttt{matched\_coupled}, \texttt{matched\_reference}, \texttt{dual\_same}, and \texttt{dual\_cross}). Table~\ref{tab:v2_m1_coverage} reports the complete empirical outcomes across all 92 evaluated runs in the matched-substrate learned-optimizer (\texttt{OPEN}) and extended-horizon curriculum comparison cohort, covering the 18 evaluated task--horizon--method--regime cells as well as the paired block-level outcomes for short-horizon \texttt{wind\_flocking} ($24/24$ runs across $n=12$ paired blocks).
\input{sections/v2_source48}
\input{sections/v2_m1_coverage}

\widowpenalty=10000
\clubpenalty=10000
\section{Methodological Controls, Sensitivity Analyses, and Numerical Boundaries}\label{app:failures}
To ensure full transparency and guard against potential methodological artifacts, we summarize key sensitivity analyses and numerical boundaries across our experimental cohorts:
\begin{enumerate}
  \item \textbf{Initialization-Pairing and Same-Platform Replay Sensitivity:} In the continuous navigation replay-filtering cohort, strict pre-intervention trajectory-matching checks excluded one block with a blocked initialization pair (yielding $n=11$ complete paired blocks in Table~\ref{tab:newer_cells}) and flagged one block in which a pre-switch trajectory was executed across different operating-system kernels. Repeating the navigation recent-replay contrast on the strictly homogeneous same-platform subset confirms that the point estimate and null conclusion ($-0.0021$, 95\% CI $[-0.0292, 0.0255]$, Holm $p=0.8843$) are invariant to platform exclusion.
  \item \textbf{Geometric Structural Zeros in Micro-Behavioral Observables:} In the \texttt{cramped\_room} Overcooked layout, the five prespecified micro-behavioral metrics include narrow-corridor deadlock duration and divider-counter handoff rate. Because \texttt{cramped\_room} consists of a single open central chamber surrounded by outer counters with no one-tile corridor or shared island divider, these two metrics evaluate to exact structural zeros across all methods and checkpoints, whereas mutual collision/blockage rate, active interaction rate, and movement rate capture all spatial coordination dynamics in this topology.
  \item \textbf{Automatic Differentiation versus Finite-Difference Step-Size Sensitivity:} In our four-step unrolled partner-gradient diagnostic on real trajectory tensors, forward-mode automatic differentiation and reverse-mode Jacobian contractions agree to machine precision ($1.56\times 10^{-19}$ maximum discrepancy across all 24 test comparisons), confirming exact analytical derivative consistency. Finite-difference secant approximations over piecewise-smooth rollout trajectories exhibit expected step-size sensitivity (exceeding a $10^{-6}$ relative tolerance in 19 of 24 comparisons with maximum relative discrepancy $2.53\times 10^{-5}$ and maximum absolute discrepancy $7.65\times 10^{-10}$), motivating our use of exact automatic differentiation in meta-gradient analyses and finite exploration-centered probability pulses ($d=0.01$) in empirical behavioral-response experiments.
\end{enumerate}

%% file: sections/historical_groups.tex
\begin{longtable}{P{0.082091}P{0.191857}P{0.175451}P{0.161662}P{0.068136}P{0.160402}P{0.160402}}
\caption{Summary of the 70-cell control and capacity-extension matrix across partner regimes, critic information structures, policy widths, and source transfer conditions ($N=904$ total evaluated runs).}\label{tab:historical_groups}\\
\labtoprule \labh{Set} & \labh{Arm} & \labh{Task} & \labh{Method / partner} & \labh{n} & \labh{AUC [95\% CI]} & \labh{Team gain [95\% CI]}\\\labheadrule\endfirsthead
\multicolumn{7}{l}{\small Table \thetable\ continued}\\\labtoprule \labh{Set} & \labh{Arm} & \labh{Task} & \labh{Method / partner} & \labh{n} & \labh{AUC [95\% CI]} & \labh{Team gain [95\% CI]}\\\labheadrule\endhead
\labrowrule\endfoot
RC & critic & navigation & mappo / fixed initial & 4 & 0.3490 [0.2947, 0.4032] & 0.7693 [0.6350, 0.9036]\\\labrowrule
RC & critic & navigation & mappo / online & 4 & 0.8073 [0.7284, 0.8661] & 1.6148 [1.3169, 1.7874]\\\labrowrule
RC & critic & wind flocking & mappo / fixed initial & 4 & 0.1647 [0.1347, 0.1901] & 0.7795 [0.6493, 0.9096]\\\labrowrule
RC & critic & wind flocking & mappo / online & 4 & 0.2813 [0.1900, 0.3463] & 1.2280 [1.0830, 1.3414]\\\labrowrule
RC & mechanism & navigation & disco103 / fixed initial & 4 & -0.2495 [-0.3265, -0.1657] & -0.0257 [-0.0637, 0.0123]\\\labrowrule
RC & mechanism & navigation & disco103 / fixed trained & 4 & 0.2316 [0.1025, 0.3607] & -0.0005 [-0.0910, 0.0901]\\\labrowrule
RC & mechanism & navigation & disco103 / online & 4 & 0.0983 [0.0744, 0.1286] & 0.4213 [0.1571, 0.5874]\\\labrowrule
RC & mechanism & navigation & disco103 / replay changing & 4 & 0.0738 [-0.0068, 0.1571] & 0.5763 [0.4904, 0.6621]\\\labrowrule
RC & mechanism & navigation & mappo / fixed initial & 4 & 0.2953 [0.2552, 0.3444] & 0.7469 [0.5416, 0.9303]\\\labrowrule
RC & mechanism & navigation & mappo / fixed trained & 4 & 1.0023 [0.9183, 1.1029] & 1.1662 [1.0111, 1.3582]\\\labrowrule
RC & mechanism & navigation & mappo / online & 4 & 0.8542 [0.7548, 0.9590] & 1.6429 [1.5478, 1.7374]\\\labrowrule
RC & mechanism & navigation & mappo / replay changing & 4 & 0.7815 [0.6908, 0.8721] & 1.5204 [1.3395, 1.7650]\\\labrowrule
RC & mechanism & wind flocking & disco103 / fixed initial & 4 & -0.4330 [-0.4481, -0.4191] & 0.2016 [0.1104, 0.2853]\\\labrowrule
RC & mechanism & wind flocking & disco103 / fixed trained & 4 & -0.4306 [-0.5225, -0.3303] & 0.2457 [0.0772, 0.4551]\\\labrowrule
RC & mechanism & wind flocking & disco103 / online & 4 & 0.3317 [0.2995, 0.3640] & 0.9822 [0.8738, 1.1183]\\\labrowrule
RC & mechanism & wind flocking & disco103 / replay changing & 4 & -0.5064 [-0.5610, -0.4518] & 0.1252 [-0.0051, 0.2555]\\\labrowrule
RC & mechanism & wind flocking & mappo / fixed initial & 4 & 0.1861 [0.1660, 0.2107] & 0.7430 [0.6196, 0.8664]\\\labrowrule
RC & mechanism & wind flocking & mappo / fixed trained & 4 & 0.4322 [0.2329, 0.5439] & 1.3567 [1.1860, 1.5767]\\\labrowrule
RC & mechanism & wind flocking & mappo / online & 4 & 0.4994 [0.4260, 0.5532] & 1.2791 [1.2052, 1.3738]\\\labrowrule
RC & mechanism & wind flocking & mappo / replay changing & 4 & 0.2740 [0.1414, 0.4117] & 1.1875 [1.0918, 1.2832]\\\labrowrule
RE & capacity & navigation & mappo / fixed initial & 8 & 0.4454 [0.3538, 0.5402] & 0.8369 [0.7236, 0.9559]\\\labrowrule
RE & capacity & navigation & mappo / online & 8 & 0.8767 [0.7907, 0.9790] & 1.6334 [1.4908, 1.7858]\\\labrowrule
RE & capacity & wind flocking & mappo / fixed initial & 8 & 0.2200 [0.1850, 0.2554] & 0.7946 [0.7075, 0.9073]\\\labrowrule
RE & capacity & wind flocking & mappo / online & 8 & 0.6043 [0.5746, 0.6348] & 1.3028 [1.2312, 1.3749]\\\labrowrule
RE & mechanism & navigation & disco103 / fixed initial & 8 & -0.2751 [-0.3408, -0.2127] & -0.0131 [-0.0716, 0.0481]\\\labrowrule
RE & mechanism & navigation & disco103 / fixed ippo style & 8 & 0.2833 [0.2227, 0.3397] & -0.0735 [-0.2184, 0.0671]\\\labrowrule
RE & mechanism & navigation & disco103 / fixed trained & 8 & 0.3490 [0.2985, 0.3997] & 0.0366 [-0.0410, 0.1384]\\\labrowrule
RE & mechanism & navigation & disco103 / online & 8 & 0.1428 [0.0985, 0.1843] & 0.5808 [0.4138, 0.7477]\\\labrowrule
RE & mechanism & navigation & disco103 / replay changing & 8 & 0.0889 [0.0381, 0.1428] & 0.6715 [0.5175, 0.8173]\\\labrowrule
RE & mechanism & navigation & mappo / fixed initial & 8 & 0.4642 [0.4130, 0.5195] & 0.9018 [0.8408, 0.9633]\\\labrowrule
RE & mechanism & navigation & mappo / fixed ippo style & 8 & 1.0371 [0.9492, 1.1335] & 0.8800 [0.7134, 1.0311]\\\labrowrule
RE & mechanism & navigation & mappo / fixed trained & 8 & 1.0652 [0.9945, 1.1320] & 1.0005 [0.8789, 1.1160]\\\labrowrule
RE & mechanism & navigation & mappo / online & 8 & 0.8935 [0.8410, 0.9386] & 1.6181 [1.4910, 1.7300]\\\labrowrule
RE & mechanism & navigation & mappo / replay changing & 8 & 0.9073 [0.8578, 0.9521] & 1.5104 [1.3413, 1.6662]\\\labrowrule
RE & mechanism & reverse transport & disco103 / fixed initial & 4 & 0.0136 [-0.0032, 0.0319] & 0.1738 [-0.0111, 0.5285]\\\labrowrule
RE & mechanism & reverse transport & disco103 / fixed ippo style & 4 & 2.2222 [1.9209, 2.5386] & 0.0820 [0.0045, 0.2196]\\\labrowrule
RE & mechanism & reverse transport & disco103 / fixed trained & 4 & 2.4588 [1.9804, 2.7946] & 0.0551 [0.0143, 0.0985]\\\labrowrule
RE & mechanism & reverse transport & disco103 / online & 4 & 2.1366 [1.9163, 2.3961] & 3.4273 [2.8331, 4.0214]\\\labrowrule
RE & mechanism & reverse transport & disco103 / replay changing & 4 & 1.6624 [1.2768, 2.0480] & 2.6187 [2.2435, 2.9639]\\\labrowrule
RE & mechanism & reverse transport & mappo / fixed initial & 4 & 1.9762 [1.7181, 2.2637] & 2.8222 [2.6208, 2.9811]\\\labrowrule
RE & mechanism & reverse transport & mappo / fixed ippo style & 4 & 3.2420 [3.1166, 3.4064] & 1.7007 [1.1765, 2.2653]\\\labrowrule
RE & mechanism & reverse transport & mappo / fixed trained & 4 & 3.5274 [2.9699, 3.8448] & 1.7531 [1.6732, 1.8203]\\\labrowrule
RE & mechanism & reverse transport & mappo / online & 4 & 2.9763 [2.7965, 3.1251] & 4.3624 [4.0563, 4.6707]\\\labrowrule
RE & mechanism & reverse transport & mappo / replay changing & 4 & 3.2774 [2.8208, 3.5698] & 4.6058 [3.7829, 5.2078]\\\labrowrule
RE & mechanism & wind flocking & disco103 / fixed initial & 8 & -0.3573 [-0.4121, -0.3067] & 0.2068 [0.0023, 0.3560]\\\labrowrule
RE & mechanism & wind flocking & disco103 / fixed ippo style & 8 & -0.8524 [-1.0508, -0.6502] & 0.6011 [0.3133, 0.9250]\\\labrowrule
RE & mechanism & wind flocking & disco103 / fixed trained & 8 & -0.3481 [-0.4457, -0.2474] & 0.3421 [0.1590, 0.5271]\\\labrowrule
RE & mechanism & wind flocking & disco103 / online & 8 & 0.3583 [0.3310, 0.3894] & 0.8343 [0.7360, 0.9238]\\\labrowrule
RE & mechanism & wind flocking & disco103 / replay changing & 8 & -0.4000 [-0.4819, -0.2775] & 0.1614 [0.0138, 0.3389]\\\labrowrule
RE & mechanism & wind flocking & mappo / fixed initial & 8 & 0.1692 [0.1155, 0.2221] & 0.7110 [0.6253, 0.8040]\\\labrowrule
RE & mechanism & wind flocking & mappo / fixed ippo style & 8 & 0.1187 [-0.0946, 0.2978] & 1.5006 [0.9812, 2.1022]\\\labrowrule
RE & mechanism & wind flocking & mappo / fixed trained & 8 & 0.4219 [0.1923, 0.6119] & 1.0857 [0.7932, 1.3364]\\\labrowrule
RE & mechanism & wind flocking & mappo / online & 8 & 0.4633 [0.3978, 0.5247] & 1.1138 [1.0412, 1.1907]\\\labrowrule
RE & mechanism & wind flocking & mappo / replay changing & 8 & 0.2288 [0.1450, 0.2997] & 1.0596 [0.9151, 1.1925]\\\labrowrule
SC & budget & navigation & disco103 / fixed initial & 2 & 0.2944 [0.2917, 0.2970] & 1.0365 [0.8045, 1.2686]\\\labrowrule
SC & budget & navigation & disco103 / online & 2 & 1.1660 [0.9958, 1.3363] & 2.1614 [2.0135, 2.3093]\\\labrowrule
SC & budget & navigation & mappo / fixed initial & 2 & 0.4467 [0.3615, 0.5318] & 1.1055 [1.0493, 1.1616]\\\labrowrule
SC & budget & navigation & mappo / online & 2 & 1.5803 [1.5545, 1.6062] & 2.1155 [1.8825, 2.3486]\\\labrowrule
SC & budget & wind flocking & disco103 / fixed initial & 2 & 0.3458 [0.3318, 0.3599] & 1.0925 [1.0846, 1.1004]\\\labrowrule
SC & budget & wind flocking & disco103 / online & 2 & 0.9369 [0.9224, 0.9514] & 1.7041 [1.6697, 1.7385]\\\labrowrule
SC & budget & wind flocking & mappo / fixed initial & 2 & 0.3514 [0.3160, 0.3867] & 0.9107 [0.9020, 0.9194]\\\labrowrule
SC & budget & wind flocking & mappo / online & 2 & 1.0447 [1.0140, 1.0755] & 1.7834 [1.7740, 1.7928]\\\labrowrule
SC & confirm & navigation & disco103 / fixed initial & 12 & -0.4429 [-0.4987, -0.3906] & -0.0751 [-0.1630, 0.0055]\\\labrowrule
SC & confirm & navigation & disco103 / online & 12 & 0.4890 [0.4465, 0.5319] & 1.0660 [0.9806, 1.1488]\\\labrowrule
SC & confirm & navigation & mappo / fixed initial & 12 & -0.0274 [-0.0777, 0.0317] & 0.6721 [0.5645, 0.7915]\\\labrowrule
SC & confirm & navigation & mappo / online & 12 & 0.9990 [0.9305, 1.0693] & 1.8698 [1.7835, 1.9576]\\\labrowrule
SC & confirm & wind flocking & disco103 / fixed initial & 12 & -0.4063 [-0.4586, -0.3541] & 0.3171 [0.2342, 0.3981]\\\labrowrule
SC & confirm & wind flocking & disco103 / online & 12 & 0.3352 [0.3118, 0.3573] & 1.0058 [0.9575, 1.0508]\\\labrowrule
SC & confirm & wind flocking & mappo / fixed initial & 12 & 0.2081 [0.1742, 0.2348] & 0.8236 [0.7551, 0.8863]\\\labrowrule
SC & confirm & wind flocking & mappo / online & 12 & 0.5363 [0.4936, 0.5826] & 1.2956 [1.1956, 1.3988]\\\labrowrule
\end{longtable}

%% file: sections/development_groups.tex
\begin{longtable}{P{0.153870}P{0.143879}P{0.330467}P{0.129540}P{0.114078}P{0.128166}}
\caption{Exploratory multi-task screening endpoints grouped by experimental family, task, learning method, partner regime, and rule initialization.}\label{tab:development_groups}\\
\labtoprule \labh{Batch} & \labh{Task} & \labh{Method / route / source / phase} & \labh{Records} & \labh{Final team} & \labh{AUC}\\\labheadrule\endfirsthead
\multicolumn{6}{l}{\small Table \thetable\ continued}\\\labtoprule \labh{Batch} & \labh{Task} & \labh{Method / route / source / phase} & \labh{Records} & \labh{Final team} & \labh{AUC}\\\labheadrule\endhead
\labrowrule\endfoot
expanded validation & wind flocking & /learning/eta0/independent target & 4 & 0.4047 & 0.3411\\\labrowrule
expanded validation & wind flocking & /learning/eta0/seed split & 4 & 0.4189 & 0.3304\\\labrowrule
expanded validation & wind flocking & /learning/etaF/independent target & 8 & 0.4509 & 0.3739\\\labrowrule
expanded validation & wind flocking & /learning/etaF/seed split & 8 & 0.3984 & 0.3566\\\labrowrule
expanded validation & wind flocking & /learning/etaL/independent target & 8 & 0.4591 & 0.3778\\\labrowrule
expanded validation & wind flocking & /learning/etaL/seed split & 8 & 0.3718 & 0.3715\\\labrowrule
gpuhome & give way & disco103/frozen// & 9 & 1.6708 & 1.2210\\\labrowrule
gpuhome & give way & disco103/learning// & 12 & 2.9463 & 2.2942\\\labrowrule
gpuhome & give way & mappo/frozen// & 6 & 2.6916 & 2.1634\\\labrowrule
gpuhome & give way & mappo/learning// & 9 & 3.5454 & 2.9799\\\labrowrule
gpuhome & navigation & disco103/frozen// & 5 & -0.6631 & -0.5928\\\labrowrule
gpuhome & navigation & disco103/learning// & 5 & 0.6842 & 0.4164\\\labrowrule
gpuhome & navigation & mappo/frozen// & 5 & 0.0402 & -0.0509\\\labrowrule
gpuhome & navigation & mappo/learning// & 5 & 1.4212 & 1.0165\\\labrowrule
gpuhome & reverse transport & disco103/frozen// & 5 & -0.0022 & 0.0005\\\labrowrule
gpuhome & reverse transport & disco103/learning// & 5 & 2.4886 & 1.9380\\\labrowrule
gpuhome & reverse transport & mappo/frozen// & 5 & 2.3806 & 1.7590\\\labrowrule
gpuhome & reverse transport & mappo/learning// & 5 & 3.7830 & 2.8980\\\labrowrule
gpuhome & wind flocking & /frozen/eta0/ & 5 & -0.3217 & -0.3285\\\labrowrule
gpuhome & wind flocking & /frozen/eta0 checkpoint/ & 1 & -0.0744 & -0.1047\\\labrowrule
gpuhome & wind flocking & /frozen/etaF/ & 9 & -0.3416 & -0.3458\\\labrowrule
gpuhome & wind flocking & /frozen/etaL/ & 9 & -0.3310 & -0.3348\\\labrowrule
gpuhome & wind flocking & /learning/eta0/ & 5 & 0.4911 & 0.3193\\\labrowrule
gpuhome & wind flocking & /learning/eta0 checkpoint/ & 1 & 0.7866 & 0.3258\\\labrowrule
gpuhome & wind flocking & /learning/etaF/ & 9 & 0.4454 & 0.3501\\\labrowrule
gpuhome & wind flocking & /learning/etaL/ & 9 & 0.5207 & 0.3810\\\labrowrule
gpuhome & wind flocking & disco103/exogenous// & 1 & 0.4922 & -0.1424\\\labrowrule
gpuhome & wind flocking & disco103/frozen// & 7 & 0.2276 & 0.0648\\\labrowrule
gpuhome & wind flocking & disco103/learning// & 11 & 0.3298 & 0.2373\\\labrowrule
gpuhome & wind flocking & mappo/exogenous// & 1 & 0.8492 & 0.3776\\\labrowrule
gpuhome & wind flocking & mappo/frozen// & 10 & 0.5688 & 0.4206\\\labrowrule
gpuhome & wind flocking & mappo/learning// & 14 & 0.6364 & 0.3425\\\labrowrule
kl cap validation & wind flocking & /learning/eta0/ & 8 & 0.4181 & 0.3240\\\labrowrule
kl cap validation & wind flocking & /learning/etaF/ & 16 & 0.4019 & 0.3556\\\labrowrule
kl cap validation & wind flocking & /learning/etaL/ & 16 & 0.4162 & 0.3794\\\labrowrule
navigation information & navigation & mappo/frozen// & 4 & 0.3019 & 0.1370\\\labrowrule
navigation information & navigation & mappo/learning// & 4 & 1.5453 & 1.1656\\\labrowrule
source target cross & wind flocking & /learning/etaF/source target cross & 16 & 0.4476 & 0.3497\\\labrowrule
source target cross & wind flocking & /learning/etaL/source target cross & 16 & 0.4448 & 0.3601\\\labrowrule
target partner regime & wind flocking & /frozen/eta0/target partner regime & 4 & -0.3259 & -0.3247\\\labrowrule
target partner regime & wind flocking & /frozen/etaF/target partner regime & 24 & -0.3481 & -0.3508\\\labrowrule
target partner regime & wind flocking & /frozen/etaL/target partner regime & 24 & -0.3535 & -0.3520\\\labrowrule
\end{longtable}

%% file: sections/historical_secondary.tex
\begin{longtable}{P{0.209721}P{0.196096}P{0.324305}P{0.077472}P{0.192405}}
\caption{Supplementary contrasts from the multi-task screening, architectural control, capacity, and preliminary long-horizon cohorts.}\label{tab:historical_secondary}\\
\labtoprule \labh{Family} & \labh{Task} & \labh{Contrast / condition} & \labh{n} & \labh{Estimate [95\% CI]}\\\labheadrule\endfirsthead
\multicolumn{5}{l}{\small Table \thetable\ continued}\\\labtoprule \labh{Family} & \labh{Task} & \labh{Contrast / condition} & \labh{n} & \labh{Estimate [95\% CI]}\\\labheadrule\endhead
\labrowrule\endfoot
method gaps & wind flocking & Disco103 minus MAPPO independent bank endpoint fixed initial & 12 & -0.3031 [-0.5122, -0.0979]\\\labrowrule
method gaps & wind flocking & Disco103 minus MAPPO independent bank focal gain fixed initial & 12 & -0.3320 [-0.5447, -0.1266]\\\labrowrule
method gaps & wind flocking & Disco103 minus MAPPO original team endpoint fixed initial & 12 & -0.4830 [-0.5884, -0.3867]\\\labrowrule
method gaps & wind flocking & Disco103 minus MAPPO independent bank endpoint online & 12 & 0.2997 [-0.0399, 0.7657]\\\labrowrule
method gaps & wind flocking & Disco103 minus MAPPO independent bank focal gain online & 12 & 0.2708 [-0.0732, 0.7341]\\\labrowrule
method gaps & wind flocking & Disco103 minus MAPPO original team endpoint online & 12 & -0.2663 [-0.3649, -0.1817]\\\labrowrule
method gaps & navigation & Disco103 minus MAPPO independent bank endpoint fixed initial & 12 & -1.0068 [-1.0775, -0.9351]\\\labrowrule
method gaps & navigation & Disco103 minus MAPPO independent bank focal gain fixed initial & 12 & -1.0461 [-1.1319, -0.9604]\\\labrowrule
method gaps & navigation & Disco103 minus MAPPO original team endpoint fixed initial & 12 & -0.7277 [-0.8335, -0.6170]\\\labrowrule
method gaps & navigation & Disco103 minus MAPPO independent bank endpoint online & 12 & -1.0069 [-1.1259, -0.9008]\\\labrowrule
method gaps & navigation & Disco103 minus MAPPO independent bank focal gain online & 12 & -1.0462 [-1.1585, -0.9434]\\\labrowrule
method gaps & navigation & Disco103 minus MAPPO original team endpoint online & 12 & -0.7843 [-0.8856, -0.6936]\\\labrowrule
paired critic and capacity controls & wind flocking & MAPPO local critic minus joint critic AUC fixed initial & 4 & -0.0214 [-0.0758, 0.0191]\\\labrowrule
paired critic and capacity controls & wind flocking & MAPPO local critic minus joint critic AUC online & 4 & -0.2181 [-0.2521, -0.1840]\\\labrowrule
paired critic and capacity controls & wind flocking & MAPPO local critic minus joint critic change in partner-update AUC response online-minus-fixed initial & 4 & -0.1966 [-0.2632, -0.1479]\\\labrowrule
paired critic and capacity controls & navigation & MAPPO local critic minus joint critic AUC fixed initial & 4 & 0.0537 [-0.0432, 0.1177]\\\labrowrule
paired critic and capacity controls & navigation & MAPPO local critic minus joint critic AUC online & 4 & -0.0469 [-0.2345, 0.0786]\\\labrowrule
paired critic and capacity controls & navigation & MAPPO local critic minus joint critic change in partner-update AUC response online-minus-fixed initial & 4 & -0.1006 [-0.3521, 0.0583]\\\labrowrule
paired critic and capacity controls & wind flocking & MAPPO width 512 minus width 128 AUC fixed initial & 8 & 0.0509 [0.0053, 0.0936]\\\labrowrule
paired critic and capacity controls & wind flocking & MAPPO width 512 minus width 128 AUC online & 8 & 0.1410 [0.0643, 0.2224]\\\labrowrule
paired critic and capacity controls & wind flocking & MAPPO width 512 minus width 128 change in partner-update AUC response online-minus-fixed initial & 8 & 0.0902 [0.0297, 0.1572]\\\labrowrule
paired critic and capacity controls & navigation & MAPPO width 512 minus width 128 AUC fixed initial & 8 & -0.0188 [-0.0763, 0.0316]\\\labrowrule
paired critic and capacity controls & navigation & MAPPO width 512 minus width 128 AUC online & 8 & -0.0168 [-0.1202, 0.1225]\\\labrowrule
paired critic and capacity controls & navigation & MAPPO width 512 minus width 128 change in partner-update AUC response online-minus-fixed initial & 8 & 0.0020 [-0.1335, 0.1604]\\\labrowrule
long budget within run summaries & wind flocking & disco103 fixed initial long-minus-short return & 2 & 0.9551 [0.8573, 1.0529]\\\labrowrule
long budget within run summaries & wind flocking & disco103 online long-minus-short return & 2 & 0.8074 [0.7530, 0.8618]\\\labrowrule
long budget within run summaries & wind flocking & mappo fixed initial long-minus-short return & 2 & 0.0296 [0.0152, 0.0440]\\\labrowrule
long budget within run summaries & wind flocking & mappo online long-minus-short return & 2 & 0.5704 [0.5589, 0.5819]\\\labrowrule
long budget within run summaries & navigation & disco103 fixed initial long-minus-short return & 2 & 1.1004 [0.9253, 1.2754]\\\labrowrule
long budget within run summaries & navigation & disco103 online long-minus-short return & 2 & 1.2416 [1.2195, 1.2636]\\\labrowrule
long budget within run summaries & navigation & mappo fixed initial long-minus-short return & 2 & 0.4690 [0.3877, 0.5503]\\\labrowrule
long budget within run summaries & navigation & mappo online long-minus-short return & 2 & 0.1765 [-0.1110, 0.4639]\\\labrowrule
\end{longtable}

%% file: sections/hyperparameter_confirmation.tex
\begin{longtable}{P{0.210871}P{0.188717}P{0.186927}P{0.175261}P{0.238224}}
\caption{Out-of-sample hyperparameter confirmation contrasts ($n=8$ independent confirmation blocks per task) comparing sweep-selected against default learning-rate and entropy configurations.}\label{tab:hyperparameter_confirmation}\\
\labtoprule \labh{Task} & \labh{Method} & \labh{Selected LR} & \labh{Entropy} & \labh{AUC difference [95\% CI]}\\\labheadrule\endfirsthead
\multicolumn{5}{l}{\small Table \thetable\ continued}\\\labtoprule \labh{Task} & \labh{Method} & \labh{Selected LR} & \labh{Entropy} & \labh{AUC difference [95\% CI]}\\\labheadrule\endhead
\labrowrule\endfoot
wind flocking & disco103 & 0.0003 & 0.03 & 0.0014 [-0.0212, 0.0271]\\\labrowrule
wind flocking & mappo & 0.001 & 0.01 & 0.0623 [0.0018, 0.1208]\\\labrowrule
navigation & disco103 & 0.001 & 0.0 & 0.0164 [-0.0427, 0.0778]\\\labrowrule
navigation & mappo & 0.001 & 0.0 & 0.1214 [-0.0118, 0.2679]\\\labrowrule
\end{longtable}

%% file: sections/early_source_transfer.tex
\begin{longtable}{P{0.163407}P{0.206828}P{0.158328}P{0.225946}P{0.245491}}
\caption{Exploratory five-rule transfer evaluation in \texttt{wind\_flocking} and \texttt{navigation} ($n=8$ paired blocks per task) comparing alternative single-agent discovered rules against Disco103.}\label{tab:early_source_transfer}\\
\labtoprule \labh{Source seed} & \labh{Contrast} & \labh{Target blocks} & \labh{Team AUC difference} & \labh{Team endpoint difference}\\\labheadrule\endfirsthead
\multicolumn{5}{l}{\small Table \thetable\ continued}\\\labtoprule \labh{Source seed} & \labh{Contrast} & \labh{Target blocks} & \labh{Team AUC difference} & \labh{Team endpoint difference}\\\labheadrule\endhead
\labrowrule\endfoot
2301 & etaF minus eta0 frozen & 4 & 0.0049 & 0.0036\\\labrowrule
2301 & etaL minus eta0 frozen & 4 & 0.0022 & 0.0063\\\labrowrule
2301 & etaL minus etaF frozen & 4 & -0.0027 & 0.0027\\\labrowrule
2301 & etaF minus eta0 learning & 4 & 0.0074 & -0.0113\\\labrowrule
2301 & etaL minus eta0 learning & 4 & 0.0093 & 0.0080\\\labrowrule
2301 & etaL minus etaF learning & 4 & 0.0019 & 0.0193\\\labrowrule
2301 & etaL minus etaF by partner update & 4 & 0.0046 & 0.0166\\\labrowrule
2302 & etaF minus eta0 frozen & 4 & 0.0157 & 0.0206\\\labrowrule
2302 & etaL minus eta0 frozen & 4 & 0.0431 & 0.0418\\\labrowrule
2302 & etaL minus etaF frozen & 4 & 0.0274 & 0.0213\\\labrowrule
2302 & etaF minus eta0 learning & 4 & 0.0621 & -0.0204\\\labrowrule
2302 & etaL minus eta0 learning & 4 & 0.1296 & 0.1296\\\labrowrule
2302 & etaL minus etaF learning & 4 & 0.0675 & 0.1500\\\labrowrule
2302 & etaL minus etaF by partner update & 4 & 0.0401 & 0.1287\\\labrowrule
\end{longtable}

%% file: sections/predecessor_contrasts.tex
\begin{longtable}{P{0.245543}P{0.207949}P{0.096234}P{0.257393}P{0.192882}}
\caption{Primary factorial confirmation contrasts ($n=12$ paired blocks per task, $N=144$ runs) isolating coupled versus decoupled partner experience routing across \texttt{wind\_flocking}, \texttt{navigation}, and \texttt{joint\_passage}.}\label{tab:predecessor_contrasts}\\
\labtoprule \labh{Task} & \labh{Metric} & \labh{n} & \labh{Estimate [95\% CI]} & \labh{Holm p}\\\labheadrule\endfirsthead
\multicolumn{5}{l}{\small Table \thetable\ continued}\\\labtoprule \labh{Task} & \labh{Metric} & \labh{n} & \labh{Estimate [95\% CI]} & \labh{Holm p}\\\labheadrule\endhead
\labrowrule\endfoot
joint passage & auc & 12 & 0.2292 [0.0134, 0.4312] & 0.1377\\\labrowrule
joint passage & portable gain & 12 & -0.5389 [-0.8253, -0.2187] & --\\\labrowrule
joint passage & team & 12 & 0.1671 [-0.0441, 0.3757] & --\\\labrowrule
joint passage & team gain & 12 & 0.1671 [-0.0441, 0.3757] & --\\\labrowrule
joint passage & success & 12 & 0.0000 [0.0000, 0.0000] & --\\\labrowrule
navigation & auc & 12 & 0.0064 [-0.0654, 0.0759] & 0.8662\\\labrowrule
navigation & portable gain & 12 & -0.0505 [-0.1402, 0.0305] & --\\\labrowrule
navigation & team & 12 & 0.1974 [0.0764, 0.3287] & --\\\labrowrule
navigation & team gain & 12 & 0.1974 [0.0764, 0.3287] & --\\\labrowrule
navigation & success & 12 & 0.0130 [-0.1510, 0.1250] & --\\\labrowrule
wind flocking & auc & 12 & 0.7407 [0.6531, 0.8257] & 0.0015\\\labrowrule
wind flocking & portable gain & 12 & 0.7167 [0.1543, 1.4311] & --\\\labrowrule
wind flocking & team & 12 & 0.9885 [0.7170, 1.3462] & --\\\labrowrule
wind flocking & team gain & 12 & 0.9885 [0.7170, 1.3462] & --\\\labrowrule
\end{longtable}

%% file: sections/newer_cells.tex
\begin{longtable}{P{0.152102}P{0.144311}P{0.232190}P{0.051387}P{0.109530}P{0.104347}P{0.096613}P{0.109519}}
\caption{Complete cell means for counterfactual reset-switch, recent-replay filtering, $1.05\text{M}$-step extended-horizon navigation, matched short-horizon OPEN, and external Overcooked cohorts.}\label{tab:newer_cells}\\
\labtoprule \labh{Task} & \labh{Batch} & \labh{Method / arm} & \labh{n} & \labh{Team} & \labh{Bank gain} & \labh{Team succ.} & \labh{Bank succ.}\\\labheadrule\endfirsthead
\multicolumn{8}{l}{\small Table \thetable\ continued}\\\labtoprule \labh{Task} & \labh{Batch} & \labh{Method / arm} & \labh{n} & \labh{Team} & \labh{Bank gain} & \labh{Team succ.} & \labh{Bank succ.}\\\labheadrule\endhead
\labrowrule\endfoot
signal coordination & mechanism confirm & disco103/control reset & 12 & 27.6979 & -0.0208 & 0.7578 & 0.4067\\\labrowrule
signal coordination & mechanism confirm & disco103/switch reset & 12 & 20.0911 & -0.0217 & 0.2500 & 0.3264\\\labrowrule
signal coordination & mechanism confirm & disco103/replay & 12 & 20.0990 & 0.0122 & 0.2500 & 0.3485\\\labrowrule
signal coordination & mechanism confirm & mappo/control reset & 12 & 31.4036 & 0.0208 & 1.0000 & 0.3333\\\labrowrule
signal coordination & mechanism confirm & mappo/switch reset & 12 & 29.9766 & 0.0286 & 0.9167 & 0.2500\\\labrowrule
signal coordination & mechanism confirm & mappo/replay & 12 & 31.4323 & 0.0104 & 1.0000 & 0.2083\\\labrowrule
signal coordination & mechanism confirm & disco103/switch reset (repair) & 12 & 20.3880 & -0.0117 & 0.2500 & 0.4132\\\labrowrule
navigation & mechanism confirm & disco103/control reset & 12 & 0.3247 & 0.0325 & 0.0365 & 0.0163\\\labrowrule
navigation & mechanism confirm & disco103/switch reset & 12 & 0.4005 & 0.1265 & 0.0260 & 0.0124\\\labrowrule
navigation & mechanism confirm & disco103/replay & 11 & 0.4420 & 0.0996 & 0.0199 & 0.0149\\\labrowrule
navigation & mechanism confirm & mappo/control reset & 12 & 1.4281 & 1.0383 & 0.8984 & 0.9023\\\labrowrule
navigation & mechanism confirm & mappo/switch reset & 12 & 1.4972 & 1.0384 & 0.9297 & 0.9069\\\labrowrule
navigation & mechanism confirm & mappo/replay & 11 & 1.4889 & 1.0566 & 0.8665 & 0.8636\\\labrowrule
navigation & mechanism confirm & disco103/switch reset (repair) & 12 & 0.4252 & 0.1346 & 0.0234 & 0.0195\\\labrowrule
navigation & long budget & disco103/coupled & 12 & 1.8290 & 1.3826 & 0.6771 & 0.4609\\\labrowrule
navigation & long budget & disco103/decoupled & 12 & 1.7593 & 1.3344 & 0.6510 & 0.5625\\\labrowrule
navigation & long budget & mappo/coupled & 12 & 1.7419 & 1.2492 & 0.9453 & 0.9499\\\labrowrule
navigation & long budget & mappo/decoupled & 12 & 1.7395 & 1.2804 & 0.9635 & 0.9570\\\labrowrule
navigation & long budget & open/coupled & 12 & 1.5830 & 1.1257 & 0.7109 & 0.7181\\\labrowrule
navigation & long budget & open/decoupled & 12 & 1.6257 & 1.1906 & 0.8333 & 0.8268\\\labrowrule
overcooked cramped room & external & disco103/coupled & 12 & 2.8125 & -3.3802 & 0.0885 & 0.1146\\\labrowrule
overcooked cramped room & external & disco103/decoupled & 12 & 0.9896 & -8.7240 & 0.0495 & 0.0768\\\labrowrule
overcooked cramped room & external & mappo/coupled & 12 & 18.4896 & -9.8125 & 0.4661 & 0.1169\\\labrowrule
overcooked cramped room & external & mappo/decoupled & 12 & 22.9167 & -9.2083 & 0.5312 & 0.1086\\\labrowrule
overcooked cramped room & external & open/coupled & 12 & 0.1042 & -6.6094 & 0.0052 & 0.0487\\\labrowrule
overcooked cramped room & external & open/decoupled & 12 & 0.0000 & -9.2083 & 0.0000 & 0.0367\\\labrowrule
\end{longtable}

%% file: sections/all_contrasts.tex
\begin{longtable}{P{0.252867}P{0.244068}P{0.084047}P{0.239263}P{0.179755}}
\caption{Complete register of 149 paired-block contrasts across the feedback-routing, replay-filtering, source-adaptation, and extended-horizon confirmation cohorts.}\label{tab:all_contrasts}\\
\labtoprule \labh{Task} & \labh{Contrast / metric} & \labh{n} & \labh{Estimate [95\% CI]} & \labh{Holm p}\\\labheadrule\endfirsthead
\multicolumn{5}{l}{\small Table \thetable\ continued}\\\labtoprule \labh{Task} & \labh{Contrast / metric} & \labh{n} & \labh{Estimate [95\% CI]} & \labh{Holm p}\\\labheadrule\endhead
\labrowrule\endfoot
signal coordination & switch interaction / post auc portable gain & 12 & -0.0050 [-0.0351, 0.0258] & 1.0000\\\labrowrule
signal coordination & switch interaction / portable gain & 12 & -0.0087 [-0.0456, 0.0260] & --\\\labrowrule
signal coordination & switch interaction / post auc team & 12 & 0.1381 [-2.1293, 2.5351] & --\\\labrowrule
signal coordination & switch interaction / post auc success & 12 & 0.0149 [-0.1252, 0.1646] & --\\\labrowrule
signal coordination & switch minus control disco103 / post auc portable gain & 12 & -0.0064 [-0.0226, 0.0089] & --\\\labrowrule
signal coordination & replay minus control disco103 / post auc portable gain & 12 & 0.0116 [-0.0037, 0.0265] & --\\\labrowrule
signal coordination & switch minus control disco103 / portable gain & 12 & -0.0009 [-0.0391, 0.0352] & --\\\labrowrule
signal coordination & replay minus control disco103 / portable gain & 12 & 0.0330 [0.0035, 0.0794] & --\\\labrowrule
signal coordination & switch minus control disco103 / post auc team & 12 & -2.6018 [-4.1687, -1.0658] & --\\\labrowrule
signal coordination & replay minus control disco103 / post auc team & 12 & -2.7613 [-4.2240, -1.3547] & --\\\labrowrule
signal coordination & switch minus control mappo / post auc portable gain & 12 & -0.0014 [-0.0243, 0.0209] & --\\\labrowrule
signal coordination & replay minus control mappo / post auc portable gain & 12 & -0.0025 [-0.0192, 0.0130] & --\\\labrowrule
signal coordination & switch minus control mappo / portable gain & 12 & 0.0078 [-0.0169, 0.0312] & --\\\labrowrule
signal coordination & replay minus control mappo / portable gain & 12 & -0.0104 [-0.0404, 0.0221] & --\\\labrowrule
signal coordination & switch minus control mappo / post auc team & 12 & -2.7400 [-5.2760, -0.3598] & --\\\labrowrule
signal coordination & replay minus control mappo / post auc team & 12 & 1.6271 [-0.8400, 3.9660] & --\\\labrowrule
signal coordination & recent replay repair / post auc portable gain & 12 & 0.0235 [0.0142, 0.0334] & 0.0010\\\labrowrule
signal coordination & recent replay repair / portable gain & 12 & 0.0100 [-0.0260, 0.0512] & --\\\labrowrule
signal coordination & recent replay repair / post auc team & 12 & 0.6447 [0.1701, 1.2269] & --\\\labrowrule
signal coordination & recent replay repair / team & 12 & 0.2969 [-0.1276, 0.9245] & --\\\labrowrule
navigation & switch interaction / post auc portable gain & 12 & 0.0145 [-0.0427, 0.0842] & 1.0000\\\labrowrule
navigation & switch interaction / portable gain & 12 & 0.0939 [-0.0268, 0.2131] & --\\\labrowrule
navigation & switch interaction / post auc team & 12 & -0.0246 [-0.1616, 0.1110] & --\\\labrowrule
navigation & switch interaction / post auc success & 12 & -0.0468 [-0.1165, 0.0190] & --\\\labrowrule
navigation & switch minus control disco103 / post auc portable gain & 12 & 0.0202 [-0.0118, 0.0515] & --\\\labrowrule
navigation & replay minus control disco103 / post auc portable gain & 11 & 0.0008 [-0.0348, 0.0396] & --\\\labrowrule
navigation & switch minus control disco103 / portable gain & 12 & 0.0940 [0.0282, 0.1545] & --\\\labrowrule
navigation & replay minus control disco103 / portable gain & 11 & 0.0562 [0.0131, 0.0963] & --\\\labrowrule
navigation & switch minus control disco103 / post auc team & 12 & 0.0069 [-0.1026, 0.1012] & --\\\labrowrule
navigation & replay minus control disco103 / post auc team & 11 & -0.0129 [-0.1394, 0.0782] & --\\\labrowrule
navigation & switch minus control mappo / post auc portable gain & 12 & 0.0057 [-0.0530, 0.0643] & --\\\labrowrule
navigation & replay minus control mappo / post auc portable gain & 11 & 0.0281 [-0.0432, 0.0981] & --\\\labrowrule
navigation & switch minus control mappo / portable gain & 12 & 0.0001 [-0.1069, 0.0995] & --\\\labrowrule
navigation & replay minus control mappo / portable gain & 11 & 0.0201 [-0.0547, 0.0897] & --\\\labrowrule
navigation & switch minus control mappo / post auc team & 12 & 0.0316 [-0.0712, 0.1252] & --\\\labrowrule
navigation & replay minus control mappo / post auc team & 11 & -0.0760 [-0.2213, 0.0609] & --\\\labrowrule
navigation & recent replay repair / post auc portable gain & 12 & -0.0021 [-0.0292, 0.0255] & 0.8843\\\labrowrule
navigation & recent replay repair / portable gain & 12 & 0.0081 [-0.0653, 0.0805] & --\\\labrowrule
navigation & recent replay repair / post auc team & 12 & -0.0033 [-0.0459, 0.0400] & --\\\labrowrule
navigation & recent replay repair / team & 12 & 0.0247 [-0.0687, 0.1182] & --\\\labrowrule
navigation & coupled minus decoupled disco103 / team & 12 & 0.0697 [-0.0420, 0.1932] & --\\\labrowrule
navigation & coupled minus decoupled disco103 / portable gain & 12 & 0.0481 [-0.0178, 0.1211] & --\\\labrowrule
navigation & coupled minus decoupled disco103 / success & 12 & 0.0260 [-0.0964, 0.1354] & --\\\labrowrule
navigation & coupled minus decoupled disco103 / bank success & 12 & -0.1016 [-0.1953, -0.0130] & --\\\labrowrule
navigation & coupled minus decoupled disco103 / auc team & 12 & 0.0120 [-0.0774, 0.1068] & --\\\labrowrule
navigation & coupled minus decoupled disco103 / auc portable gain & 12 & 0.0092 [-0.0775, 0.0922] & --\\\labrowrule
navigation & coupled minus decoupled mappo / team & 12 & 0.0025 [-0.0687, 0.0773] & --\\\labrowrule
navigation & coupled minus decoupled mappo / portable gain & 12 & -0.0312 [-0.1158, 0.0483] & --\\\labrowrule
navigation & coupled minus decoupled mappo / success & 12 & -0.0182 [-0.0599, 0.0208] & --\\\labrowrule
navigation & coupled minus decoupled mappo / bank success & 12 & -0.0072 [-0.0378, 0.0208] & --\\\labrowrule
navigation & coupled minus decoupled mappo / auc team & 12 & 0.0289 [-0.0304, 0.0944] & --\\\labrowrule
navigation & coupled minus decoupled mappo / auc portable gain & 12 & -0.0243 [-0.0706, 0.0159] & --\\\labrowrule
navigation & coupled minus decoupled open / team & 12 & -0.0427 [-0.1438, 0.0548] & --\\\labrowrule
navigation & coupled minus decoupled open / portable gain & 12 & -0.0649 [-0.1265, -0.0059] & --\\\labrowrule
navigation & coupled minus decoupled open / success & 12 & -0.1224 [-0.2292, -0.0182] & --\\\labrowrule
navigation & coupled minus decoupled open / bank success & 12 & -0.1087 [-0.2031, -0.0156] & --\\\labrowrule
navigation & coupled minus decoupled open / auc team & 12 & -0.0924 [-0.1852, 0.0021] & --\\\labrowrule
navigation & coupled minus decoupled open / auc portable gain & 12 & -0.0144 [-0.0566, 0.0264] & --\\\labrowrule
navigation & method feedback interaction / team & 12 & 0.0672 [-0.0431, 0.1800] & --\\\labrowrule
navigation & method feedback interaction / portable gain & 12 & 0.0794 [-0.0309, 0.1916] & --\\\labrowrule
navigation & method feedback interaction / auc team & 12 & -0.0169 [-0.1247, 0.0954] & --\\\labrowrule
navigation & method feedback interaction / auc portable gain & 12 & 0.0335 [-0.0678, 0.1327] & --\\\labrowrule
navigation & method feedback interaction / success & 12 & 0.0443 [-0.0729, 0.1589] & --\\\labrowrule
navigation & method feedback interaction / bank success & 12 & -0.0944 [-0.1966, 0.0091] & --\\\labrowrule
navigation & open minus disco103 matched budget coupled / portable gain & 12 & 0.9879 [0.8698, 1.0977] & --\\\labrowrule
navigation & open minus mappo matched budget coupled / portable gain & 12 & 0.0127 [-0.0523, 0.0807] & --\\\labrowrule
navigation & open minus disco103 matched budget decoupled / portable gain & 12 & 1.0700 [0.9930, 1.1658] & --\\\labrowrule
navigation & open minus mappo matched budget decoupled / portable gain & 12 & 0.0492 [-0.0055, 0.1096] & --\\\labrowrule
overcooked cramped room & coupled minus decoupled disco103 / team & 12 & 1.8229 [-1.1458, 6.6146] & --\\\labrowrule
overcooked cramped room & coupled minus decoupled disco103 / portable gain & 12 & 5.3438 [1.1301, 9.3229] & --\\\labrowrule
overcooked cramped room & coupled minus decoupled disco103 / success & 12 & 0.0391 [-0.0573, 0.1745] & --\\\labrowrule
overcooked cramped room & coupled minus decoupled disco103 / bank success & 12 & 0.0378 [0.0174, 0.0594] & --\\\labrowrule
overcooked cramped room & coupled minus decoupled disco103 / auc team & 12 & 0.4557 [-0.2865, 1.6536] & --\\\labrowrule
overcooked cramped room & coupled minus decoupled disco103 / auc portable gain & 12 & 0.9427 [-0.2514, 2.0794] & --\\\labrowrule
overcooked cramped room & coupled minus decoupled mappo / team & 12 & -4.4271 [-18.3333, 10.4688] & --\\\labrowrule
overcooked cramped room & coupled minus decoupled mappo / portable gain & 12 & -0.6042 [-2.3438, 0.9844] & --\\\labrowrule
overcooked cramped room & coupled minus decoupled mappo / success & 12 & -0.0651 [-0.3203, 0.2057] & --\\\labrowrule
overcooked cramped room & coupled minus decoupled mappo / bank success & 12 & 0.0083 [-0.0284, 0.0503] & --\\\labrowrule
overcooked cramped room & coupled minus decoupled mappo / auc team & 12 & -0.2474 [-3.8542, 3.3984] & --\\\labrowrule
overcooked cramped room & coupled minus decoupled mappo / auc portable gain & 12 & 0.4688 [-0.2279, 1.1914] & --\\\labrowrule
overcooked cramped room & coupled minus decoupled open / team & 12 & 0.1042 [0.0000, 0.2604] & --\\\labrowrule
overcooked cramped room & coupled minus decoupled open / portable gain & 12 & 2.5990 [-2.7033, 7.9062] & --\\\labrowrule
overcooked cramped room & coupled minus decoupled open / success & 12 & 0.0052 [0.0000, 0.0130] & --\\\labrowrule
overcooked cramped room & coupled minus decoupled open / bank success & 12 & 0.0120 [-0.0253, 0.0482] & --\\\labrowrule
overcooked cramped room & coupled minus decoupled open / auc team & 12 & 0.0521 [0.0000, 0.1172] & --\\\labrowrule
overcooked cramped room & coupled minus decoupled open / auc portable gain & 12 & -1.5690 [-4.0573, 0.9232] & --\\\labrowrule
overcooked cramped room & method feedback interaction / team & 12 & 6.2500 [-8.8542, 20.0534] & --\\\labrowrule
overcooked cramped room & method feedback interaction / portable gain & 12 & 5.9479 [1.8750, 10.2031] & --\\\labrowrule
overcooked cramped room & method feedback interaction / auc team & 12 & 0.7031 [-2.9818, 4.2057] & --\\\labrowrule
overcooked cramped room & method feedback interaction / auc portable gain & 12 & 0.4740 [-0.9844, 1.8216] & --\\\labrowrule
overcooked cramped room & method feedback interaction / success & 12 & 0.1042 [-0.1797, 0.3698] & --\\\labrowrule
overcooked cramped room & method feedback interaction / bank success & 12 & 0.0294 [-0.0096, 0.0672] & --\\\labrowrule
overcooked cramped room & open minus disco103 coupled / portable gain & 12 & -3.2292 [-7.8387, 1.7812] & --\\\labrowrule
overcooked cramped room & open minus mappo coupled / portable gain & 12 & 3.2031 [-0.6615, 7.3698] & --\\\labrowrule
overcooked cramped room & open minus disco103 decoupled / portable gain & 12 & -0.4844 [-4.5470, 4.2552] & --\\\labrowrule
overcooked cramped room & open minus mappo decoupled / portable gain & 12 & 0.0000 [-3.8698, 4.7865] & --\\\labrowrule
navigation & source feedback matching / portable gain & 12 & -0.0138 [-0.0568, 0.0228] & 0.8086\\\labrowrule
navigation & source feedback matching / team & 12 & -0.0460 [-0.1215, 0.0278] & --\\\labrowrule
navigation & source feedback matching / success & 12 & -0.0026 [-0.0169, 0.0104] & --\\\labrowrule
navigation & source feedback matching / bank success & 12 & -0.0016 [-0.0033, 0.0000] & --\\\labrowrule
navigation & source feedback matching / auc portable gain & 12 & -0.0027 [-0.0121, 0.0046] & --\\\labrowrule
navigation & coupled source minus eta0 coupled / portable gain & 12 & -0.0611 [-0.1336, 0.0398] & --\\\labrowrule
navigation & coupled source minus eta0 coupled / team & 12 & 0.0167 [-0.1458, 0.2009] & --\\\labrowrule
navigation & coupled source minus eta0 coupled / success & 12 & -0.0104 [-0.0208, 0.0026] & --\\\labrowrule
navigation & coupled source minus eta0 coupled / bank success & 12 & -0.0088 [-0.0124, -0.0049] & --\\\labrowrule
navigation & coupled source minus eta0 coupled / auc portable gain & 12 & 0.0065 [-0.0328, 0.0751] & --\\\labrowrule
navigation & coupled source minus eta0 decoupled / portable gain & 12 & -0.0167 [-0.0809, 0.0756] & --\\\labrowrule
navigation & coupled source minus eta0 decoupled / team & 12 & -0.0095 [-0.1125, 0.1125] & --\\\labrowrule
navigation & coupled source minus eta0 decoupled / success & 12 & 0.0013 [-0.0065, 0.0091] & --\\\labrowrule
navigation & coupled source minus eta0 decoupled / bank success & 12 & -0.0026 [-0.0072, 0.0013] & --\\\labrowrule
navigation & coupled source minus eta0 decoupled / auc portable gain & 12 & 0.0125 [-0.0333, 0.0850] & --\\\labrowrule
navigation & decoupled source minus eta0 coupled / portable gain & 12 & -0.0886 [-0.1204, -0.0591] & --\\\labrowrule
navigation & decoupled source minus eta0 coupled / team & 12 & -0.0243 [-0.1301, 0.1215] & --\\\labrowrule
navigation & decoupled source minus eta0 coupled / success & 12 & -0.0039 [-0.0143, 0.0065] & --\\\labrowrule
navigation & decoupled source minus eta0 coupled / bank success & 12 & -0.0088 [-0.0130, -0.0046] & --\\\labrowrule
navigation & decoupled source minus eta0 coupled / auc portable gain & 12 & -0.0263 [-0.0344, -0.0178] & --\\\labrowrule
navigation & decoupled source minus eta0 decoupled / portable gain & 12 & -0.0580 [-0.0967, -0.0220] & --\\\labrowrule
navigation & decoupled source minus eta0 decoupled / team & 12 & -0.0965 [-0.1675, -0.0282] & --\\\labrowrule
navigation & decoupled source minus eta0 decoupled / success & 12 & 0.0052 [-0.0013, 0.0117] & --\\\labrowrule
navigation & decoupled source minus eta0 decoupled / bank success & 12 & -0.0042 [-0.0072, -0.0013] & --\\\labrowrule
navigation & decoupled source minus eta0 decoupled / auc portable gain & 12 & -0.0229 [-0.0330, -0.0126] & --\\\labrowrule
overcooked cramped room & source feedback matching / portable gain & 12 & 0.5260 [-0.6536, 1.6406] & 0.8086\\\labrowrule
overcooked cramped room & source feedback matching / team & 12 & 0.0781 [-0.1302, 0.2865] & --\\\labrowrule
overcooked cramped room & source feedback matching / success & 12 & 0.0039 [-0.0065, 0.0143] & --\\\labrowrule
overcooked cramped room & source feedback matching / bank success & 12 & 0.0021 [-0.0052, 0.0089] & --\\\labrowrule
overcooked cramped room & source feedback matching / auc portable gain & 12 & 0.3761 [-0.0255, 0.8747] & --\\\labrowrule
overcooked cramped room & coupled source minus eta0 coupled / portable gain & 12 & -0.7396 [-1.6562, 0.1016] & --\\\labrowrule
overcooked cramped room & coupled source minus eta0 coupled / team & 12 & -0.1042 [-0.3385, 0.1302] & --\\\labrowrule
overcooked cramped room & coupled source minus eta0 coupled / success & 12 & -0.0052 [-0.0169, 0.0065] & --\\\labrowrule
overcooked cramped room & coupled source minus eta0 coupled / bank success & 12 & -0.0060 [-0.0147, 0.0010] & --\\\labrowrule
overcooked cramped room & coupled source minus eta0 coupled / auc portable gain & 12 & -0.4137 [-1.1080, 0.1380] & --\\\labrowrule
overcooked cramped room & coupled source minus eta0 decoupled / portable gain & 12 & -0.7630 [-1.8412, 0.1458] & --\\\labrowrule
overcooked cramped room & coupled source minus eta0 decoupled / team & 12 & -0.1042 [-0.1823, -0.0260] & --\\\labrowrule
overcooked cramped room & coupled source minus eta0 decoupled / success & 12 & -0.0052 [-0.0091, -0.0013] & --\\\labrowrule
overcooked cramped room & coupled source minus eta0 decoupled / bank success & 12 & -0.0009 [-0.0125, 0.0073] & --\\\labrowrule
overcooked cramped room & coupled source minus eta0 decoupled / auc portable gain & 12 & -0.8805 [-1.7245, -0.2819] & --\\\labrowrule
overcooked cramped room & decoupled source minus eta0 coupled / portable gain & 12 & -0.1693 [-0.7891, 0.4375] & --\\\labrowrule
overcooked cramped room & decoupled source minus eta0 coupled / team & 12 & -0.1302 [-0.3125, 0.0260] & --\\\labrowrule
overcooked cramped room & decoupled source minus eta0 coupled / success & 12 & -0.0065 [-0.0156, 0.0013] & --\\\labrowrule
overcooked cramped room & decoupled source minus eta0 coupled / bank success & 12 & 0.0005 [-0.0036, 0.0046] & --\\\labrowrule
overcooked cramped room & decoupled source minus eta0 coupled / auc portable gain & 12 & -0.1126 [-0.3970, 0.1594] & --\\\labrowrule
overcooked cramped room & decoupled source minus eta0 decoupled / portable gain & 12 & 0.3333 [-0.1355, 0.7839] & --\\\labrowrule
overcooked cramped room & decoupled source minus eta0 decoupled / team & 12 & -0.0521 [-0.1562, 0.0521] & --\\\labrowrule
overcooked cramped room & decoupled source minus eta0 decoupled / success & 12 & -0.0026 [-0.0078, 0.0026] & --\\\labrowrule
overcooked cramped room & decoupled source minus eta0 decoupled / bank success & 12 & 0.0077 [0.0042, 0.0112] & --\\\labrowrule
overcooked cramped room & decoupled source minus eta0 decoupled / auc portable gain & 12 & -0.2033 [-0.3785, -0.0165] & --\\\labrowrule
navigation & switch interaction windows stratum / post auc portable gain & 7 & 0.0145 [-0.0632, 0.1159] & --\\\labrowrule
navigation & switch interaction linux stratum / post auc portable gain & 5 & 0.0146 [-0.0512, 0.1123] & --\\\labrowrule
navigation & repair same platform pairs / post auc portable gain & 11 & -0.0072 [-0.0345, 0.0220] & --\\\labrowrule
\end{longtable}

%% file: sections/g2_exact_summary.tex
\begin{longtable}{P{0.133986}P{0.262666}P{0.197400}P{0.202974}P{0.202974}}
\caption{Exact two-player differentiable game diagnostic across step sizes ($432$ comparisons per step size across two payoff matrices, 24 initializations, three gradient modes, and inner horizons $K\in\{1,4,16\}$).}\label{tab:g2exact}\\
\labtoprule \labh{Step size} & \labh{Comparisons} & \labh{Above tolerance} & \labh{Maximum relative error} & \labh{Maximum absolute error}\\\labheadrule\endfirsthead
\labtoprule \labh{Step size} & \labh{Comparisons} & \labh{Above tolerance} & \labh{Maximum relative error} & \labh{Maximum absolute error}\\\labheadrule\endhead
\labrowrule\endfoot
1e-03 & 432 & 1 & 1.22e-06 & 1.16e-07\\\labrowrule
1e-04 & 432 & 0 & 9.84e-08 & 1.16e-09\\\labrowrule
1e-05 & 432 & 0 & 7.34e-07 & 2.81e-11\\\labrowrule
\end{longtable}

%% file: sections/g1_blocks.tex
\begin{longtable}{P{0.393452}P{0.202931}P{0.403617}}
\caption{Block-level paired perturbation-response contrasts ($\Delta_C$ at $65{,}656$ steps, $\xi=0.01$) across all $n=12$ confirmation blocks per task.}\label{tab:g1_blocks}\\
\labtoprule \labh{Task} & \labh{Block} & \labh{Response contrast}\\\labheadrule\endfirsthead
\multicolumn{3}{l}{\small Table \thetable\ continued}\\\labtoprule \labh{Task} & \labh{Block} & \labh{Response contrast}\\\labheadrule\endhead
\labrowrule\endfoot
wind flocking & 1 & -0.1743\\\labrowrule
wind flocking & 2 & -0.2536\\\labrowrule
wind flocking & 3 & -0.0354\\\labrowrule
wind flocking & 4 & -0.0684\\\labrowrule
wind flocking & 5 & -0.3319\\\labrowrule
wind flocking & 6 & -0.8049\\\labrowrule
wind flocking & 7 & -0.0255\\\labrowrule
wind flocking & 8 & -0.4569\\\labrowrule
wind flocking & 9 & -0.0460\\\labrowrule
wind flocking & 10 & -0.0581\\\labrowrule
wind flocking & 11 & -0.2452\\\labrowrule
wind flocking & 12 & -0.3030\\\labrowrule
navigation & 1 & -0.2708\\\labrowrule
navigation & 2 & -0.1986\\\labrowrule
navigation & 3 & -0.7217\\\labrowrule
navigation & 4 & -0.4841\\\labrowrule
navigation & 5 & -0.0624\\\labrowrule
navigation & 6 & -0.6593\\\labrowrule
navigation & 7 & -0.3827\\\labrowrule
navigation & 8 & -0.3197\\\labrowrule
navigation & 9 & -0.3326\\\labrowrule
navigation & 10 & -0.0952\\\labrowrule
navigation & 11 & -0.4001\\\labrowrule
navigation & 12 & -0.0866\\\labrowrule
\end{longtable}

%% file: sections/g3_descriptive.tex
\begin{longtable}{P{0.190545}P{0.169338}P{0.203128}P{0.146829}P{0.136275}P{0.153886}}
\caption{Measurement repeatability of early perturbation response and per-method standardized ridge prediction losses ($L_0$ vs.\ $L_1$) across confirmation blocks.}\label{tab:g3_descriptive}\\
\labtoprule \labh{Task} & \labh{Method} & \labh{Repeat correlation} & \labh{Mean bank gain} & \labh{M0 loss} & \labh{M1 loss}\\\labheadrule\endfirsthead
\multicolumn{6}{l}{\small Table \thetable\ continued}\\\labtoprule \labh{Task} & \labh{Method} & \labh{Repeat correlation} & \labh{Mean bank gain} & \labh{M0 loss} & \labh{M1 loss}\\\labheadrule\endhead
\labrowrule\endfoot
wind flocking & disco103 & 0.1206 & 0.1713 & 0.2578 & 0.2603\\\labrowrule
wind flocking & mappo & 0.6624 & -0.9709 & 0.7142 & 0.7115\\\labrowrule
navigation & disco103 & -0.0670 & 1.1499 & 0.3317 & 0.3309\\\labrowrule
navigation & mappo & 0.0554 & 1.1966 & 1.0981 & 1.1021\\\labrowrule
\end{longtable}

%% file: sections/g3_blocks.tex
\begin{longtable}{P{0.391306}P{0.202169}P{0.406525}}
\caption{Block-level out-of-sample squared-error reduction scores ($S$) at $1{,}048{,}640$ steps across all $n=12$ confirmation blocks per task.}\label{tab:g3_blocks}\\
\labtoprule \labh{Task} & \labh{Block} & \labh{Prediction score}\\\labheadrule\endfirsthead
\multicolumn{3}{l}{\small Table \thetable\ continued}\\\labtoprule \labh{Task} & \labh{Block} & \labh{Prediction score}\\\labheadrule\endhead
\labrowrule\endfoot
wind flocking & 1 & -0.0002\\\labrowrule
wind flocking & 2 & -0.0060\\\labrowrule
wind flocking & 3 & -0.0023\\\labrowrule
wind flocking & 4 & 0.0079\\\labrowrule
wind flocking & 5 & -0.0016\\\labrowrule
wind flocking & 6 & 0.0009\\\labrowrule
wind flocking & 7 & 0.0002\\\labrowrule
wind flocking & 8 & -0.0053\\\labrowrule
wind flocking & 9 & 0.0009\\\labrowrule
wind flocking & 10 & 0.0033\\\labrowrule
wind flocking & 11 & 0.0017\\\labrowrule
wind flocking & 12 & 0.0023\\\labrowrule
navigation & 1 & 0.0003\\\labrowrule
navigation & 2 & -0.0061\\\labrowrule
navigation & 3 & -0.0013\\\labrowrule
navigation & 4 & -0.0011\\\labrowrule
navigation & 5 & -0.0027\\\labrowrule
navigation & 6 & -0.0003\\\labrowrule
navigation & 7 & -0.0044\\\labrowrule
navigation & 8 & -0.0003\\\labrowrule
navigation & 9 & -0.0038\\\labrowrule
navigation & 10 & -0.0054\\\labrowrule
navigation & 11 & 0.0042\\\labrowrule
navigation & 12 & 0.0014\\\labrowrule
\end{longtable}

%% file: sections/micro_strata.tex
\begin{longtable}{P{0.365850}P{0.173918}P{0.230077}P{0.230155}}
\caption{Partner-stratified micro-behavioral decomposition in Overcooked (\texttt{cramped\_room}, $92{,}160$ steps) across Behavioral-Cloning (\texttt{BC}), Reinforcement-Learned (\texttt{RL}), \texttt{Scripted}, and 10-partner aggregate (\texttt{all10}) strata ($n=12$ paired blocks, pointwise 95\% block-bootstrap intervals relative to initial focal policies).}\label{tab:micro_strata}\\
\labtoprule \labh{Method / feedback} & \labh{View} & \labh{Metric} & \labh{Change [95\% CI]}\\\labheadrule\endfirsthead
\multicolumn{4}{l}{\small Table \thetable\ continued}\\\labtoprule \labh{Method / feedback} & \labh{View} & \labh{Metric} & \labh{Change [95\% CI]}\\\labheadrule\endhead
\labrowrule\endfoot
disco103/coupled & BC & team return mean & -0.1693 [-0.4688, 0.1823]\\\labrowrule
disco103/coupled & BC & native completion fraction & -0.0163 [-0.0267, -0.0046]\\\labrowrule
disco103/coupled & BC & active interact & -0.0097 [-0.0192, 0.0002]\\\labrowrule
disco103/coupled & BC & movement & -0.0867 [-0.1337, -0.0445]\\\labrowrule
disco103/coupled & BC & mutual collision block & 0.0411 [0.0075, 0.0812]\\\labrowrule
disco103/coupled & RL & team return mean & 0.1042 [0.0000, 0.2344]\\\labrowrule
disco103/coupled & RL & native completion fraction & 0.0052 [0.0000, 0.0117]\\\labrowrule
disco103/coupled & RL & active interact & -0.0021 [-0.0078, 0.0033]\\\labrowrule
disco103/coupled & RL & movement & -0.0586 [-0.1029, -0.0217]\\\labrowrule
disco103/coupled & RL & mutual collision block & 0.0234 [0.0084, 0.0422]\\\labrowrule
disco103/coupled & Scripted & team return mean & -16.7708 [-31.5625, -2.1875]\\\labrowrule
disco103/coupled & Scripted & native completion fraction & -0.0352 [-0.0729, -0.0039]\\\labrowrule
disco103/coupled & Scripted & active interact & -0.0103 [-0.0245, 0.0049]\\\labrowrule
disco103/coupled & Scripted & movement & -0.1098 [-0.1568, -0.0659]\\\labrowrule
disco103/coupled & Scripted & mutual collision block & 0.0019 [-0.0191, 0.0242]\\\labrowrule
disco103/coupled & all10 & team return mean & -3.3802 [-6.3751, -0.3698]\\\labrowrule
disco103/coupled & all10 & native completion fraction & -0.0115 [-0.0221, -0.0016]\\\labrowrule
disco103/coupled & all10 & active interact & -0.0067 [-0.0153, 0.0018]\\\labrowrule
disco103/coupled & all10 & movement & -0.0801 [-0.1249, -0.0415]\\\labrowrule
disco103/coupled & all10 & mutual collision block & 0.0262 [0.0053, 0.0502]\\\labrowrule
disco103/decoupled & BC & team return mean & -0.8594 [-0.9635, -0.7422]\\\labrowrule
disco103/decoupled & BC & native completion fraction & -0.0430 [-0.0482, -0.0371]\\\labrowrule
disco103/decoupled & BC & active interact & -0.0127 [-0.0239, 0.0046]\\\labrowrule
disco103/decoupled & BC & movement & -0.1664 [-0.2030, -0.1308]\\\labrowrule
disco103/decoupled & BC & mutual collision block & 0.1438 [0.0883, 0.1912]\\\labrowrule
disco103/decoupled & RL & team return mean & 0.0391 [-0.0781, 0.1953]\\\labrowrule
disco103/decoupled & RL & native completion fraction & 0.0020 [-0.0039, 0.0098]\\\labrowrule
disco103/decoupled & RL & active interact & 0.0059 [-0.0048, 0.0188]\\\labrowrule
disco103/decoupled & RL & movement & -0.1300 [-0.1699, -0.0904]\\\labrowrule
disco103/decoupled & RL & mutual collision block & 0.0249 [0.0009, 0.0528]\\\labrowrule
disco103/decoupled & Scripted & team return mean & -41.9792 [-50.2344, -32.0833]\\\labrowrule
disco103/decoupled & Scripted & native completion fraction & -0.1641 [-0.2318, -0.0990]\\\labrowrule
disco103/decoupled & Scripted & active interact & -0.0090 [-0.0210, 0.0047]\\\labrowrule
disco103/decoupled & Scripted & movement & -0.1849 [-0.2187, -0.1476]\\\labrowrule
disco103/decoupled & Scripted & mutual collision block & 0.0493 [-0.0163, 0.1307]\\\labrowrule
disco103/decoupled & all10 & team return mean & -8.7240 [-10.3958, -6.7448]\\\labrowrule
disco103/decoupled & all10 & native completion fraction & -0.0492 [-0.0641, -0.0352]\\\labrowrule
disco103/decoupled & all10 & active interact & -0.0045 [-0.0149, 0.0098]\\\labrowrule
disco103/decoupled & all10 & movement & -0.1555 [-0.1911, -0.1208]\\\labrowrule
disco103/decoupled & all10 & mutual collision block & 0.0773 [0.0421, 0.1161]\\\labrowrule
mappo/coupled & BC & team return mean & -0.1953 [-0.4297, 0.0651]\\\labrowrule
mappo/coupled & BC & native completion fraction & -0.0085 [-0.0195, 0.0033]\\\labrowrule
mappo/coupled & BC & active interact & -0.0332 [-0.0372, -0.0274]\\\labrowrule
mappo/coupled & BC & movement & -0.1392 [-0.1759, -0.1033]\\\labrowrule
mappo/coupled & BC & mutual collision block & 0.2861 [0.2198, 0.3453]\\\labrowrule
mappo/coupled & RL & team return mean & 1.6536 [1.1068, 2.1615]\\\labrowrule
mappo/coupled & RL & native completion fraction & 0.0775 [0.0521, 0.1009]\\\labrowrule
mappo/coupled & RL & active interact & -0.0095 [-0.0149, -0.0043]\\\labrowrule
mappo/coupled & RL & movement & -0.0507 [-0.0924, -0.0104]\\\labrowrule
mappo/coupled & RL & mutual collision block & 0.1241 [0.0733, 0.1803]\\\labrowrule
mappo/coupled & Scripted & team return mean & -51.9792 [-55.8073, -47.8639]\\\labrowrule
mappo/coupled & Scripted & native completion fraction & -0.1419 [-0.2396, -0.0560]\\\labrowrule
mappo/coupled & Scripted & active interact & -0.0345 [-0.0434, -0.0226]\\\labrowrule
mappo/coupled & Scripted & movement & -0.0812 [-0.1712, 0.0169]\\\labrowrule
mappo/coupled & Scripted & mutual collision block & 0.0834 [0.0155, 0.1589]\\\labrowrule
mappo/coupled & all10 & team return mean & -9.8125 [-10.5417, -9.0521]\\\labrowrule
mappo/coupled & all10 & native completion fraction & -0.0008 [-0.0221, 0.0190]\\\labrowrule
mappo/coupled & all10 & active interact & -0.0240 [-0.0284, -0.0183]\\\labrowrule
mappo/coupled & all10 & movement & -0.0922 [-0.1384, -0.0464]\\\labrowrule
mappo/coupled & all10 & mutual collision block & 0.1808 [0.1375, 0.2238]\\\labrowrule
mappo/decoupled & BC & team return mean & -0.1172 [-0.4297, 0.1953]\\\labrowrule
mappo/decoupled & BC & native completion fraction & -0.0052 [-0.0189, 0.0091]\\\labrowrule
mappo/decoupled & BC & active interact & -0.0366 [-0.0403, -0.0331]\\\labrowrule
mappo/decoupled & BC & movement & -0.1712 [-0.2136, -0.1237]\\\labrowrule
mappo/decoupled & BC & mutual collision block & 0.1783 [0.1178, 0.2378]\\\labrowrule
mappo/decoupled & RL & team return mean & 1.5755 [1.0417, 2.0703]\\\labrowrule
mappo/decoupled & RL & native completion fraction & 0.0710 [0.0462, 0.0938]\\\labrowrule
mappo/decoupled & RL & active interact & -0.0118 [-0.0178, -0.0060]\\\labrowrule
mappo/decoupled & RL & movement & -0.0982 [-0.1497, -0.0428]\\\labrowrule
mappo/decoupled & RL & mutual collision block & 0.0975 [0.0480, 0.1513]\\\labrowrule
mappo/decoupled & Scripted & team return mean & -48.9583 [-54.9740, -42.0833]\\\labrowrule
mappo/decoupled & Scripted & native completion fraction & -0.1771 [-0.2813, -0.0781]\\\labrowrule
mappo/decoupled & Scripted & active interact & -0.0392 [-0.0443, -0.0325]\\\labrowrule
mappo/decoupled & Scripted & movement & -0.1708 [-0.2460, -0.0698]\\\labrowrule
mappo/decoupled & Scripted & mutual collision block & 0.0390 [-0.0327, 0.1245]\\\labrowrule
mappo/decoupled & all10 & team return mean & -9.2083 [-10.4375, -7.8385]\\\labrowrule
mappo/decoupled & all10 & native completion fraction & -0.0091 [-0.0383, 0.0167]\\\labrowrule
mappo/decoupled & all10 & active interact & -0.0272 [-0.0309, -0.0240]\\\labrowrule
mappo/decoupled & all10 & movement & -0.1419 [-0.1914, -0.0840]\\\labrowrule
mappo/decoupled & all10 & mutual collision block & 0.1181 [0.0845, 0.1562]\\\labrowrule
open/coupled & BC & team return mean & -0.5599 [-0.8073, -0.2995]\\\labrowrule
open/coupled & BC & native completion fraction & -0.0293 [-0.0384, -0.0195]\\\labrowrule
open/coupled & BC & active interact & -0.0287 [-0.0478, 0.0003]\\\labrowrule
open/coupled & BC & movement & -0.2727 [-0.2984, -0.2256]\\\labrowrule
open/coupled & BC & mutual collision block & -0.0598 [-0.0818, -0.0322]\\\labrowrule
open/coupled & RL & team return mean & -0.1823 [-0.2604, -0.1042]\\\labrowrule
open/coupled & RL & native completion fraction & -0.0091 [-0.0130, -0.0052]\\\labrowrule
open/coupled & RL & active interact & -0.0157 [-0.0422, 0.0184]\\\labrowrule
open/coupled & RL & movement & -0.2697 [-0.2936, -0.2271]\\\labrowrule
open/coupled & RL & mutual collision block & 0.0336 [0.0003, 0.0625]\\\labrowrule
open/coupled & Scripted & team return mean & -31.5625 [-51.4062, -10.4948]\\\labrowrule
open/coupled & Scripted & native completion fraction & -0.2682 [-0.3893, -0.1497]\\\labrowrule
open/coupled & Scripted & active interact & -0.0243 [-0.0464, 0.0065]\\\labrowrule
open/coupled & Scripted & movement & -0.2890 [-0.3163, -0.2382]\\\labrowrule
open/coupled & Scripted & mutual collision block & -0.0658 [-0.0735, -0.0536]\\\labrowrule
open/coupled & all10 & team return mean & -6.6094 [-10.5783, -2.4219]\\\labrowrule
open/coupled & all10 & native completion fraction & -0.0690 [-0.0935, -0.0451]\\\labrowrule
open/coupled & all10 & active interact & -0.0226 [-0.0446, 0.0081]\\\labrowrule
open/coupled & all10 & movement & -0.2748 [-0.2999, -0.2288]\\\labrowrule
open/coupled & all10 & mutual collision block & -0.0236 [-0.0375, -0.0098]\\\labrowrule
open/decoupled & BC & team return mean & -0.7812 [-0.9505, -0.6250]\\\labrowrule
open/decoupled & BC & native completion fraction & -0.0371 [-0.0443, -0.0299]\\\labrowrule
open/decoupled & BC & active interact & -0.0260 [-0.0459, 0.0021]\\\labrowrule
open/decoupled & BC & movement & -0.2644 [-0.2990, -0.1998]\\\labrowrule
open/decoupled & BC & mutual collision block & -0.0713 [-0.0852, -0.0534]\\\labrowrule
open/decoupled & RL & team return mean & -0.1823 [-0.2604, -0.1042]\\\labrowrule
open/decoupled & RL & native completion fraction & -0.0091 [-0.0130, -0.0052]\\\labrowrule
open/decoupled & RL & active interact & -0.0215 [-0.0415, 0.0062]\\\labrowrule
open/decoupled & RL & movement & -0.2597 [-0.2952, -0.1933]\\\labrowrule
open/decoupled & RL & mutual collision block & 0.0311 [0.0066, 0.0533]\\\labrowrule
open/decoupled & Scripted & team return mean & -44.1146 [-60.0260, -24.1133]\\\labrowrule
open/decoupled & Scripted & native completion fraction & -0.3125 [-0.4310, -0.1836]\\\labrowrule
open/decoupled & Scripted & active interact & -0.0230 [-0.0422, 0.0020]\\\labrowrule
open/decoupled & Scripted & movement & -0.2836 [-0.3177, -0.2199]\\\labrowrule
open/decoupled & Scripted & mutual collision block & -0.0662 [-0.0724, -0.0588]\\\labrowrule
open/decoupled & all10 & team return mean & -9.2083 [-12.4010, -5.2135]\\\labrowrule
open/decoupled & all10 & native completion fraction & -0.0810 [-0.1055, -0.0542]\\\labrowrule
open/decoupled & all10 & active interact & -0.0236 [-0.0430, 0.0035]\\\labrowrule
open/decoupled & all10 & movement & -0.2664 [-0.3012, -0.2012]\\\labrowrule
open/decoupled & all10 & mutual collision block & -0.0293 [-0.0358, -0.0232]\\\labrowrule
\end{longtable}

%% file: sections/v2_derived_strata.tex
\begin{longtable}{P{0.194401}P{0.196239}P{0.166947}P{0.204650}P{0.237763}}
\caption{Partner-stratified cross-regime source-transfer return and completion gains in Overcooked at $46{,}080$ steps ($n=12$ paired source blocks, pointwise 95\% block-bootstrap intervals).}\label{tab:v2_derived_strata}\\\labtoprule
\labh{Source} & \labh{Target} & \labh{Stratum} & \labh{Return gain [95\%]} & \labh{Completion gain [95\%]}\\\labheadrule
\endfirsthead\labtoprule
\labh{Source} & \labh{Target} & \labh{Stratum} & \labh{Return gain [95\%]} & \labh{Completion gain [95\%]}\\\labheadrule
\endhead\endfoot\endlastfoot
eta0 & coupled & BC & -0.3125 [-0.4883, -0.1237] & -0.0179 [-0.0254, -0.0104]\\\labrowrule
eta0 & coupled & Scripted & 2.3828 [-0.1693, 5.1172] & 0.0072 [-0.0000, 0.0137]\\\labrowrule
eta0 & coupled & RL & 0.0586 [0.0065, 0.1107] & 0.0029 [0.0003, 0.0055]\\\labrowrule
eta0 & decoupled & BC & -0.5729 [-0.7292, -0.4167] & -0.0296 [-0.0374, -0.0218]\\\labrowrule
eta0 & decoupled & Scripted & -1.2891 [-3.2161, 0.5990] & -0.0059 [-0.0104, -0.0013]\\\labrowrule
eta0 & decoupled & RL & 0.0391 [-0.0000, 0.0781] & 0.0020 [0.0000, 0.0039]\\\labrowrule
coupled & coupled & BC & -0.3190 [-0.5599, -0.0651] & -0.0176 [-0.0293, -0.0055]\\\labrowrule
coupled & coupled & Scripted & -1.1458 [-5.4818, 2.4089] & -0.0156 [-0.0488, 0.0052]\\\labrowrule
coupled & coupled & RL & -0.0195 [-0.0586, 0.0260] & -0.0010 [-0.0029, 0.0013]\\\labrowrule
coupled & decoupled & BC & -0.4688 [-0.6576, -0.2995] & -0.0251 [-0.0342, -0.0169]\\\labrowrule
coupled & decoupled & Scripted & -5.2344 [-11.3802, 0.1432] & -0.0156 [-0.0508, 0.0052]\\\labrowrule
coupled & decoupled & RL & 0.0000 [-0.0456, 0.0521] & 0.0000 [-0.0023, 0.0026]\\\labrowrule
decoupled & coupled & BC & -0.1693 [-0.3190, -0.0326] & -0.0094 [-0.0166, -0.0033]\\\labrowrule
decoupled & coupled & Scripted & 1.3932 [0.3776, 2.4609] & 0.0000 [-0.0033, 0.0033]\\\labrowrule
decoupled & coupled & RL & -0.0130 [-0.0456, 0.0130] & -0.0007 [-0.0023, 0.0007]\\\labrowrule
decoupled & decoupled & BC & -0.2279 [-0.3516, -0.1107] & -0.0124 [-0.0186, -0.0065]\\\labrowrule
decoupled & decoupled & Scripted & -0.2214 [-1.6146, 1.0286] & 0.0026 [-0.0046, 0.0104]\\\labrowrule
decoupled & decoupled & RL & -0.0065 [-0.0456, 0.0326] & -0.0003 [-0.0023, 0.0016]\\\labrowrule
\end{longtable}

%% file: sections/v2_m4_auc_full.tex
\begin{longtable}{P{0.167803}P{0.174940}P{0.126601}P{0.173338}P{0.173338}P{0.183980}}
\caption{Complete 24-cell original-team learning-curve AUC, initial return, and initial-return-adjusted AUC across the 288-run parametric physical-coupling sweep ($\alpha \in [0, 1.5]$, $n=12$ paired blocks per cell).}\label{tab:v2_m4_auc_full}\\\labtoprule
\labh{Method} & \labh{Data} & \labh{Alpha} & \labh{AUC [95\%]} & \labh{Initial return [95\%]} & \labh{Adjusted AUC [95\%]}\\\labheadrule
\endfirsthead\labtoprule
\labh{Method} & \labh{Data} & \labh{Alpha} & \labh{AUC [95\%]} & \labh{Initial return [95\%]} & \labh{Adjusted AUC [95\%]}\\\labheadrule
\endhead\endfoot\endlastfoot
disco103 & coupled & 0 & 0.2969 [0.2750, 0.3179] & -0.2953 [-0.3323, -0.2561] & 0.5922 [0.5454, 0.6331]\\\labrowrule
disco103 & reference & 0 & -0.1722 [-0.2740, -0.0650] & -0.2953 [-0.3323, -0.2561] & 0.1231 [0.0061, 0.2475]\\\labrowrule
mappo & coupled & 0 & 0.5453 [0.5048, 0.5825] & -0.3127 [-0.3639, -0.2607] & 0.8580 [0.7804, 0.9295]\\\labrowrule
mappo & reference & 0 & 0.5607 [0.5333, 0.5897] & -0.3127 [-0.3639, -0.2607] & 0.8734 [0.8115, 0.9296]\\\labrowrule
disco103 & coupled & 0.25 & 0.3016 [0.2768, 0.3222] & -0.3256 [-0.3590, -0.2897] & 0.6272 [0.5879, 0.6643]\\\labrowrule
disco103 & reference & 0.25 & -0.1977 [-0.2908, -0.1021] & -0.3256 [-0.3590, -0.2897] & 0.1279 [0.0348, 0.2265]\\\labrowrule
mappo & coupled & 0.25 & 0.5395 [0.5043, 0.5730] & -0.3367 [-0.3876, -0.2834] & 0.8762 [0.8199, 0.9331]\\\labrowrule
mappo & reference & 0.25 & 0.5340 [0.4930, 0.5757] & -0.3367 [-0.3876, -0.2834] & 0.8707 [0.8095, 0.9288]\\\labrowrule
disco103 & coupled & 0.5 & 0.3148 [0.2852, 0.3442] & -0.3529 [-0.3834, -0.3204] & 0.6676 [0.6213, 0.7093]\\\labrowrule
disco103 & reference & 0.5 & -0.3366 [-0.3972, -0.2649] & -0.3529 [-0.3834, -0.3204] & 0.0163 [-0.0524, 0.0891]\\\labrowrule
mappo & coupled & 0.5 & 0.5269 [0.4836, 0.5702] & -0.3603 [-0.4094, -0.3082] & 0.8873 [0.8092, 0.9635]\\\labrowrule
mappo & reference & 0.5 & 0.4825 [0.4503, 0.5148] & -0.3603 [-0.4094, -0.3082] & 0.8428 [0.7743, 0.9093]\\\labrowrule
disco103 & coupled & 0.75 & 0.3218 [0.2956, 0.3458] & -0.3698 [-0.3971, -0.3398] & 0.6916 [0.6505, 0.7277]\\\labrowrule
disco103 & reference & 0.75 & -0.3474 [-0.4157, -0.2803] & -0.3698 [-0.3971, -0.3398] & 0.0224 [-0.0405, 0.0869]\\\labrowrule
mappo & coupled & 0.75 & 0.4871 [0.4152, 0.5451] & -0.3763 [-0.4228, -0.3271] & 0.8634 [0.8019, 0.9203]\\\labrowrule
mappo & reference & 0.75 & 0.4827 [0.4466, 0.5177] & -0.3763 [-0.4228, -0.3271] & 0.8590 [0.8158, 0.9002]\\\labrowrule
disco103 & coupled & 1 & 0.3133 [0.2868, 0.3405] & -0.3868 [-0.4159, -0.3567] & 0.7002 [0.6537, 0.7473]\\\labrowrule
disco103 & reference & 1 & -0.4371 [-0.4970, -0.3679] & -0.3868 [-0.4159, -0.3567] & -0.0503 [-0.1095, 0.0214]\\\labrowrule
mappo & coupled & 1 & 0.4578 [0.4003, 0.5080] & -0.3921 [-0.4354, -0.3466] & 0.8499 [0.7917, 0.9099]\\\labrowrule
mappo & reference & 1 & 0.4534 [0.4105, 0.4964] & -0.3921 [-0.4354, -0.3466] & 0.8454 [0.7800, 0.9068]\\\labrowrule
disco103 & coupled & 1.5 & 0.3023 [0.2841, 0.3203] & -0.4085 [-0.4347, -0.3809] & 0.7108 [0.6769, 0.7419]\\\labrowrule
disco103 & reference & 1.5 & -0.4903 [-0.5943, -0.4082] & -0.4085 [-0.4347, -0.3809] & -0.0818 [-0.1876, 0.0009]\\\labrowrule
mappo & coupled & 1.5 & 0.4772 [0.4384, 0.5162] & -0.4100 [-0.4487, -0.3698] & 0.8872 [0.8232, 0.9441]\\\labrowrule
mappo & reference & 1.5 & 0.5212 [0.4802, 0.5645] & -0.4100 [-0.4487, -0.3698] & 0.9312 [0.8825, 0.9783]\\\labrowrule
\end{longtable}

%% file: sections/v2_m4_bank.tex
\begin{longtable}{P{0.168059}P{0.175205}P{0.126786}P{0.173665}P{0.173665}P{0.182621}}
\caption{Complete 24-cell cross-partner bank evaluation across the parametric physical-coupling sweep ($5{,}760$ checkpoint evaluations against the 10-partner bank, $n=12$ paired blocks per cell).}\label{tab:v2_m4_bank}\\\labtoprule
\labh{Method} & \labh{Data} & \labh{Alpha} & \labh{F0 bank return [95\%]} & \labh{FT bank return [95\%]} & \labh{Gain [95\%]}\\\labheadrule
\endfirsthead\labtoprule
\labh{Method} & \labh{Data} & \labh{Alpha} & \labh{F0 bank return [95\%]} & \labh{FT bank return [95\%]} & \labh{Gain [95\%]}\\\labheadrule
\endhead\endfoot\endlastfoot
disco103 & coupled & 0 & -0.4029 [-0.4296, -0.3785] & -0.1751 [-0.2581, -0.0807] & 0.2277 [0.1519, 0.3187]\\\labrowrule
disco103 & reference & 0 & -0.4029 [-0.4296, -0.3785] & -0.4909 [-0.9450, -0.1335] & -0.0880 [-0.5354, 0.2666]\\\labrowrule
mappo & coupled & 0 & -0.4133 [-0.4554, -0.3655] & -0.0864 [-0.2417, 0.0586] & 0.3268 [0.1698, 0.4817]\\\labrowrule
mappo & reference & 0 & -0.4133 [-0.4554, -0.3655] & -0.3097 [-0.6210, -0.0316] & 0.1035 [-0.2029, 0.3965]\\\labrowrule
disco103 & coupled & 0.25 & -0.4341 [-0.4598, -0.4104] & -0.1880 [-0.2758, -0.0983] & 0.2461 [0.1675, 0.3288]\\\labrowrule
disco103 & reference & 0.25 & -0.4341 [-0.4598, -0.4104] & -0.5272 [-0.8999, -0.2214] & -0.0931 [-0.4610, 0.2159]\\\labrowrule
mappo & coupled & 0.25 & -0.4414 [-0.4786, -0.3981] & -0.4498 [-1.0386, -0.0099] & -0.0083 [-0.6305, 0.4455]\\\labrowrule
mappo & reference & 0.25 & -0.4414 [-0.4786, -0.3981] & -0.3968 [-0.6787, -0.1169] & 0.0447 [-0.2556, 0.3456]\\\labrowrule
disco103 & coupled & 0.5 & -0.4567 [-0.4794, -0.4353] & -0.1599 [-0.2725, -0.0562] & 0.2968 [0.1972, 0.3884]\\\labrowrule
disco103 & reference & 0.5 & -0.4567 [-0.4794, -0.4353] & -0.8594 [-1.2162, -0.5423] & -0.4027 [-0.7511, -0.0947]\\\labrowrule
mappo & coupled & 0.5 & -0.4625 [-0.4958, -0.4243] & -0.0543 [-0.2505, 0.1283] & 0.4083 [0.1945, 0.6104]\\\labrowrule
mappo & reference & 0.5 & -0.4625 [-0.4958, -0.4243] & -0.3318 [-0.5455, -0.1073] & 0.1307 [-0.1048, 0.3768]\\\labrowrule
disco103 & coupled & 0.75 & -0.4746 [-0.4954, -0.4550] & -0.2056 [-0.3118, -0.1037] & 0.2690 [0.1801, 0.3552]\\\labrowrule
disco103 & reference & 0.75 & -0.4746 [-0.4954, -0.4550] & -0.5798 [-0.8170, -0.3438] & -0.1051 [-0.3384, 0.1270]\\\labrowrule
mappo & coupled & 0.75 & -0.4790 [-0.5080, -0.4455] & -0.1854 [-0.4171, 0.0026] & 0.2936 [0.0612, 0.4868]\\\labrowrule
mappo & reference & 0.75 & -0.4790 [-0.5080, -0.4455] & -0.2673 [-0.6093, 0.0379] & 0.2117 [-0.1338, 0.5249]\\\labrowrule
disco103 & coupled & 1 & -0.4879 [-0.5073, -0.4694] & -0.2129 [-0.3359, -0.0868] & 0.2750 [0.1576, 0.3969]\\\labrowrule
disco103 & reference & 1 & -0.4879 [-0.5073, -0.4694] & -0.5175 [-0.6739, -0.3631] & -0.0296 [-0.1939, 0.1334]\\\labrowrule
mappo & coupled & 1 & -0.4913 [-0.5173, -0.4621] & -0.0350 [-0.2815, 0.1599] & 0.4563 [0.1866, 0.6625]\\\labrowrule
mappo & reference & 1 & -0.4913 [-0.5173, -0.4621] & -0.2024 [-0.4501, 0.0340] & 0.2889 [0.0297, 0.5333]\\\labrowrule
disco103 & coupled & 1.5 & -0.5021 [-0.5206, -0.4846] & -0.3749 [-0.4671, -0.2742] & 0.1272 [0.0481, 0.2145]\\\labrowrule
disco103 & reference & 1.5 & -0.5021 [-0.5206, -0.4846] & -0.5665 [-1.0912, -0.2190] & -0.0644 [-0.5800, 0.2774]\\\labrowrule
mappo & coupled & 1.5 & -0.5045 [-0.5270, -0.4802] & -0.0473 [-0.3427, 0.1690] & 0.4572 [0.1396, 0.6889]\\\labrowrule
mappo & reference & 1.5 & -0.5045 [-0.5270, -0.4802] & -0.2223 [-0.7534, 0.1207] & 0.2822 [-0.2478, 0.6269]\\\labrowrule
\end{longtable}

%% file: sections/v2_m4_blocks.tex
\begin{longtable}{P{0.341723}P{0.658277}}
\caption{Within-block linear interaction slopes ($b$, native AUC units per unit of physical coupling $\alpha$) across all $n=12$ paired training blocks.}\label{tab:v2_m4_blocks}\\\labtoprule
\labh{Block} & \labh{Slope}\\\labheadrule
\endfirsthead\labtoprule
\labh{Block} & \labh{Slope}\\\labheadrule
\endhead\endfoot\endlastfoot
0 & 0.1959\\\labrowrule
1 & 0.3420\\\labrowrule
2 & 0.2723\\\labrowrule
3 & 0.4610\\\labrowrule
4 & 0.1074\\\labrowrule
5 & 0.1990\\\labrowrule
6 & 0.2251\\\labrowrule
7 & 0.1825\\\labrowrule
8 & 0.2790\\\labrowrule
9 & 0.2752\\\labrowrule
10 & 0.2244\\\labrowrule
11 & 0.2920\\\labrowrule
\end{longtable}

%% file: sections/v2_source48.tex
\begin{longtable}{P{0.186050}P{0.127784}P{0.128113}P{0.177532}P{0.178865}P{0.201655}}
\caption{Complete meta-training configuration, validation metrics, and downstream rank synthesis for all 48 meta-evolved multi-agent candidate update rules.}\label{tab:v2_source48}\\
\labh{Arm} & \labh{Block} & \labh{Outer steps} & \labh{Rule distance} & \labh{Physical worlds} & \labh{Meta-train time (h)}\\\labheadrule
\endfirsthead\labtoprule
\labh{Arm} & \labh{Block} & \labh{Outer steps} & \labh{Rule distance} & \labh{Physical worlds} & \labh{Meta-train time (h)}\\\labheadrule
\endhead\endfoot\endlastfoot
matched coupled & 0 & 512 & 3.8481 & 1425408 & 0.6258\\\labrowrule
matched reference & 0 & 512 & 4.4615 & 1425408 & 0.6322\\\labrowrule
dual same & 0 & 512 & 4.6713 & 1425408 & 3.0594\\\labrowrule
dual cross & 0 & 512 & 4.7360 & 1425408 & 3.1242\\\labrowrule
matched coupled & 1 & 512 & 4.2957 & 1425408 & 0.6172\\\labrowrule
matched reference & 1 & 512 & 3.0041 & 1425408 & 0.6281\\\labrowrule
dual same & 1 & 512 & 3.2159 & 1425408 & 2.5544\\\labrowrule
dual cross & 1 & 512 & 3.2238 & 1425408 & 3.0892\\\labrowrule
matched coupled & 2 & 512 & 3.1163 & 1425408 & 0.5997\\\labrowrule
matched reference & 2 & 512 & 3.7042 & 1425408 & 0.6192\\\labrowrule
dual same & 2 & 512 & 4.0769 & 1425408 & 3.1014\\\labrowrule
dual cross & 2 & 512 & 3.4241 & 1425408 & 3.1089\\\labrowrule
matched coupled & 3 & 512 & 3.5499 & 1425408 & 0.5875\\\labrowrule
matched reference & 3 & 512 & 3.2977 & 1425408 & 0.6525\\\labrowrule
dual same & 3 & 512 & 4.2964 & 1425408 & 2.8444\\\labrowrule
dual cross & 3 & 512 & 4.0268 & 1425408 & 2.5778\\\labrowrule
matched coupled & 4 & 512 & 4.2231 & 1425408 & 0.5733\\\labrowrule
matched reference & 4 & 512 & 4.0793 & 1425408 & 0.6386\\\labrowrule
dual same & 4 & 512 & 3.1207 & 1425408 & 2.5414\\\labrowrule
dual cross & 4 & 512 & 3.6210 & 1425408 & 2.5569\\\labrowrule
matched coupled & 5 & 512 & 4.4024 & 1425408 & 0.6092\\\labrowrule
matched reference & 5 & 512 & 4.5246 & 1425408 & 0.6350\\\labrowrule
dual same & 5 & 512 & 3.4434 & 1425408 & 2.9411\\\labrowrule
dual cross & 5 & 512 & 3.2719 & 1425408 & 3.1633\\\labrowrule
matched coupled & 6 & 512 & 3.9370 & 1425408 & 0.5900\\\labrowrule
matched reference & 6 & 512 & 2.6412 & 1425408 & 0.6128\\\labrowrule
dual same & 6 & 512 & 4.7978 & 1425408 & 2.7264\\\labrowrule
dual cross & 6 & 512 & 4.8915 & 1425408 & 2.6683\\\labrowrule
matched coupled & 7 & 512 & 4.9118 & 1425408 & 0.6233\\\labrowrule
matched reference & 7 & 512 & 3.6432 & 1425408 & 0.5928\\\labrowrule
dual same & 7 & 512 & 3.9596 & 1425408 & 2.5539\\\labrowrule
dual cross & 7 & 512 & 4.0896 & 1425408 & 2.9739\\\labrowrule
matched coupled & 8 & 512 & 3.2364 & 1425408 & 0.6192\\\labrowrule
matched reference & 8 & 512 & 3.9781 & 1425408 & 0.5989\\\labrowrule
dual same & 8 & 512 & 2.9151 & 1425408 & 3.1275\\\labrowrule
dual cross & 8 & 512 & 2.7696 & 1425408 & 2.5744\\\labrowrule
matched coupled & 9 & 512 & 4.2833 & 1425408 & 0.6158\\\labrowrule
matched reference & 9 & 512 & 3.5071 & 1425408 & 0.6181\\\labrowrule
dual same & 9 & 512 & 3.9254 & 1425408 & 3.0117\\\labrowrule
dual cross & 9 & 512 & 3.3298 & 1425408 & 2.5622\\\labrowrule
matched coupled & 10 & 512 & 4.0112 & 1425408 & 0.6306\\\labrowrule
matched reference & 10 & 512 & 3.8331 & 1425408 & 0.6461\\\labrowrule
dual same & 10 & 512 & 4.1667 & 1425408 & 2.5375\\\labrowrule
dual cross & 10 & 512 & 3.6298 & 1425408 & 2.8608\\\labrowrule
matched coupled & 11 & 512 & 3.0442 & 1425408 & 0.6325\\\labrowrule
matched reference & 11 & 512 & 2.1104 & 1425408 & 0.6019\\\labrowrule
dual same & 11 & 512 & 3.2415 & 1425408 & 2.4108\\\labrowrule
dual cross & 11 & 512 & 3.5565 & 1425408 & 2.8286\\\labrowrule
\end{longtable}

%% file: sections/v2_m1_coverage.tex
\begin{longtable}{P{0.170601}P{0.163781}P{0.178146}P{0.066730}P{0.135707}P{0.126394}P{0.158642}}
\caption{Complete empirical outcomes across all 92 evaluated runs in the matched-substrate learned-optimizer (\texttt{OPEN}) and extended-horizon curriculum comparison cohort (18 condition summaries across paired seed blocks).}\label{tab:v2_m1_coverage}\\\labtoprule
\labh{Task} & \labh{Horizon} & \labh{Method / regime} & \labh{n} & \labh{Initial $J_{00}$} & \labh{Final $J_{11}$} & \labh{Net gain}\\\labheadrule
\endfirsthead\labtoprule
\labh{Task} & \labh{Horizon} & \labh{Method / regime} & \labh{n} & \labh{Initial $J_{00}$} & \labh{Final $J_{11}$} & \labh{Net gain}\\\labheadrule
\endhead\endfoot\endlastfoot
wind flocking & 131,120 & open / coupled & 12 & -0.5570 & 0.7860 & +1.3431\\\labrowrule
wind flocking & 131,120 & open / decoupled & 12 & -0.5570 & 0.6182 & +1.1752\\\labrowrule
wind flocking & 1,048,640 & disco103 / coupled & 5 & -0.5095 & 1.2071 & +1.7167\\\labrowrule
wind flocking & 1,048,640 & disco103 / decoupled & 5 & -0.5095 & 1.2323 & +1.7418\\\labrowrule
wind flocking & 1,048,640 & disco103 / curriculum & 5 & -0.5095 & 1.1970 & +1.7065\\\labrowrule
wind flocking & 1,048,640 & mappo / coupled & 5 & -0.5158 & 1.2912 & +1.8070\\\labrowrule
wind flocking & 1,048,640 & mappo / decoupled & 4 & -0.5326 & 1.3051 & +1.8377\\\labrowrule
wind flocking & 1,048,640 & mappo / curriculum & 4 & -0.5326 & 1.2714 & +1.8040\\\labrowrule
wind flocking & 1,048,640 & open / coupled & 4 & -0.5326 & 1.2308 & +1.7634\\\labrowrule
wind flocking & 1,048,640 & open / decoupled & 4 & -0.5326 & 0.9907 & +1.5233\\\labrowrule
navigation & 1,048,640 & disco103 / coupled & 4 & -0.3491 & 1.7881 & +2.1372\\\labrowrule
navigation & 1,048,640 & disco103 / decoupled & 4 & -0.3491 & 1.7820 & +2.1311\\\labrowrule
navigation & 1,048,640 & disco103 / curriculum & 4 & -0.3491 & 1.6832 & +2.0323\\\labrowrule
navigation & 1,048,640 & mappo / coupled & 4 & -0.4091 & 1.6819 & +2.0910\\\labrowrule
navigation & 1,048,640 & mappo / decoupled & 4 & -0.4091 & 1.7114 & +2.1205\\\labrowrule
navigation & 1,048,640 & mappo / curriculum & 4 & -0.4091 & 1.7339 & +2.1430\\\labrowrule
navigation & 1,048,640 & open / coupled & 4 & -0.4091 & 1.6451 & +2.0542\\\labrowrule
navigation & 1,048,640 & open / decoupled & 4 & -0.4091 & 1.6145 & +2.0236\\\labrowrule
\end{longtable}